\pdfoutput=1
\documentclass[a4paper,fleqn,usenatbib]{mnras}

\usepackage{newtxtext,newtxmath}

\usepackage[T1]{fontenc}
\usepackage{ae,aecompl}

\usepackage{graphicx}	
\usepackage{amsmath}	
\usepackage{amssymb}	
\usepackage{longtable}
\usepackage{lscape}

\title{Optical and radio properties of extragalactic radio sources with recurrent jet activity}

\author[A. Ku\'zmicz et al.]{
A. Ku\'zmicz,$^{1,2,3}$\thanks{E-mail: cygnus@oa.uj.edu.pl}
M. Jamrozy,$^{2}$
D. Kozie{\l}-Wierzbowska$^{2}$
and M. We\.zgowiec$^{2}$
\\
$^{1}$Center for Theoretical Physics, Polish Academy of Sciences, Al. Lotnik\'ow 32/46, 02-668 Warsaw, Poland\\
$^{2}$Astronomical Observatory, Jagiellonian University, ul. Orla 171, 30-244 Krakow, Poland\\
$^{3}$Queen Jadwiga Astronomical Observatory in Rzepiennik Biskupi, 33-163 Rzepiennik Strzy\.zewski, Poland\\
}

\date{Accepted XXX. Received YYY; in original form ZZZ}

\pubyear{2016}

\begin{document}
\label{firstpage}
\pagerange{\pageref{firstpage}--\pageref{lastpage}}
\maketitle

\begin{abstract}
We present a sample of 74 radio sources with recurrent jet activity. The sample consists of 67 galaxies, 2 quasars, and 5 unidentified sources, selected from the published data as well as newly recognized. The sample's redshift range is 0.002 $<$ z $<$ 0.7 and the size of inner and outer structures varies from 0.02 to 4248\,kpc. We analyse the optical and radio properties of the sample and compare them with the characteristics of ordinary one-off FRII radio sources. With the help of stellar population modelling we derive black hole masses and stellar masses of host galaxies of 35 restarting radio sources, finding that the black hole masses in restarting radio sources are comparable to those of typical single cycle FRII radio sources. The obtained median values of log M$\rm_{BH}$ are 8.58 M$_{\odot}$ and 8.62 M$_{\odot}$, respectively. Unlike the black hole masses, the stellar masses in restarting radio sources tend to be smaller than in the FRII sources. Although the stellar populations of the hosts of recurrent activity sources are dominated by old stars, a significant fraction of young stars can be observed as well. Based on the SDSS photometric observations, we also analyse the morphology of the host galaxies and obtained significantly smaller concentration indices for the restarting radio sources when compared to the classical FRII hosts. This effect can be interpreted as a result of frequent merger events in the history of host galaxies of restarting radio sources.  
\end{abstract}




\section{Introduction}
The classic extragalactic radio sources have been investigated for much over a half of century; for instance, \cite{Baade1954} found an optical counterpart for the Cygnus A source as long ago as 1954). The most widely assumed underlying model of radio sources comprises a fast rotating supermassive black hole (SMBH) accompanied by an accretion disk. The transfer of disk energy to the external lobes is accomplished through powerful relativistic jets \citep{Scheuer1974}. The jet activity phases are rather brief ($\sim$10$^8$\,yrs) in comparison with the entire life time of the host galaxy ($\sim$10$^{10}$\,yrs). However, the activity of a jet can restart after a period of silence. This phenomenon depends strictly on the physical conditions in the vicinity of the central SMBH. The presence of sources of peculiar character, classified neither as FRI nor FRII (\citealt{Fanaroff1974}), could indicate a recurring AGN activity. The active galactic nuclei in general can be classified into a number of separate categories; among others, the radio galaxies with two or more pairs of lobes extending from the core along the same axis, called double-double radio sources (DDRS; \citealt{Schoenmakers2000}). This group is rather sparse, with just about 45 objects recognized to date (for a review and references, see \citealt{Saikia2009, Nandi2012}). There are DDRSs of much different sizes, from merely scores of parsecs up to over one megaparsec. Besides the axial symmetric sources there is  another type of recurrent activity radio galaxies. These are objects exhibiting large-scale diffuse relic radio emission that is due to an earlier cycle of activity. Such relics can be seen around compact powerful young radio sources (e.g. \citealt{Baum1990}). A separate group are so-called `X-shaped' radio sources, which are also regarded as recurrent radio sources (e.g. \citealt{Merritt2002}; \citealt{Saripalli2009}). We are not going to analyse these sources here, as their optical properties were studied already by \cite{Mezcua2011,Mezcua2012}.

The fact that in some radio galaxies two or even three pairs of lobes can be observed implies that the time required for the jet flow to cease is shorter than that for the outer lobes to fade. Thus, as the energy transported by the jets is able to last through the quiescent phase within their extended lobes, the radio galaxies can hold memory of the previous AGN activity. \cite{Konar2013} determined the duration of the inactive period for some DDRSs to be of the order of several million years. Interestingly enough, this time is too long to be explained in the frame of the currently existing theoretical models that describe properties of an AGN (e.g. accretion disk instability with the idle time of about 10$^4$ years; \citealt{Czerny2009}). On the other hand, it is too short for a new disk to form. Therefore, alternative theories have been formulated. The interruptions in the jet production mechanism could have been possibly caused by refuelling of the central engine. Liu et al. \citep{Liu2004,Liu2003,Liu2012} suggested that this kind of objects may arise due to interactions of SMBH binaries with the accretion disk. In such a picture, the secondary black hole (BH) drifts towards the centre, disrupting the inner parts of the accretion disk. After the both black holes merge, the gap widens, effectively stopping the process of jet formation. Such a behaviour could be expected in the case of the blazar OJ287 \citep{Valtonen2008}, which possesses two SMBHs, the smaller of which orbits the larger companion, surrounded by an accretion disk, with a period of about 11 years. The jet activity may start again at a later time, triggered by new matter flowing into the inner region. Consequently, a DDRS is created. The accepted belief that the multiple pair of lobes are produced by recurrent jet activity was questioned by \cite{Wagner2011}, who noted that such objects could also be produced by random emergence of radio plasma from an inhomogeneous, porous ISM. However, in this scenario we cannot expect an ideally symmetric interstellar medium (ISM) on both sides of the central AGN. This in turn would make it difficult to explain the existence of DDRSs with linearly symmetric inner lobes.

The AGNs with repeated jet activity could affect both the parent galaxy and the diffuse extended radio structure. Therefore, some properties of the objects with multiple activity cycles can be different from the radio galaxies with single activity.
 
\cite{Wagner2011} found that there is a considerable transfer of energy and momentum to the ISM, and that the jets with powers of $10^{43}$ -- $10^{46}$ erg s$^{-1}$ can inhibit star formation in the galactic core. This would influence the evolution of the galaxy. Therefore, the repeated jet activity could have implications for feedback processes of the parent galaxy.

\cite{Subrahmanyan1996} postulated that recurrent activity influences the linear size of radio sources. Therefore, at least some of the giant radio galaxies could become quite large because their central AGNs experienced several jet activity cycles. The jets of restarted cycles move not in the original intergalactic medium (IGM) but in a less dense environment of the primordial lobes, which allows them to reach further distances in a shorter time. Nevertheless, the mechanism underlying formation of the inner lobes is not yet fully understood. There are two plausible models suggested in the literature that could explain this. In the first model, called the `classical FRII model', the inner lobes are formed in the same way as the lobes of typical `single-cycle' radio galaxies. The inner lobes propagate through a denser medium than could be expected from purely synchrotron-emitting plasma, because thermally emitting material is drawn into the outer lobes of DDRSs during their growth and quiescent phases \citep{Kaiser2000}. The second model, called the `bow-shock model', assumes that the inner lobes are created by reaccelerating of the outer cocoon particles at the bow-shock created by almost ballistically moving jet heads \citep{Brocksopp2007,Brocksopp2011,Safouris2008}. Both the `classical FRII' and the `bow-shock' models may need to be employed simultaneously to fully describe the dynamics and the structure of inner lobes.

Surprisingly enough, only recently it was realised that the jet activity can be a multiple phenomenon. The main reason for this is that such objects are not too frequent among radio galaxies. Furthermore, it is not to easy to recognize a restarting radio source, since the radio structures usually come from separate activity cycles and have significantly different surface brightnesses, angular extents, and spectral profiles. The ideal instance of a DDRS would be an object with bright detached pairs of lobes but most often the structures produced by consecutive cycles merge together. \cite {Parma1999} selected a class of radio galaxies, which possess spectra that steepen outwards, from the core to the outer edge of lobes. Most of them are narrow-angle-tail (NAT) or wide-angle-tail (WAT) radio galaxies, while within these two types of objects we can still find sources with multiple jet activity. In order to distinguish them and understand the cycles and the phases of interruption of the jet flow, it is crucial to determine ages of charged particles in different regions within the radio lobes. In order to fully recognize the recurrent activity phenomenon a statistical study of an extended sample is essential. 

In this paper, we present the first systematic study of extragalactic radio sources with structures showing recurrent jet activity. We analyse the optical and radio properties of a large sample of restarting radio sources and compare them with characteristics of ordinary FRII radio galaxies, in order to learn more about the causes underlying episodic jet activity.
In the next section, we briefly describe the source samples used for the analysis. The observations, data reduction, and processing are described in the third section, while in the fourth section we review and discuss the results. The conclusions are presented at the end of the paper. Throughout the paper, a flat vacuum-dominated universe with $\Omega$$\rm_{m}$=0.27, $\Omega$$\rm_{\Lambda}$=0.73 and $H$$\rm_{0}$=71 km s$^{-1}$ Mpc$^{-1}$ is assumed.

\section{The sample}

The recent studies by \citet{Saikia2009} and \citet{Nandi2012} significantly extend the number of sources which show multiple jet activity. It was noticed that multi-cycle jet activity is not just specific to giant objects, but is manifested also by smaller radio galaxies (with outer structures having sizes of only several hundred kiloparsecs). Nevertheless, such objects are not too well studied and still a number of physical parameters, e.g. the duration of an inactive phase, need better estimates.

Our sample of restarting radio sources contains 74 objects, including 67 galaxies, 2 quasars, and 5 unidentified sources. The selection was based mostly on the published data, but we also included new sources, recognized recently by the authors of this paper. The redshifts for our sample objects range from 0.002 to 0.7. The only exception is J2107$+$2331, whose redshift is as high as z=2.48. The radio structures show a wide range of projected linear sizes -- from 0.02 to 876\,kpc for the inner to 48 to 4248\,kpc for the outer lobes. In our sample there are 8 newly recognized restarting radio sources: J0042-0613, J0504+3806, J0914+1006, J0924+0602, J1004+5434, J1021+1216, J1520-0546, J1528+0544. All of them were identified using maps from the 1.4\,GHz Faint Images of the Radio Sky at Twenty centimeters (FIRST; \citealt{Becker1995}) survey and the NRAO VLA Sky Survey (NVSS; \citealt{Condon1998}). The redshifts of these objects are within the range from z=0.04 to z=0.31. A half of the newly recognized restarting radio sources are giants with projected linear sizes larger than 0.7 Mpc (taking into account the sizes of the outer radio lobes). In most of the newly discovered sources the inner radio lobes have medium linear sizes, but there is one source J1528+0544 with a very compact (of 15 kpc) inner structure as well as one source J1021+1216 with very extended (876 kpc) inner doubles.
In our sample, we consider also two X-shaped radio galaxies, i.e. J0009+1244 and J1513+2607. While we do not analyse this type of sources in this paper in general, these two sources are of peculiar type. Besides the usual X-shape morphology, they possess an additional pair of lobes in their central region.
The sample presented in this paper is not complete, yet it is quite representative of the class. It covers a wide range of physical parameters, thus it is sufficiently large and diverse to allow for performing an analysis of their fundamental properties.\\
There are a lot of candidates in the literature for a restarting radio source with structures that still have to be confirmed as multi-episodic (e.g. \citealt{Nandi2012}). Most of them either do not have optical identifications, or the resolution of the available radio maps is not sufficient to properly reveal their structure. We also expect a number of new multi-cycle source detections in the next few years from deep survey observations with new low-frequency radio telescopes (e.g. the Low Frequency Array; \citealt{VanHaarlem2013}, the Long Wavelength Array; \citealt{Ellingson2009}, and the Murchison Widefield Array; \citealt{Lonsdale2009}).

We divided the radio sources from our sample according to their radio morphology into two classes: typical double-double radio sources, where two pairs of radio lobes are clearly visible (class A) and radio sources with prominent inner structure, surrounded by diffuse outer structures (class B). We classified 63 radio sources studied in this paper as class A objects, while only 11 belong to the class B.

The restarting radio sources and their basic properties are listed in Table \ref{ddrs}, where the newly recognized radio sources are denoted with boldface. The radio maps are presented in Appendix\,A -- class A radio sources and Appendix\,B -- class B.

As a comparison sample, we used~the~FRII-type radio galaxies from \citet{Koziel2011}, where the optical and radio data were provided. This sample contains 401\,FRII radio galaxies with a wide range of radio powers and sizes of radio structures. Ten of the FRII radio galaxies from the comparison sample show a double-double radio structure, and therefore were not used. Instead, they were included in the sample of restarting sources. As a result, the final comparison sample consists of 391 sources.

\begin{figure}
    \includegraphics[width=\columnwidth]{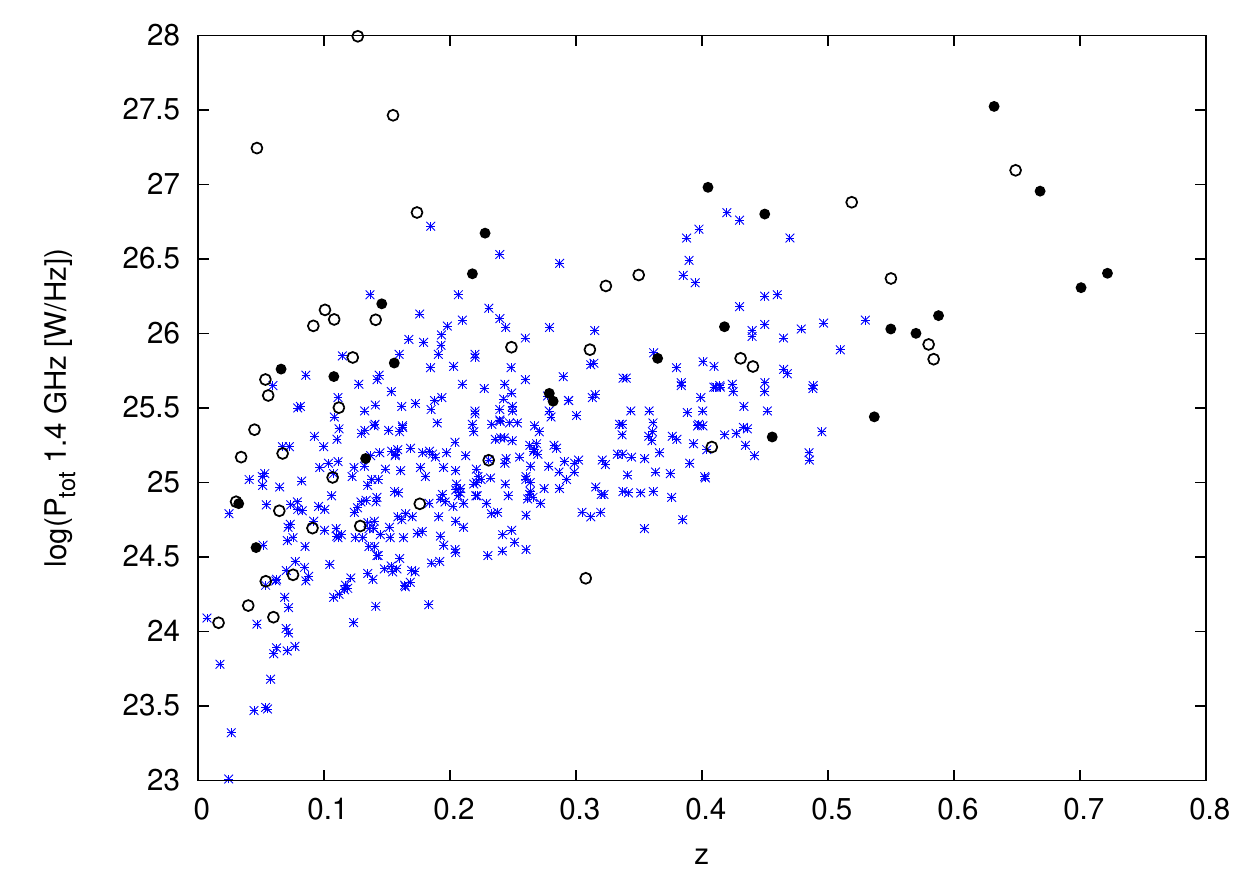} 
\caption{Sample of recurrent jet activity radio sources (black open and filled circles) and the comparison sample (blue asterisks) on the 1.4 GHz total radio 
luminosity--redshift plane. The recurrent radio sources with available spectra are marked as open black circles.}
\label{P_z}
\end{figure}

Radio sources from both the samples are presented in Figure\,\ref{P_z}, where the relation between the distribution of the 1.4~GHz total radio luminosity and the redshift is plotted. The source with exceptionally high redshift (J2107$+$2331) is not shown in this figure.

\section{Observational data}
In our analysis of recurrent jet activity sources we use both optical and radio data.

\subsection{Optical evidence}
\label{Observations}

The optical data are taken mainly from the SDSS Data Release 10 (DR10; \citealt{Ahn2014}). The photometric data were available for 53\,radio sources and the spectroscopic data were available for 30\,objects.
\\
Spectroscopic data of further four objects~(J0301+3512, J1651+0459, J1844+4533, J2048+0701) were taken from the archives of the Telescopio Nazionale Galileo (TNG) and the 5-meter Hale telescope at the Palomar Observatory. The observations and reduction procedures are described in \citet{Ho1995} and \citet{Buttiglione2009}.

Additional spectra for six objects were obtained with our dedicated spectroscopic observations using the South African Astronomical Observatory (SAAO) telescopes and the William Herschel Telescope (WHT).
Details of the WHT observations and data reduction are presented below.
Spectra of J1520$-$0546 and J1528$+$0544 were obtained with the 1.9-m SAAO telescope equipped with the Cassegrain Grating Spectrograph. The optical spectra and description of observations are provided in \citet{Machalski2007}. 

\begin{figure}
    \includegraphics[width=0.95\columnwidth]{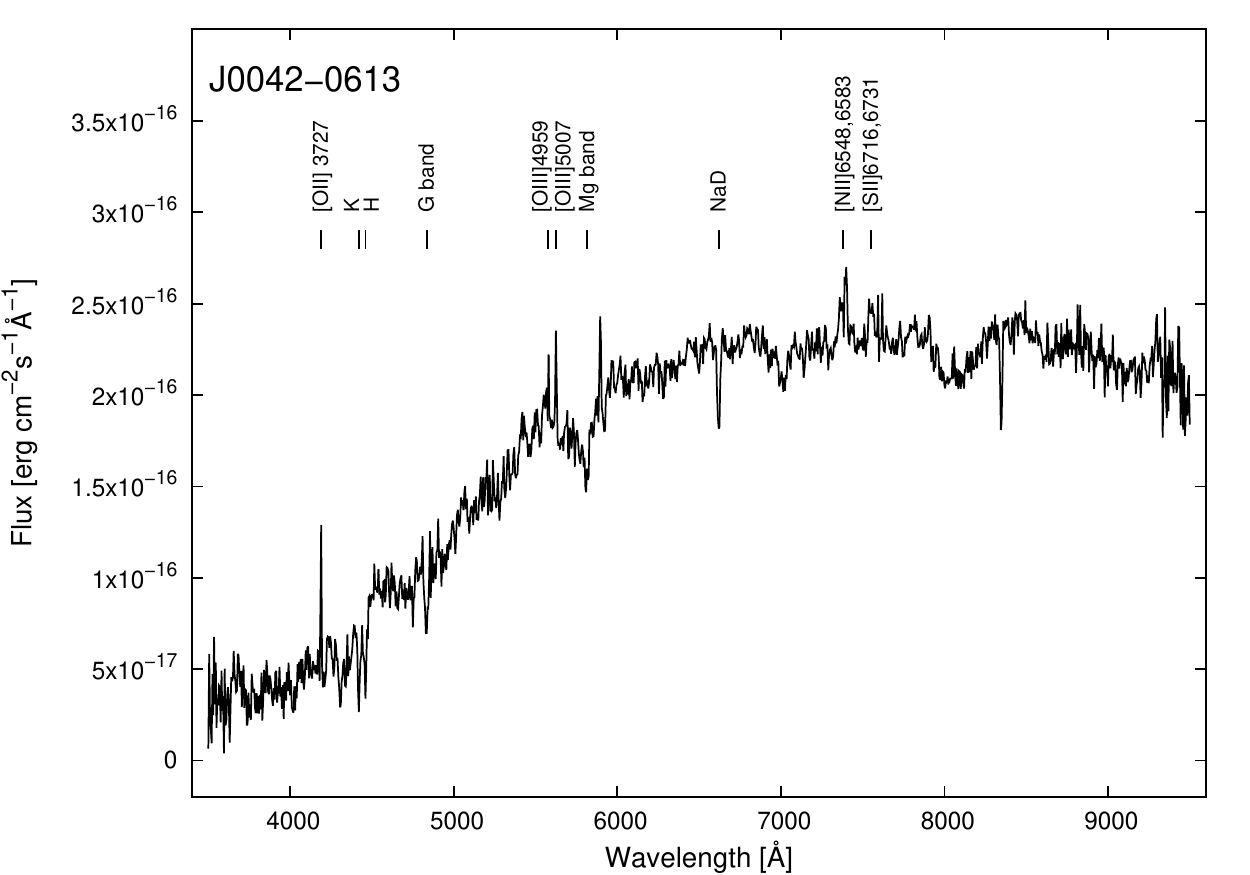}\\
    \includegraphics[width=0.95\columnwidth]{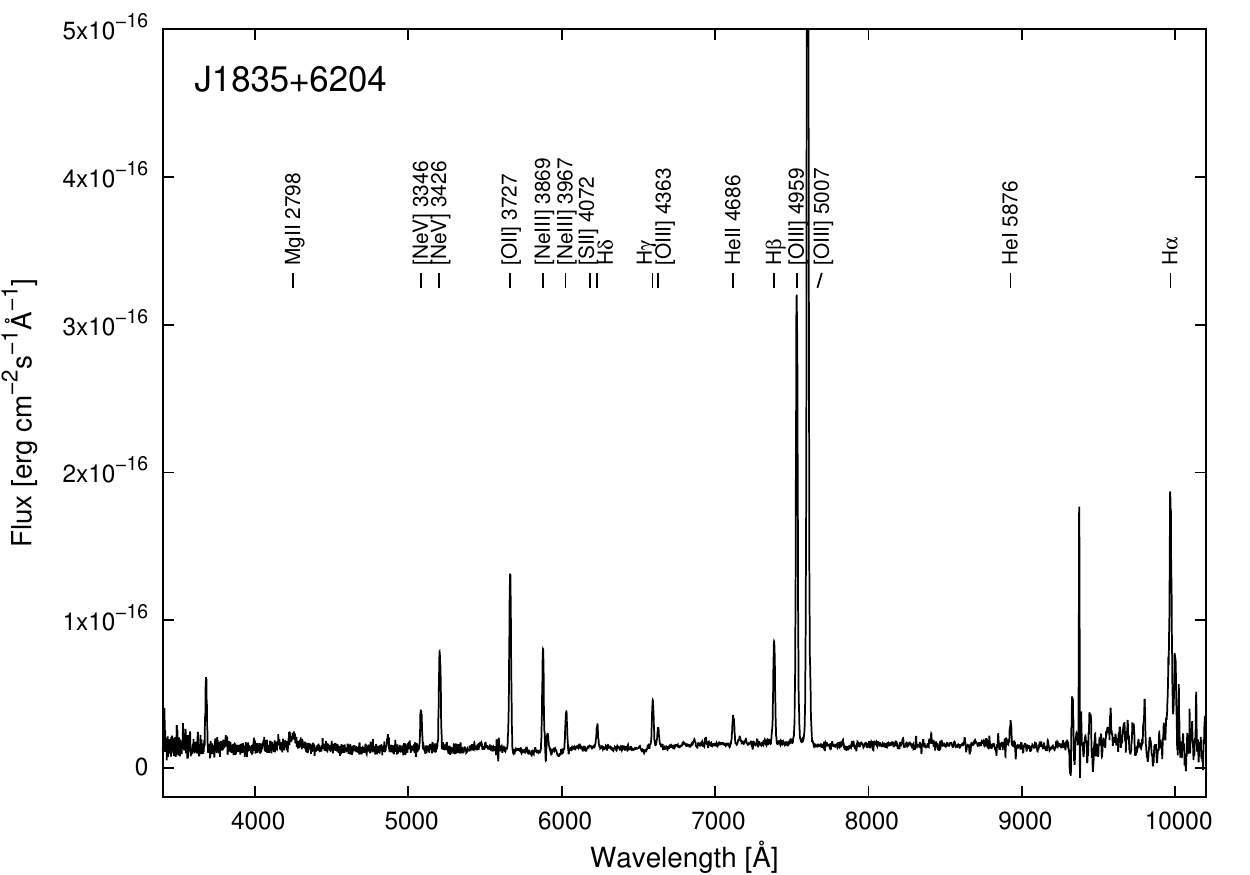}\\
    \includegraphics[width=0.95\columnwidth]{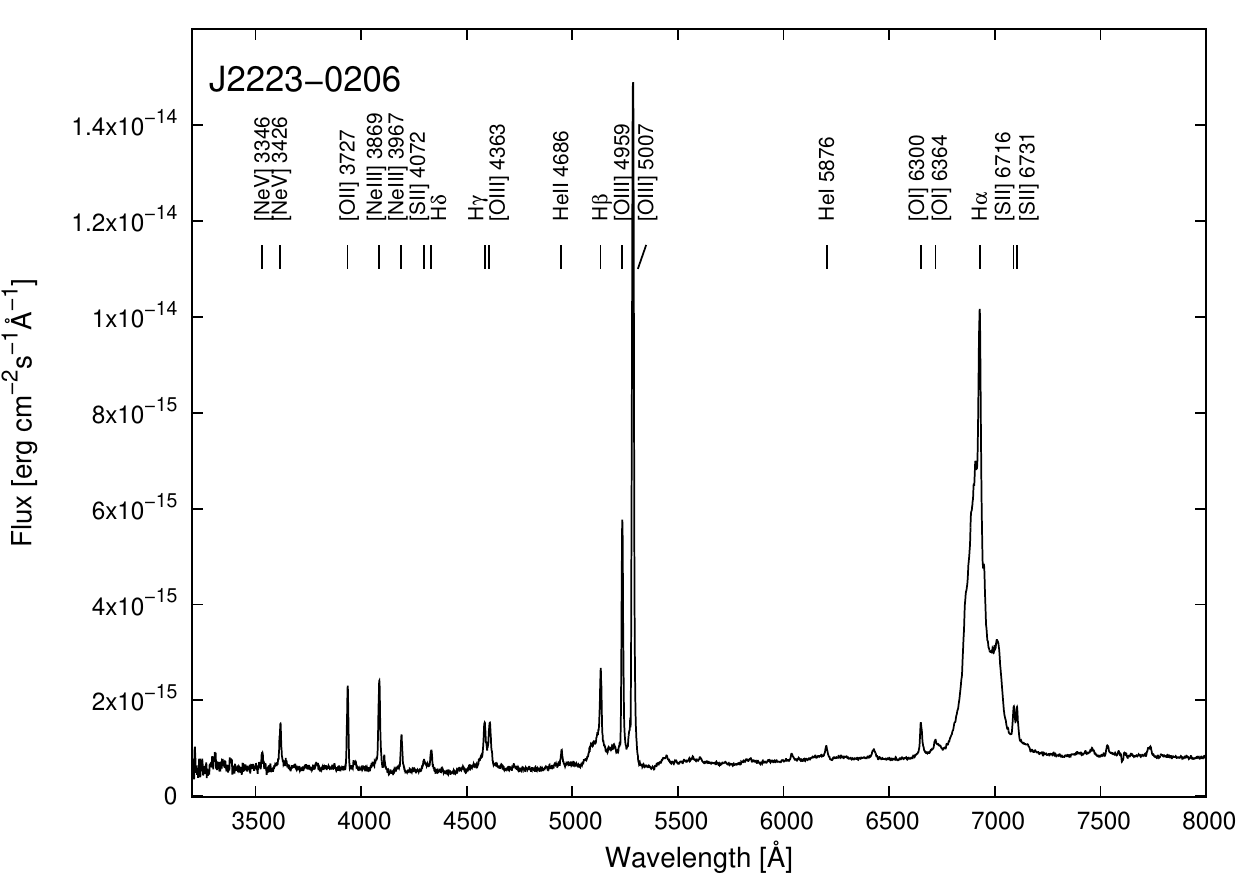}\\
    \includegraphics[width=0.95\columnwidth]{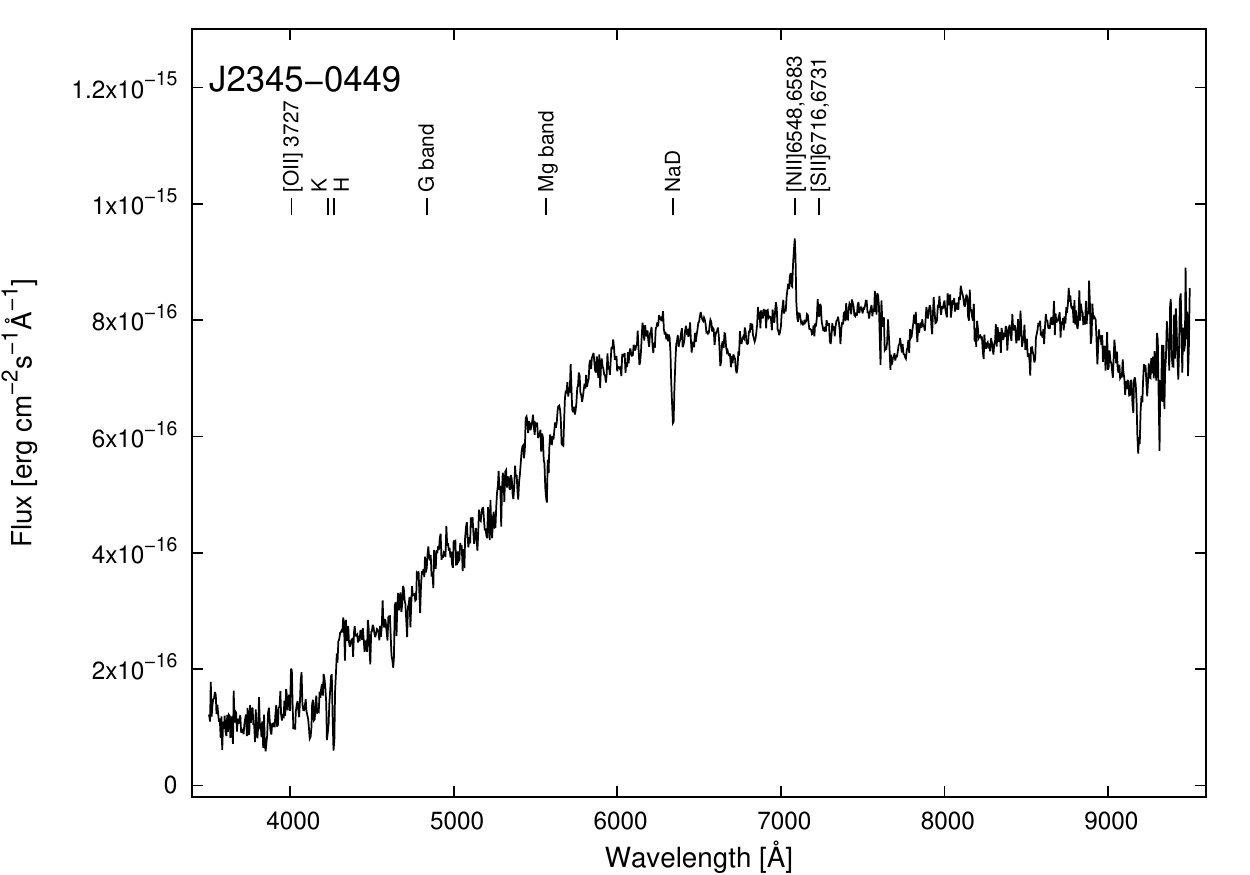}\\
\caption{The optical WHT spectra of J0042$-$0613, J1835$+$6204, J2223$-$0206, and J2345$-$0449 restarting radio galaxies.}
\label{w1}
\end{figure}

\noindent
\underline{WHT spectra:}
Spectral observations of four DDRGs, J0042$-$0613, J1835$+$6204, J2223$-$0206, and J2345$-$0449 were carried out with the WHT at the Roque de los Muchachos Observatory in La Palma, Spain. The observations were obtained in the service mode on July 29th, 2009. The long-slit spectra were taken using the Intermediate Dispersion Spectrograph and Imaging System Double-Armed Spectrograph, which permit simultaneous observations in both blue and red filters. The slit of a width of 2\arcsec was centred on the nuclei of the sources and positioned along the jet axis of the radio galaxy. The integration time was split into several exposures to reduce the influence of cosmic rays. Spectra of two spectrophotometric standards, i.e. SP2011+065 and SP0031-124 were obtained for a proper intensity calibration. Arc lamp spectra were taken before and/or after target observations to allow for an accurate wavelength calibration. The average seeing during the observing run was about 1.2\arcsec. The WHT spectra of these restarting radio galaxies have not been published to date and we present them in Figure \ref{w1}.

The WHT data were reduced using the IRAF NOAO packages{\footnote{http://iraf.noao.edu}}. Reduction steps were performed separately for each spectral range. The master bias frame was created by averaging all the bias frames obtained during the observing night and subtracted from the science frames. The master flat field frame was also created and all 2D-science frames were corrected for flat field. Then, the cosmic rays were removed from the science exposures. Wavelength calibration was performed using ArNe lamp exposures and verified by using sky lines. Next, a correction for optical distortion was applied. The contribution from the sky was determined from the sky regions at both sides of the resulting spectrum and subtracted. The 1D-spectra extraction was performed using the APALL task. The scientific exposures were flux-calibrated using exposures of the suitable spectrophotometric standard stars. In order to obtain the resulting spectra with a better S/N ratio, all the calibrated one-dimensional spectra of a given galaxy were combined.

\begin{figure}
    \includegraphics[width=\columnwidth]{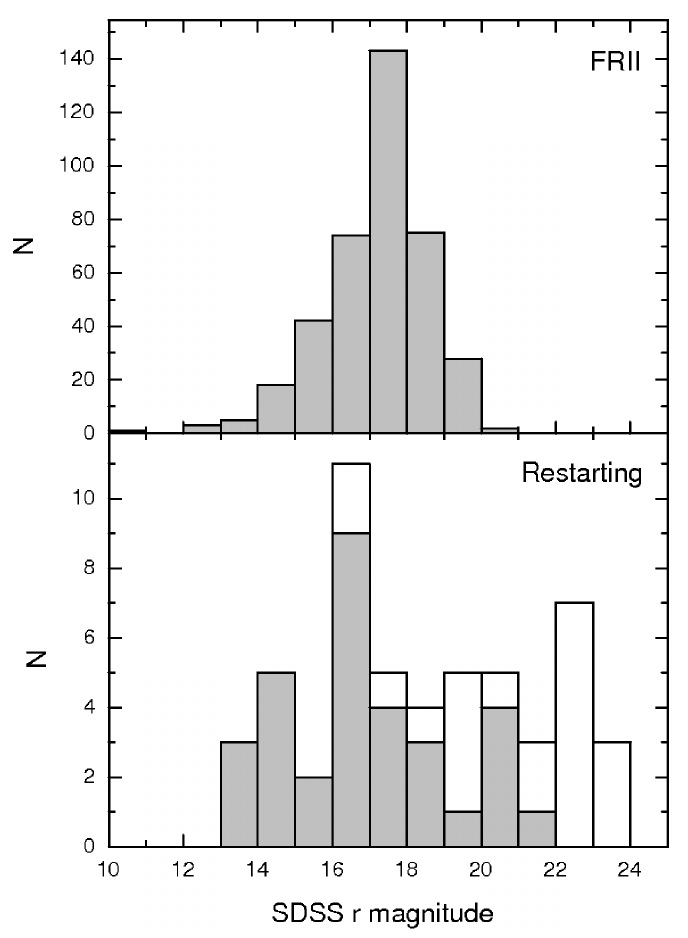}
\caption{SDSS r magnitude distribution for FRII radio galaxies ({\bf top}) and restarting radio galaxies ({\bf bottom}). The radio galaxies with available optical spectra are plotted as a solid grey boxes, both for the restarting sample and the comparison one.}
\label{mag}
\end{figure}

The spectroscopic data of good quality are available only for 37 galaxies of the 74 restarting radio sources. Despite that, the galaxies with spectroscopic data are representative for the whole sample. In Figure \ref{P_z} we marked the galaxies with available optical spectra. It can be seen that the distribution of galaxies with spectra is uniform and spans a wide range of redshifts and total radio luminosities. In Figure \ref{mag} we plotted the distributions of SDSS r magnitudes for both the samples. The sample of FRII radio galaxies contains only objects with optical spectra and is limited to r band magnitudes brighter than 22. The distribution for restarting radio galaxies is more uniform and optical spectra are available for the brighter objects in sample, but the range of magnitudes for the galaxies with spectra is similar as for the comparison sample.

\subsection{Radio maps}
In our radio analysis we measured the angular sizes and flux densities of inner and outer radio lobes using mainly maps from the 1.4\,GHz FIRST and the NVSS radio surveys. For a few sources for which no data were available in NVSS/FIRST, or for which it was not possible to obtain precise measurements of angular sizes and flux densities using NVSS/FIRST, we took the relevant parameters from the literature given in Table \ref{ddrs}.\\    

\subsection{Data analysis}

The optical spectra of restarting radio sources have been used to measure the basic properties of the hosts of AGNs. We applied the stellar population synthesis code STARLIGHT\footnote {http://www.astro.ufsc.br/starlightst} \citep{CidFernandes2005} to model the observed spectrum. Stellar continuum was fitted basing on the superposition of 150\,stellar spectra templates (extracted from the evolutionary synthesis models of \citealt{Bruzual2003}) with various ages and metallicities. Apart from continuum fitting, the STARLIGHT modelling gives the information about such parameters like star formation, chemical enrichment, and velocity dispersion.\\
To determine BH masses, we used the M$_{BH}$ - $\sigma_*$ method. It is based on a tight correlation between the BH mass and the velocity dispersion of the stars in the galactic bulge ($\sigma_*$; \citealt{Ferrarese2000}, \citealt{Gebhardt2000}), which is described by the relation
\begin{equation}
      log\rm(\frac{M_{BH}}{M_{\odot}})=\alpha+\beta log\rm(\frac{\sigma_*}{\rm 200 km s^{-1}})
\label{eq1g}
\end{equation}
where the constants are $\alpha$ = 8.13$\pm$0.05 and $\beta$ = 5.13$\pm$0.34 (\citealt{Graham2011}).

For the two quasars (J0741$+$3112 and J0935$+$0204) we used their BH mass estimates based on the H$_\beta$ mass scaling relation given by \citet{Shen2008}.\\

The radio luminosity of the inner ($P\rm_{in}$) and outer ($P\rm_{out}$) radio lobes was calculated using the following formula given by \cite{Brown2001}
\begin{equation*}
   log P(WHz^{-1})=log S(mJy)-(1+\alpha) \cdot log(1+z)
 \end{equation*}
\vspace{-0.8cm}
\begin{equation}
\hspace{3cm}+2log(D_L(Mpc))+17.08 
 \label{eq1}
 \end{equation}
\noindent
where $\alpha$ is the spectral index\footnote{throughout the paper we use the convention $S_{\nu}\sim\nu^{\alpha}$} and $D_L$ is a luminosity distance. The flux densities of the inner ($S_{\rm in}$) and outer lobes ($S_{\rm out}$) of individual sources were measured in the available maps and then listed in Table \ref{ddrs}. Following e.g. \citet{Kellermann1988}, we adopted a typical spectral index value of $\alpha=-0.75$ for the inner and $\alpha=-1$ for the outer radio structures.\\

\section{Results}

In this section, we compare some of the basic characteristics (radio morphology, physical parameters of host galaxy) of restarting radio galaxies with those of typical radio sources using the optical, radio and infrared data described above.

\subsection{Optical properties}

Using the Starlight Synthesis Code, we modelled the stellar continuum in spectra of 35\,galaxies from our sample (except for two quasars) for which spectroscopic data were available. We obtained the information about a mixture of stellar populations present in a individual galaxies. It can be expressed by the light-fraction population vector $x_j$, which gives the percentage fraction of a galaxy light (luminosity) that comes from stars with a given age and metallicity (\citealt{CidFernandes2004}).

\begin{figure}
\centering
    \includegraphics[width=\columnwidth]{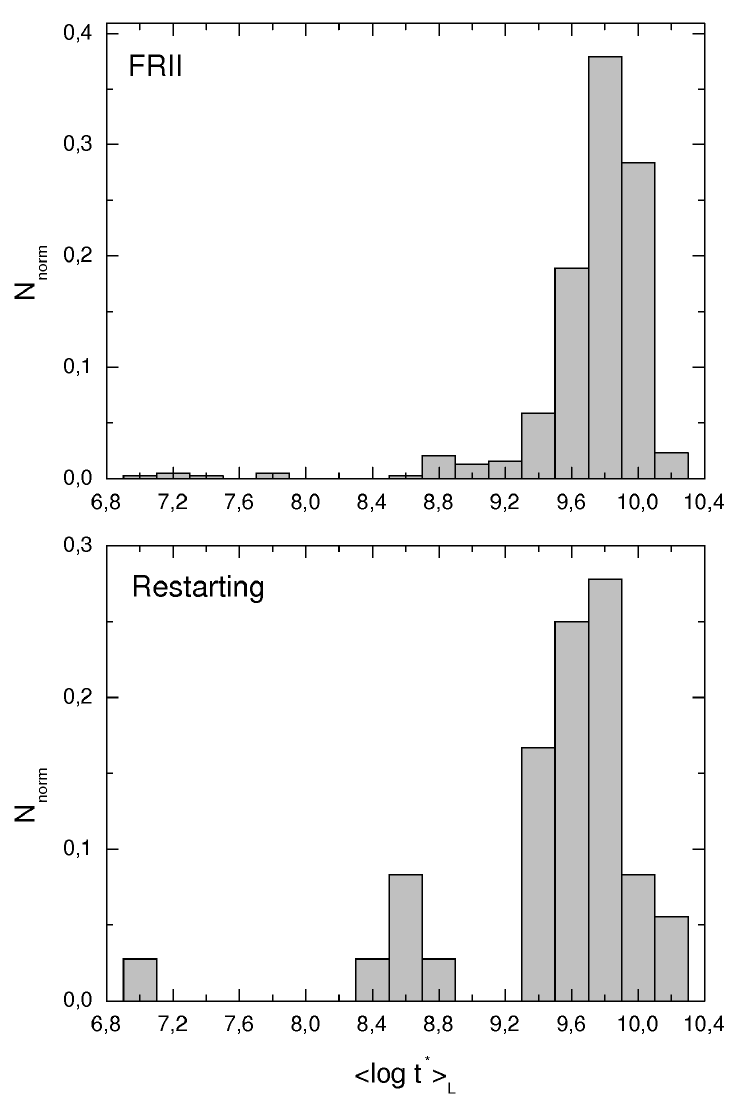} 
\caption{Distribution of light-weighted average age for all 35 restarting radio galaxies for which spectroscopic data are available ({\it bottom panel}), and single-cycle FRII radio galaxies ({\it top panel}). The normalised number, N$_{norm}$, is a number of radio galaxies in a particular bin divided by the total number of sources in the entire sample.}
\label{at_flux}. 
\end{figure}

\begin{figure}
\centering
    \includegraphics[width=\columnwidth]{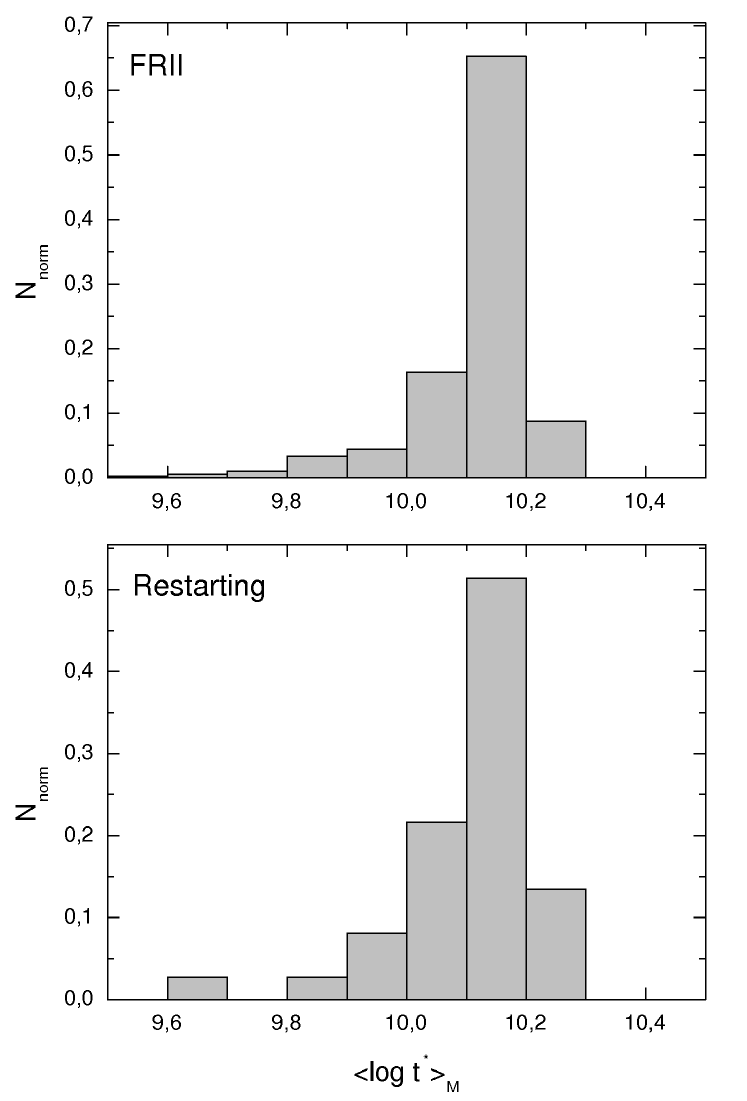} 
\caption{Distribution of mass-weighted average age for restarting radio galaxies ({\it bottom panel}), and for single-cycle FRII radio galaxies ({\it top panel}).}
\label{at_mass}
\end{figure}

According to \citet{CidFernandes2005}, we derived the light-weighted average age $\left <\log t^{\star}\right >_L$ and the mass-weighted average age $\left <\log t^{\star}\right >_M$, which are defined as:
\begin{equation}
\left <\log t^{\star}\right >_L=\sum_{j=1}^{N} x_j \log t_j \hspace{0.5cm} and \hspace{0.5cm} \left <\log t^{\star}\right>_M=\sum_{j=1}^{N} \mu_j \log t_j
\label{eq4}
\end{equation}
respectively, where $x_j$ is the modelled light-fraction population vector, and $\mu_j$ is the mass-fraction population vector obtained by using the model light-to-mass ratios. By definition, the two weighted average ages are sensitive to different stellar populations: $\left <\log t^{\star}\right >_L$ is sensitive to the presence of young stellar populations, while $\left <\log t^{\star}\right >_M$ is sensitive to the less luminous and older stellar populations. 
In Figures~\ref{at_flux} and~\ref{at_mass}, we compared the average ages $\left <\log t^{\star}\right >_L$ and $\left <\log t^{\star}\right >_M$ of the restarting radio galaxies and the comparison sample of typical FRII radio galaxies. The observed distribution of the light-weighted age is slightly different for the restarting radio galaxies and FRIIs. The $\left <\log t^{\star}\right >_L$ distribution for restarting radio galaxies peaks in the range 9.7 -- 9.8, as for the comparison sample, but there is a large fraction of younger stellar populations as well. The Kolmogorov-Smirnov two-sample test (K--S test) was performed to check whether the two samples reveal the same distribution. We obtained the probability value p=0.001, indicating that the sample of restarting radio sources and the comparison sample have statistically different distributions of $\left <\log t^{\star}\right >_L$. In the case of mass-weighted age, the observed distributions are similar. The K--S test for $\left <\log t^{\star}\right >_M$ returns a probability of 0.17 that the restarting and comparison sample radio sources have indistinguishable distributions. The results indicate that while the stellar populations in the restarting radio galaxies do include old stars, similarly as in the FRII radio galaxies, they also show a considerable amount of young stars.

\begin{figure}
    \includegraphics[width=\columnwidth]{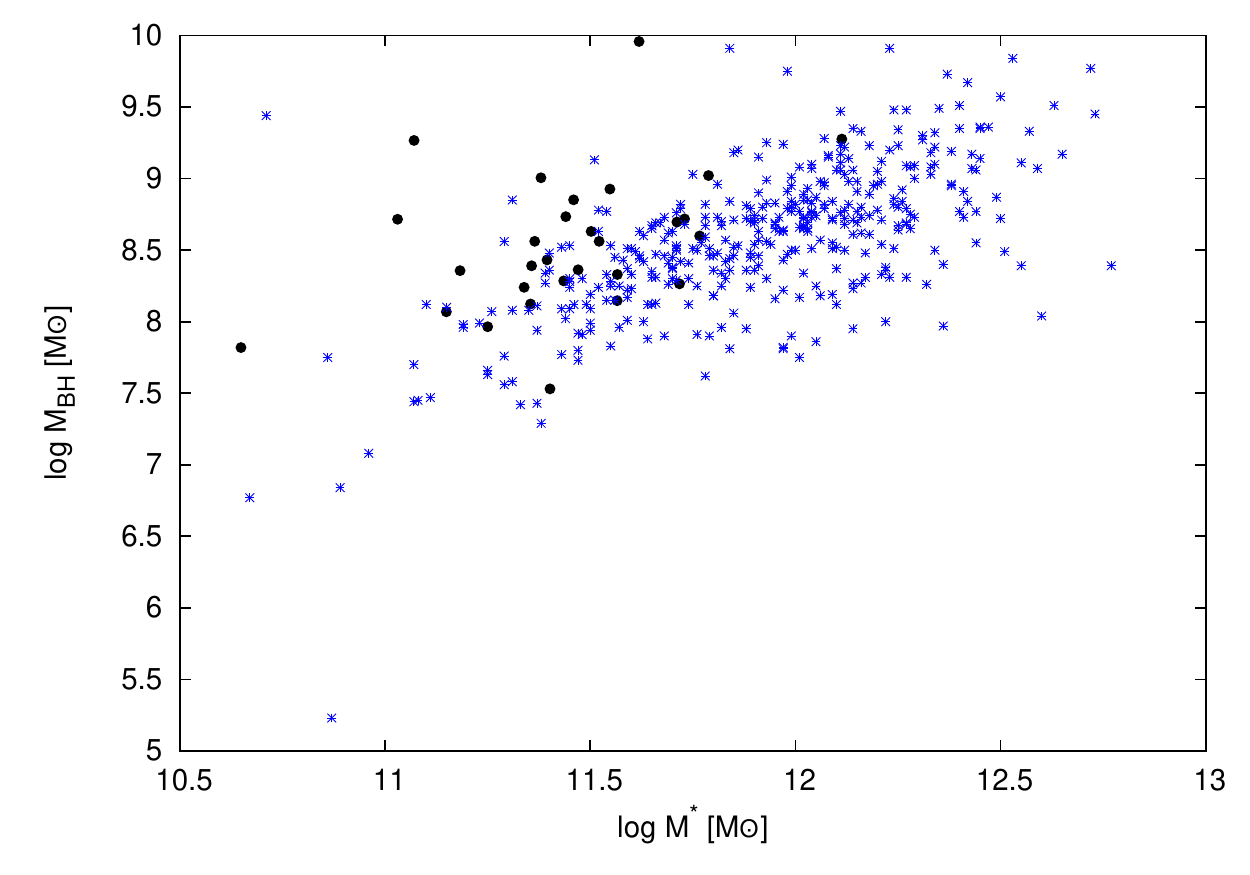} 
\caption{BH mass M$\rm_{BH}$ versus galaxy mass M$^*$. The restarting radio galaxies are denoted as black dots and the FRII radio galaxies from the comparison sample as blue asterisks.}
\label{MbhM}
\end{figure}

\begin{figure}
    \includegraphics[width=\columnwidth]{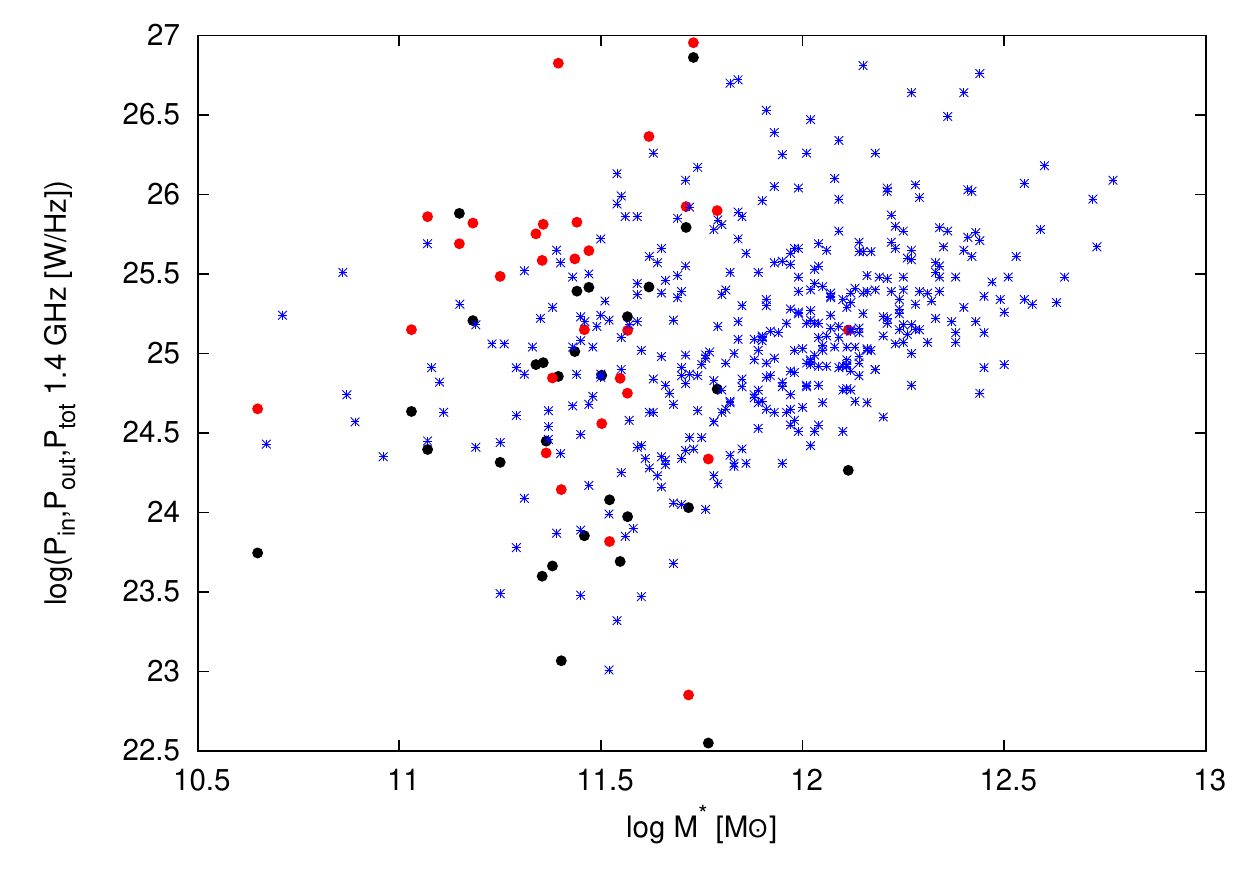} 
\caption{Relation between radio luminosity and the galaxy masses M$^*$ of the restarting and FRII galaxies. The inner radio structures are marked as black dots, the outer structures as red dots, and the FRII radio galaxies from the comparison sample are marked as blue stars. This labelling is used throughout further all similar figures in this paper.}
\label{PM}
\end{figure}

\begin{figure}
    \includegraphics[width=0.95\columnwidth]{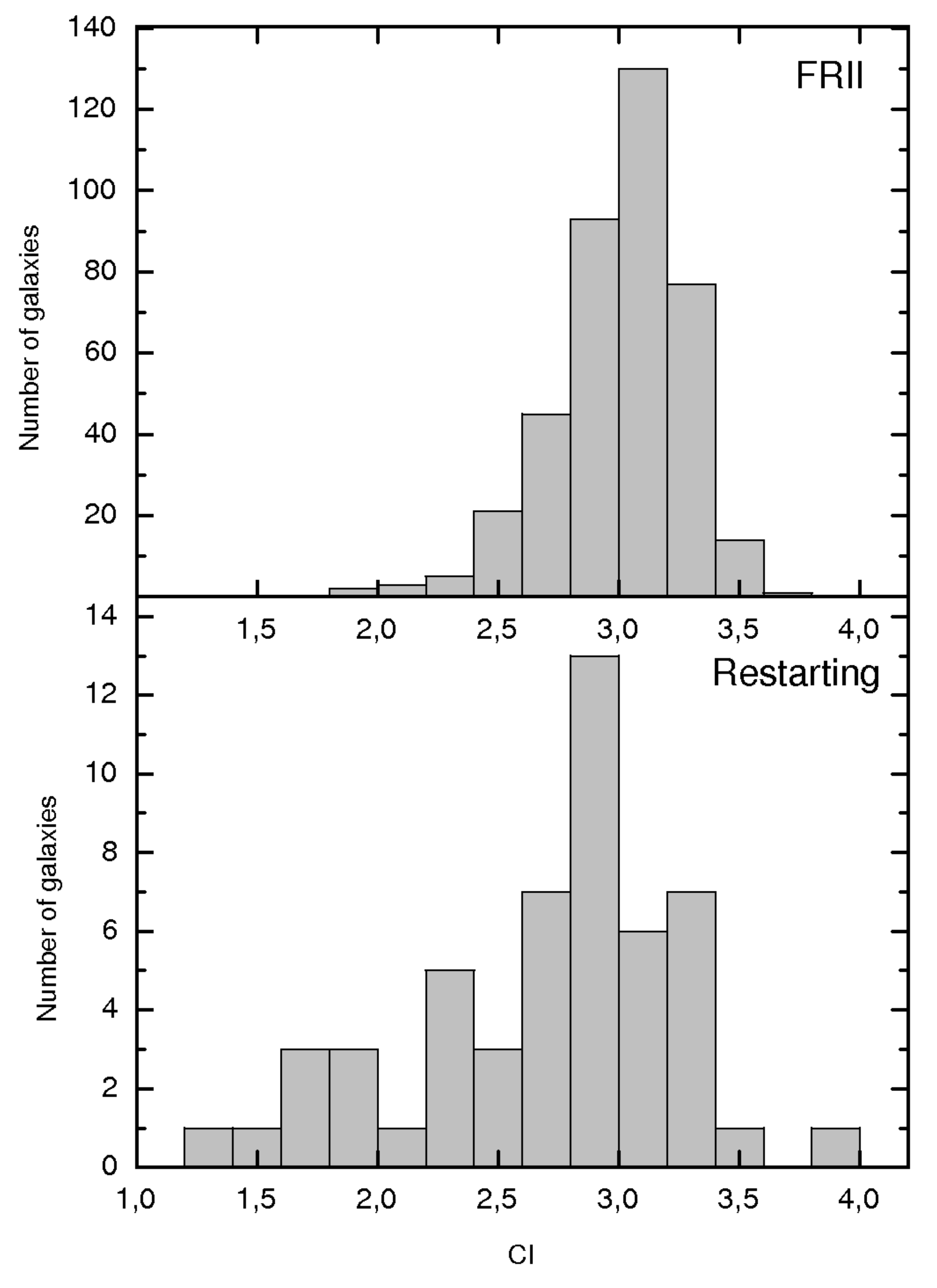} 
\caption{Distribution of concentration indices (CIs) for the estarting radio sources and the comparison sample.}
\label{CI}
\end{figure}

In Figures~\ref{MbhM} and~\ref{PM} we plotted the galaxy mass M$^*$ (the mass of stars obtained with STARLIGHT modelling) as a function of the BH mass and the 1.4\,GHz radio luminosity, respectively. It can be seen that the restarting radio galaxies have lower masses of the host galaxy when compared with the FRII radio sources. This may suggest that a larger fraction of galaxy mass in restarting sources is cumulated in gas and/or dust. Since in elliptical galaxies the amount of gas and dust is not significant when compared to spiral galaxies, its presence can be an evidence of merger events occurring in the host galaxies of sources with restarting jet activity.\\
A very useful parameter for morphological classification of galaxies is the concentration index CI, defined as the ratio of two radii R90/R50, where R90 and R50 are the radii enclosing 90\% and 50\% of the r-band Petrosian flux, respectively. According to \citet{Nakamura2003}, the CI value lower than 2.86 corresponds to late type galaxies and CI of more than 2.86 -- to the early-type galaxies. In Figure~\ref{CI} we present the distribution of concentration index for the samples of restarting and FRII radio galaxies. It can be seen that the radio sources with recurrent activity tend to have smaller CI values than the radio galaxies from the comparison sample. The difference in distribution of CI indices for restarting and comparison sample radio sources is statistically significant. The K--S test returns a probability of 0.00004 with the maximal deviation between cumulative distributions of 0.34, which indicates that the two considered samples come from different distributions. This can suggest that recurrent activity is observed more often in disturbed/amorphous galaxies.

\begin{figure}
    \includegraphics[width=\columnwidth]{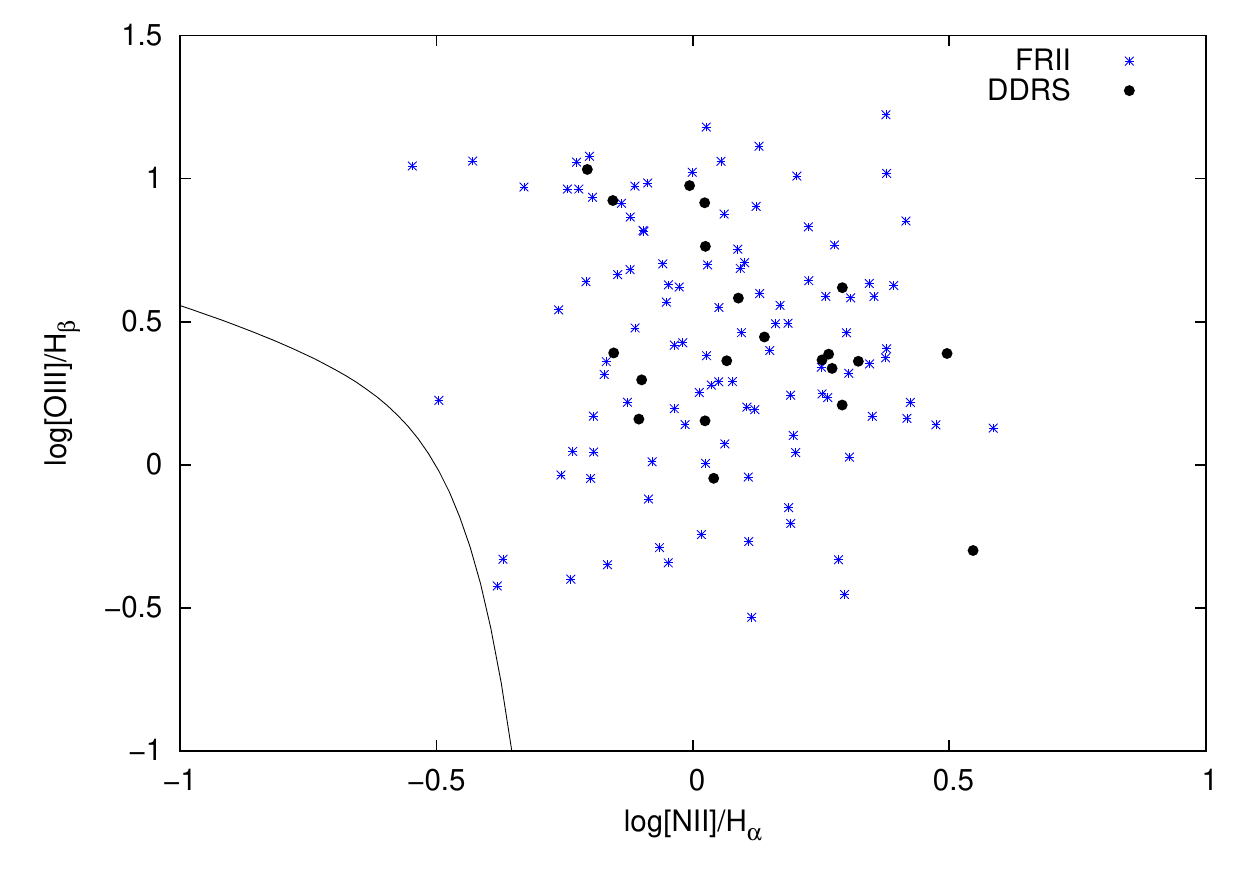} 
\caption{Diagnostic diagram: [OIII]$/$H$_\beta$ vs. [NII]$/$H$_\alpha$ for restarting radio galaxies (black dots) and FRII comparison sample (blue asterisks). The solid line indicates the division between AGNs and star-forming galaxies according to \citet{Stasinska2006}. In plotting the diagram we used only emission lines with S/N$>$3.}
\label{bpt}
\end{figure}

Moreover, in almost all ($\sim$89\%) the available spectra of restarting radio sources the emission lines were visible. Using our measurements of the line flux we constructed a diagnostic diagram proposed by \citet{Baldwin1981} and plotted the ratios [OIII]/H$_\beta$ and [NII]/H$_\alpha$. In Figure~\ref{bpt} we present diagnostic diagrams for the recurrent radio sources and the comparison sample of FRII galaxies. The diagram used only emission lines with the signal-to-noise ratio of S/N$>$3. There are 11\,sources having [OIII]/H$_\beta$$>$3 in our sample, which represents a high ionisation level. The mean value of the [OIII]/H$_\beta$ ratio for these sources is very high ($\sim$6.6). The remaining galaxies have lower [OIII]/H$_\beta$ ratios with the mean value of 2.1. The restarting radio galaxies are located in the same area on the diagram as the FRII sources, which can  suggest that the host galaxies of both types of radio sources are similar. There are no restarting radio sources near the solid line in Figure~\ref{bpt} separating the HII galaxies from AGNs. According to \citet{Stasinska2006}, the objects lying away from this line have emission lines excited mostly by an AGN source. There are also no objects in the bottom part of the diagram, where the retired galaxies, i.e. galaxies that have ceased forming stars and are ionized only by their old stellar populations \citep{Stasinska2015}, are located. 

\begin{figure}
    \includegraphics[width=\columnwidth]{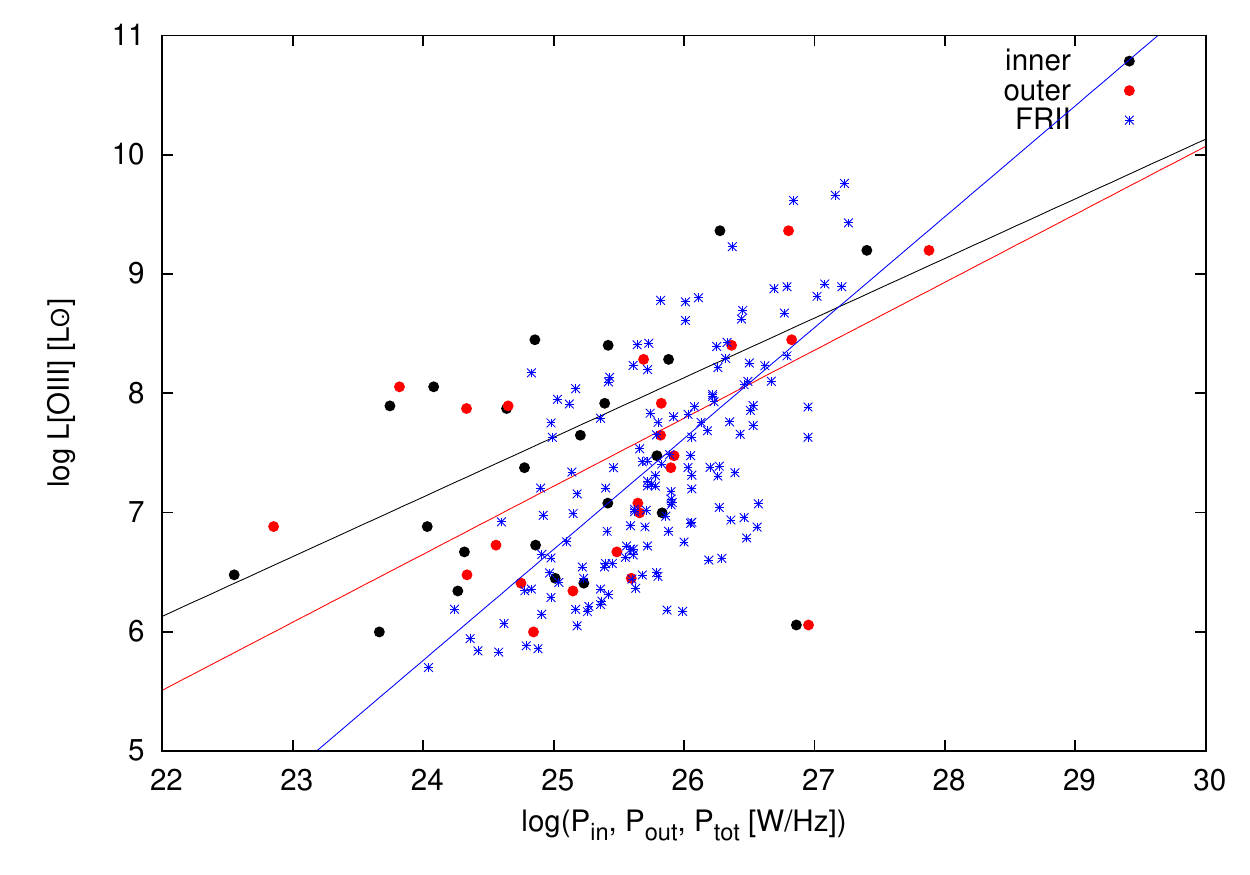} 
\caption{Radio luminosity at 1.4\,GHz versus [OIII]5007 line luminosity. For the restarting radio sources the radio luminosities of inner (black dots) and outer (red dots) radio lobes are plotted, while for the comparison sample (blue asterisks) the total radio luminosities are shown. The black and red line corresponds to the best fit for inner and outer radio lobes respectively, while the blue line corresponds to the best fit obtained for the FRII radio galaxies by \citet{Koziel2011}.}
\label{LOIIIP}
\end{figure}

We also checked the relation between the radio luminosity and the luminosity of the [OIII] emission line. It was postulated by \citet{Rawlings1989} and \citet{Baum1988} that the positive correlation can be an evidence for the physical coupling of processes that supply energy both to the emission line regions and to the extended radio structure. The correlations between optical and radio properties are generally explained by the illumination model in which the gas responsible for line emission is photoionized by UV photons coming from the central AGN. We plotted radio luminosity (separately for the inner and outer radio lobes) against the luminosity of the [OIII] emission line. The P--L[OIII] relation is shown in Figure~\ref{LOIIIP}. The linear fits to the data points are given by the following relations:
\begin{equation}
   \log L_{[OIII]}=(0.50\pm0.14)\cdot \log P_{in}-(4.87\pm3.36)
\end{equation}
\vspace{-0.5cm}
\begin{equation}
   \log L_{[OIII]}=(0.57\pm0.14)\cdot \log P_{out}-(7.03\pm3.62)
\end{equation}
and for the comparison sample: 
\begin{equation}
\log L_{[OIII]}=(0.93\pm0.08)\cdot \log P_{tot}-(16.6\pm2.1)  
\end{equation}
The correlation coefficients for the inner radio structure, outer radio structure and the comparison sample are 0.54, 0.58 and 0.68, respectively.
For the H$_\alpha$ line we obtained the following relations:
\begin{equation}
   \log L_{H_{\alpha}}=(0.38\pm0.12)\cdot \log P_{in}-(2.05\pm2.99)
\end{equation}
\vspace{-0.5cm}
\begin{equation}
   \log L_{H_{\alpha}}=(0.46\pm0.13)\cdot \log P_{out}-(4.29\pm3.27)
\end{equation}
and for comparison sample:
\begin{equation}
\log L_{H_{\alpha}}=(0.80\pm0.06)\cdot \log P_{tot}-(13.4\pm1.7)  
\end{equation}
The correlation coefficients for the inner radio structure, the outer radio structure and the comparison sample are 0.53, 0.59 and 0.73, respectively. The slopes of the fitted lines of the inner and outer radio lobes are slightly flatter than those obtained for the FRII radio galaxies.   
\begin{figure}
    \includegraphics[width=\columnwidth]{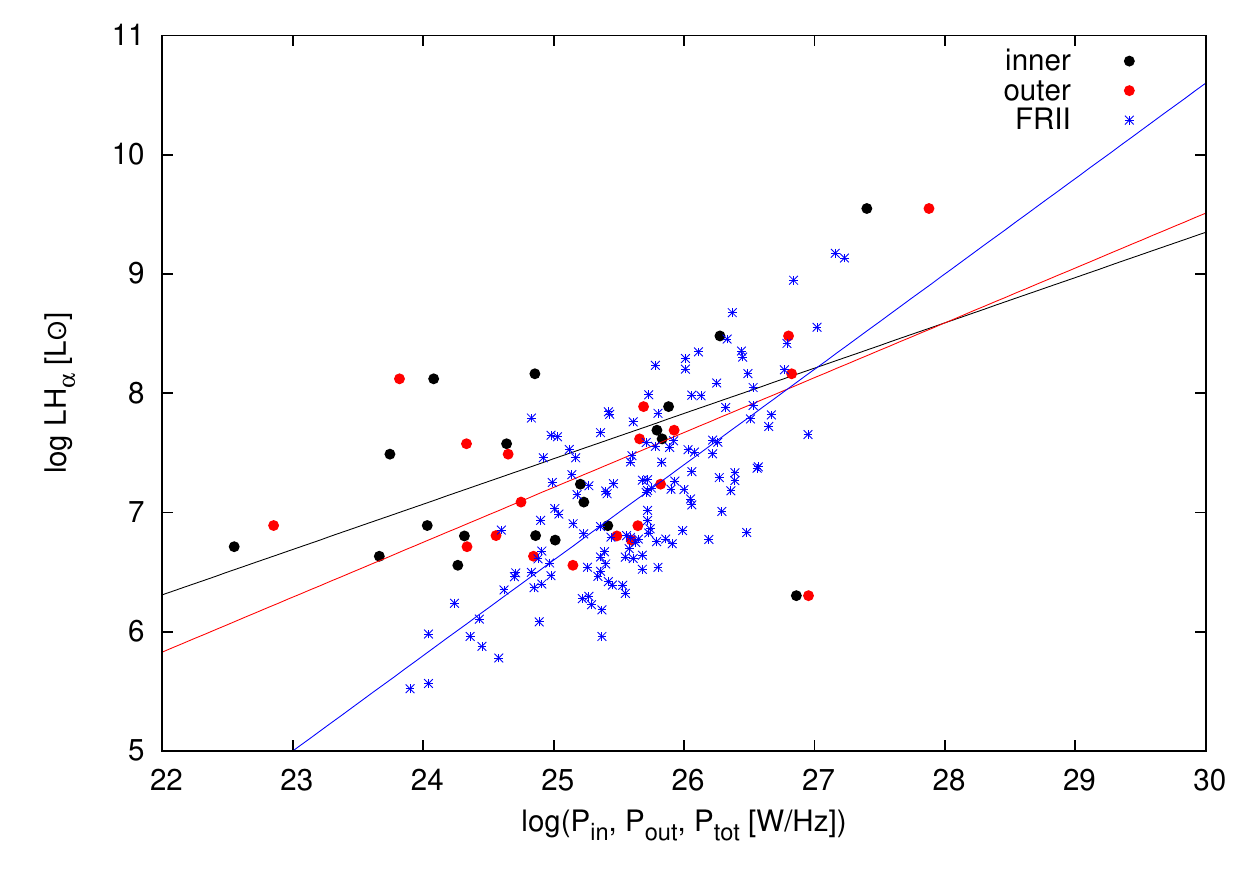} 
\caption{Radio luminosity at 1.4\,GHz versus H$_\alpha$ line luminosity. The symbols are as in Figure~\ref{LOIIIP}.}
\label{LHaP}
\end{figure}

\subsection{Radio properties}
\begin{figure}
    \includegraphics[width=\columnwidth]{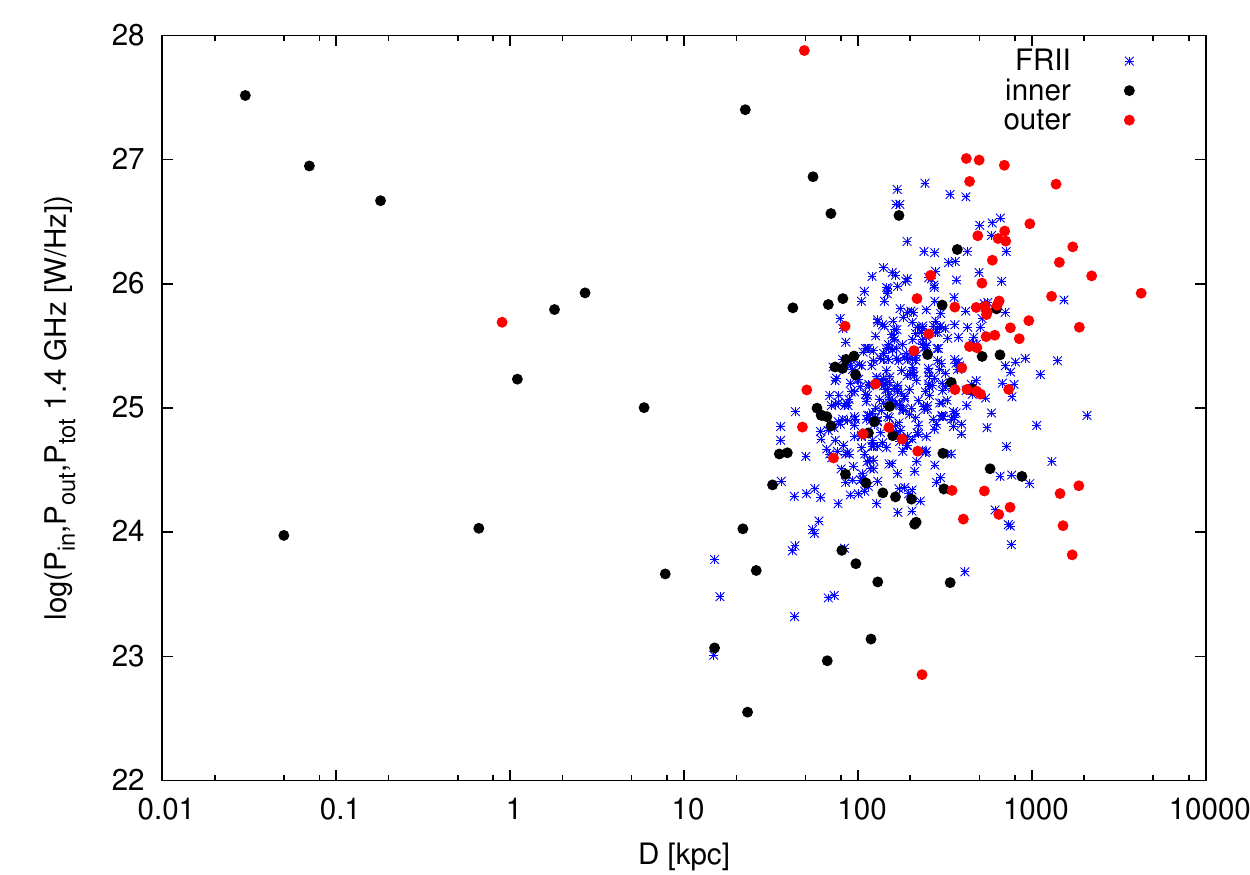} 
\caption{Luminosity--linear size diagrams. For the comparison sample we used total radio luminosity and for the restarting sources we plotted luminosity both for the inner (P$_{in}$) and the outer (P$_{out}$) structures.}
\label{P_D}
\end{figure}

In Figure~\ref{P_D} we plotted the luminosity--linear size relation (P--D diagram) for the sample of restarting and single-cycle FRII radio sources. Most restarting radio sources occupy the same region as the sources from the comparison sample. It can be also observed that the outer doubles tend to be larger than the lobes of FRII radio galaxies. This could be an observational bias resulting from the fact that it is easier to detect restarting radio sources of larger size. On the P--D diagram there are also few very compact (D$<$1kpc) restarting radio sources with high radio luminosities (log P$>$26[W/Hz]). According to the dynamical and luminosity evolution models of radio galaxies, three distinct evolutionary phases are to be expected. In the first one, when the lobes are still expanding within the host galaxy, the radio luminosity increases with the source size, which stops as soon as the synchrotron losses become dominant (at the size of about 1\,kpc). Beyond this point, the radio luminosity steadily decreases with the increasing source size (up to $\sim$ 100\,kpc) and finally enters the phase of sharply decreasing of luminosity when the inverse Compton losses, resulting from the CMB energy density, dominate the synchrotron losses (\citealt{Kaiser2007}, \citealt{An2012}). \\

\begin{figure}
    \includegraphics[width=\columnwidth]{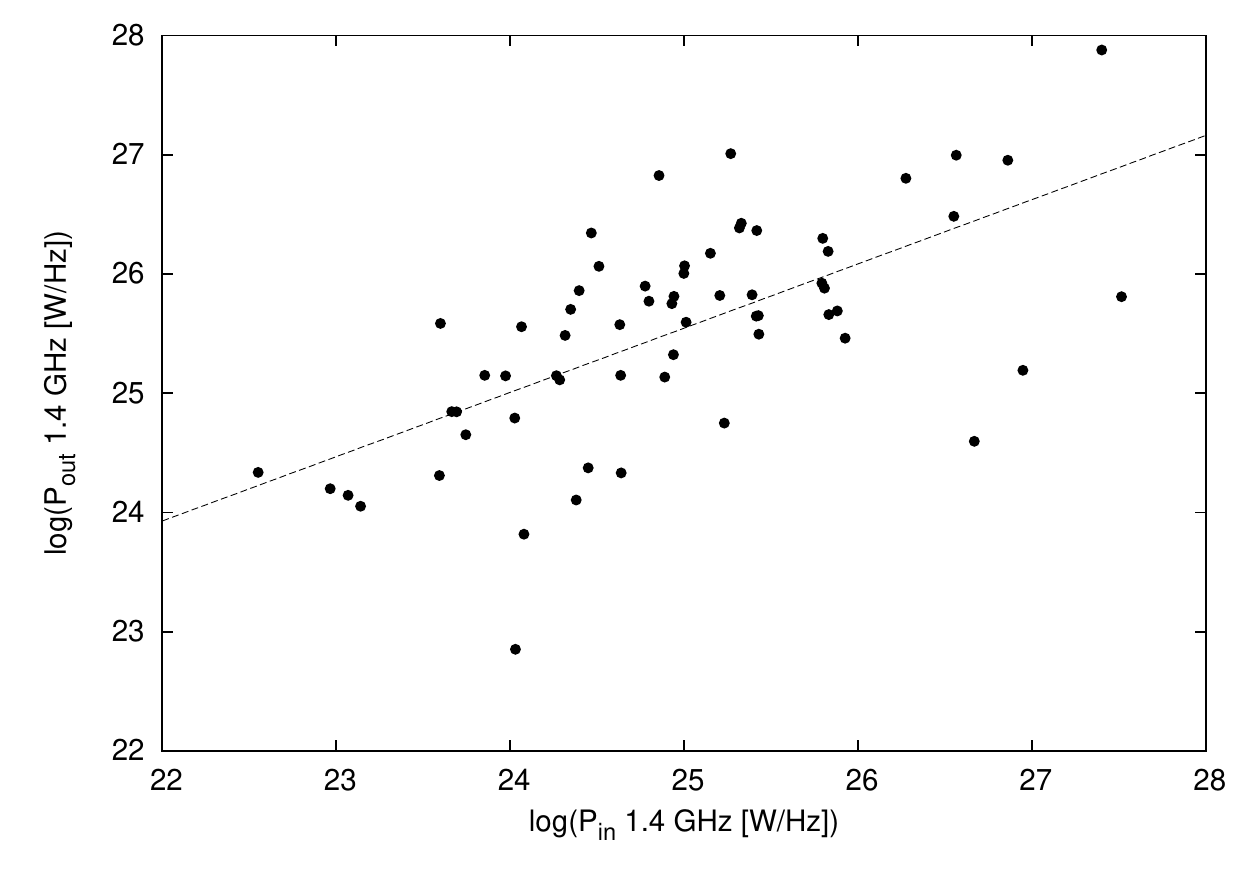} 
\caption{Relation between 1.4\,GHz radio luminosity of the inner and the outer radio lobes.}
\label{PinPout}
\end{figure}

\begin{figure}
    \includegraphics[width=\columnwidth]{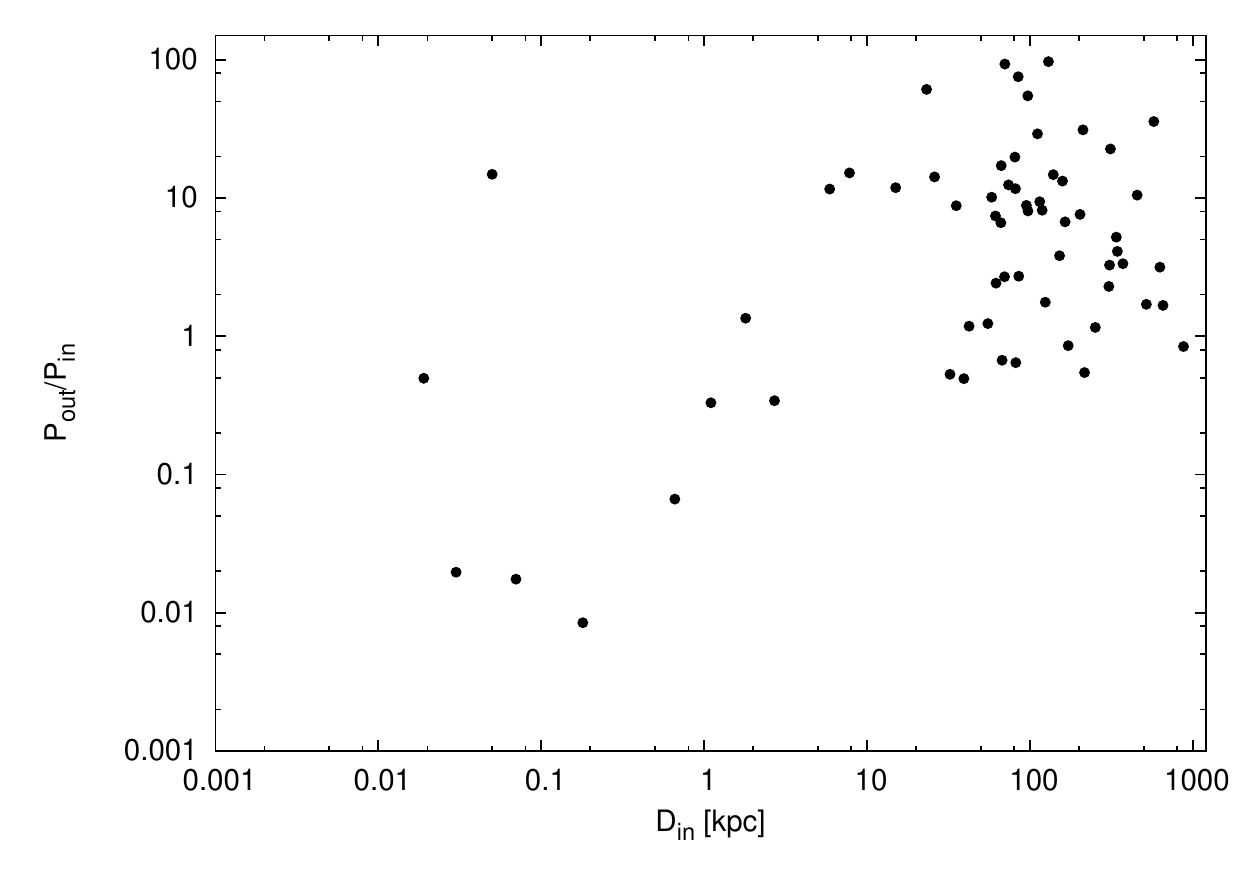} 
\caption{Ratio of 1.4\,GHz radio luminosity of outer and inner lobes against the projected linear size of inner radio lobes. 
The correlation coefficient is R=0.54.}
\label{Pratio}
\end{figure}

As shown in Figure~\ref{PinPout}, there is a significant correlation between luminosity of the outer and inner lobes. The correlation coefficient of a linear fit is 0.59.
In our sample of restarting radio galaxies we found 14\,sources, whose inner lobes are more luminous than the outer lobes (i.e. J0041$+$3224, J0111$+$3906, J0301$+$3512, J0301$+$3512, J0741$+$3112, J0821$+$2117, J0840$+$2949, J0910$+$0345, J0914$+$1006, J0943$-$0819, J1021$+$1216, J1247$+$6723 J1352$+$3126 and J1844$+$4533). This group of sources is very peculiar and we denoted them by an asterisk in Table \ref{ddrs}. Most of them have very compact inner radio structures, but some sources have larger inner lobes (over 100\,kpc). It can also be seen that the smaller is the ratio of luminosities (logP$_{out}$/logP$_{in}$), the smaller inner radio structure is observed (in log scales; Fig. \ref{Pratio}). Such a trend is found particularly for the luminosity ratios of less than one. For these sources we obtained positive correlation with a coefficient equal to 0.73. 
The higher luminosity of the inner lobes when compared to that of the outer lobes is explained by \citet{Saikia2006} as a result of more efficient radio emission in the early phase of evolution of the inner lobes when they expand within a dense interstellar medium. As the source expands and traverses a more diluted medium, the ratio P$_{out}$/P$_{in}$ usually increases with the size of inner lobes before approaching values of the order of unity. \citet{Schoenmakers2000} postulated the inverse correlation between P$_{out}$/P$_{in}$ and D$_{in}$ (basing on studies of 7\,DDRSs that have the inner doubles larger than 90\,kpc). Our results do not confirm this trend but are in agreement with the results obtained by \citet{Saikia2006}, who found that P$_{out}$/P$_{in}$$<$1 in the smallest($<$1kpc) inner doubles. They also concluded that the inverse correlation postulated by \citet{Schoenmakers2000} has a reduced level of significance. According to \citet{Nandi2012}, there could be only an upper envelope to this diagram, suggesting an inverse relation. The small number of sources on the left side of Figure~\ref{Pratio} can be caused by the selection effects. Resolving very compact inner doubles that are smaller than 10\,kpc is possible only with high-resolution radio observations.
Furthermore, we also found that nearly half of the restarting radio galaxies have brighter inner and outer radio lobes on the same side of the host galaxy (i.e. out of 41 radio galaxies for which the inner and outer lobes are well detached, the brighter inner and outer lobes are on the same side in 18 galaxies, and 23 galaxies show brighter inner and outer lobes on the opposite sides of the host galaxy). A similar trend is observed, when the length of the radio lobes is considered -- 22 and 19 sources, respectively. These findings can be explained as resulting from the combination of orientation of a radio source and an activity intermission to have occurred between the active periods. Only in an ideal case of symmetric jets and isotropic properties of the ambient medium, one could expect the same radio luminosity and equal lengths of both lobes. In reality, the orientation of radio sources in space is random and only some of them are aligned in the sky plane. Due to the latter, we expect Doppler effects to be responsible for the brightness differences. Moreover, the light-travel time effect, consisting in that radiation from the far-side lobe arrives at the observer significantly later due to its longer distance than from the near-side lobe, should be taken into account (see \citealt{Marecki2012}). 

\begin{figure}
    \includegraphics[width=\columnwidth]{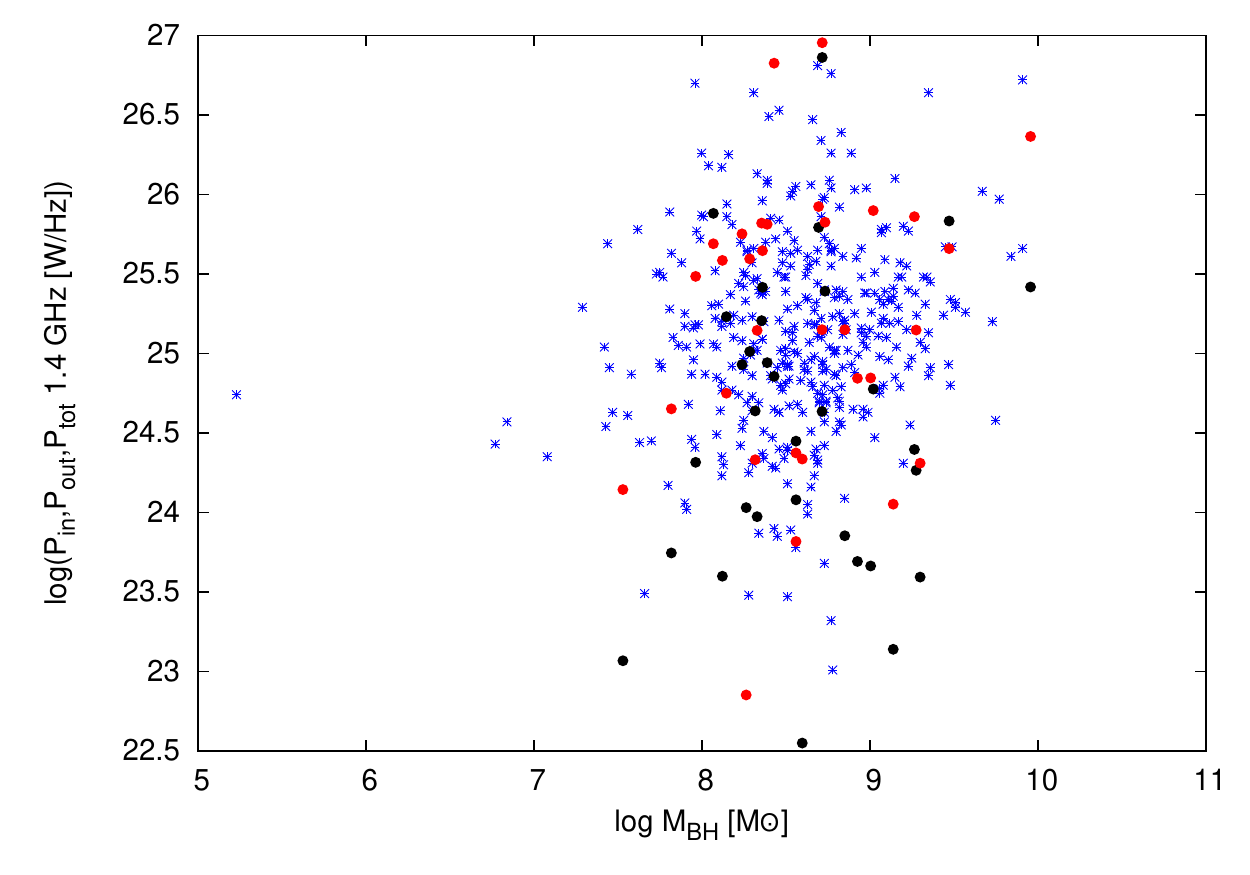} 
\caption{Relation between BH mass and radio luminosity at 1.4\,GHz.}
\label{PMBH}
\end{figure}

\begin{figure}
    \includegraphics[width=0.95\columnwidth]{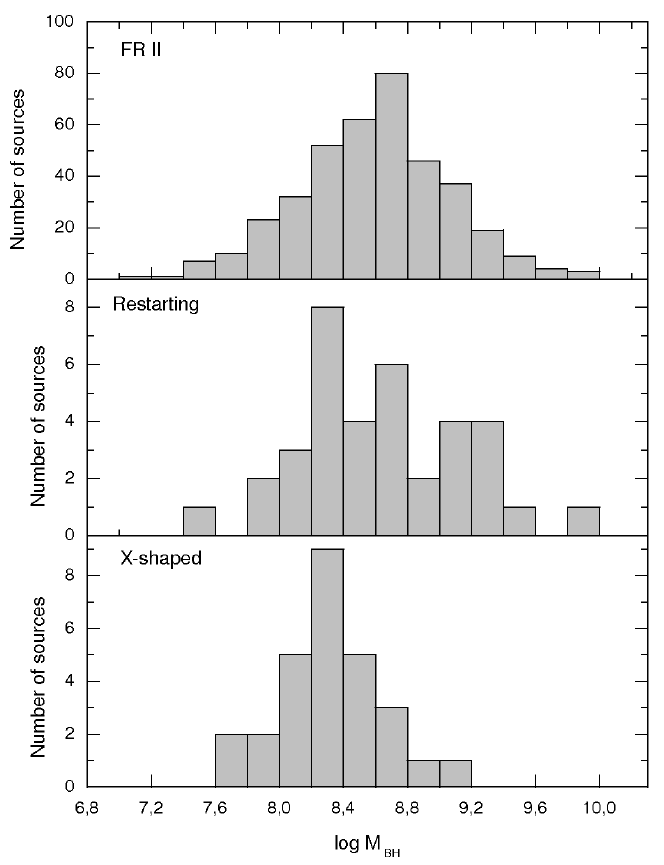} 
\caption{BH mass distribution for FRII (top panel), restarting radio galaxies (middle panel) and X-shaped radio sources (bottom panel).}
\label{MBH}
\end{figure}

In Figure~\ref{PMBH} we plotted the radio luminosity against the mass of the black hole. The distribution of sources in the P--M$\rm_{BH}$ plane is much similar for both the samples. We also compared the distributions of the black hole mass for the FRII sample, restarting radio galaxies sample, and X-shaped radio sources studied by \citet{Mezcua2011,Mezcua2012}. All the distributions are presented in Figure \ref{MBH}. In the case of FRII radio galaxies, the distribution of the black hole mass is very symmetric with a peak value of $log($M$\rm_{BH}/M_{\odot})$ ranging from 8.6 to 8.8 and the median value of 8.58. For the restarting radio galaxies the M$\rm_{BH}$ distribution does not have any pronounced maximum. However, the median value of $log($M$\rm_{BH}/M_{\odot})$ is 8.61, much the same as those of typical FRII sources. The distribution for the X-shaped radio galaxies is nearly symmetric around the peak value, which is similar to the one of the restarting radio sources. The median value of $log($M$\rm_{BH}/M_{\odot})$  for the X-shaped galaxies of $\sim$8.3 is smaller than for the other two samples. 

\subsection{Infrared properties}
\begin{figure}
    \includegraphics[width=0.99\columnwidth]{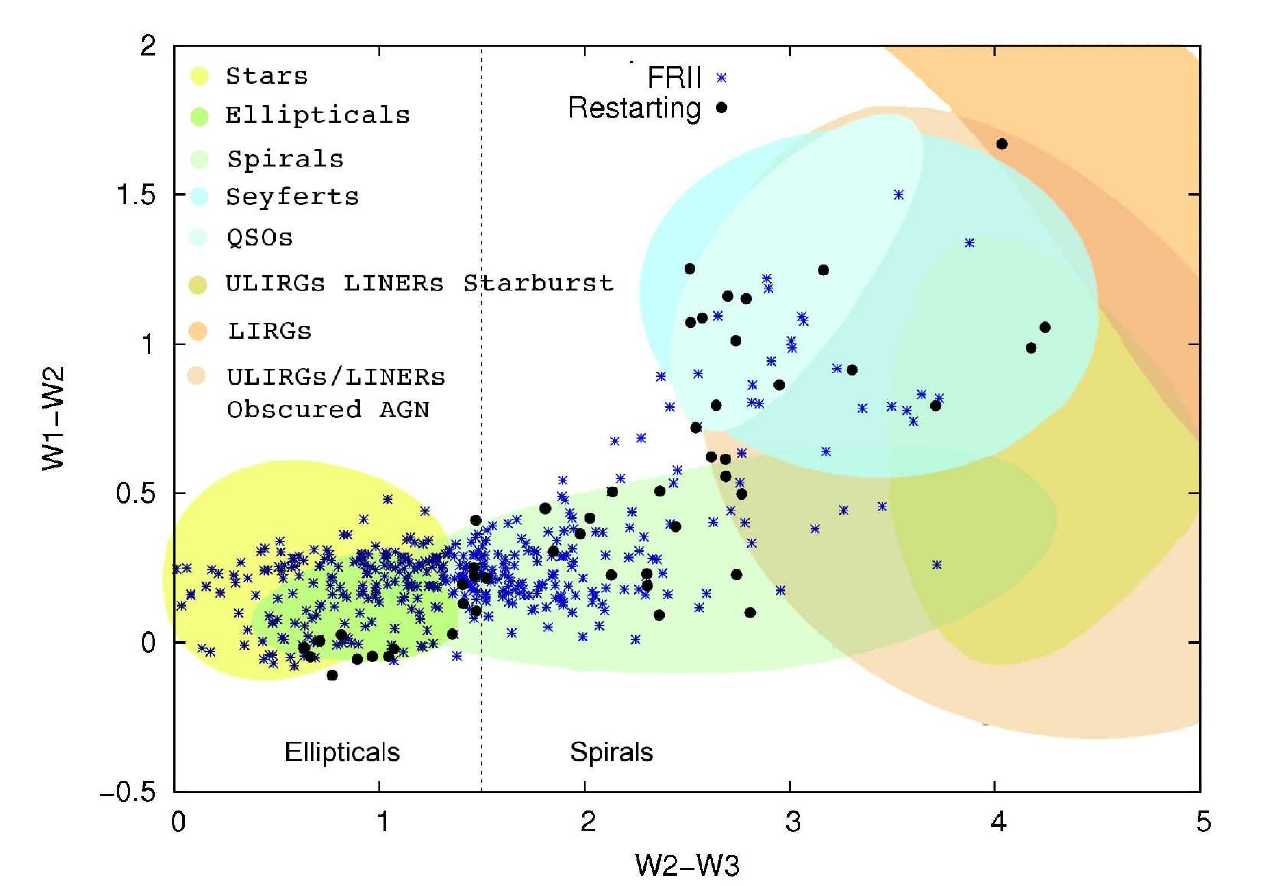} 
\caption{Colour-colour diagram for restarting and FRII radio sources. The coloured areas represent different classes of sources and the dotted line (W2-W3=1.5) shows a proposed division between elliptical and spiral galaxies  (according to \citealt{Wright2010}).}
\label{wise}
\end{figure}

Almost all the recurrent activity radio galaxies discussed in this paper were detected by the Wide-Field Infrared Survey Explorer (WISE; \citealt{Wright2010}). 
In Figure~\ref{wise} we plotted the colour-colour diagram, where the vertical axis W1-W2 corresponds to the magnitude difference between the 3.4 and the 4.6 $\mu$m bands and the horizontal axis W2-W3 corresponds to the magnitude difference between the 4.6 and the 12 $\mu$m bands. We plotted the WISE colours only for 49 restarting and 388 FRII radio sources, for which all magnitudes were above the detection limit. In Figure~\ref{wise} we marked our sources in the diagram showing the regions where the different classes of the WISE-detected sources are located (\citealt{Wright2010}).
The vertical dotted line, which, according to \citet{Wright2010}, divides elliptical and spiral galaxies, has a WISE W2-W3 colour value of +1.5 magnitude. The most powerful AGNs lie above the W1-W2 colour level of +0.6 magnitude. It can be clearly seen that most of the restarting radio sources (67\%) are located in the regions where spiral galaxies, being typically ISM-abundant, reside. In the comparison sample less than half (41\%) of FRII radio sources are in that region. Also most of the studied radio sources have W1-W2 colours lower than 0.5 and according to \citet{Izotov2014}, for galaxies with W1-W2 $<$ 0.5 the main source of radiation is the stellar and ionised gas emission. The location of radio sources in this region was explained by \citet{Assef2010} as a consequence of small Eddington ratios. In such galaxies, the mid-infrared colours are not dominated by the AGN, while they are contaminated by the host. Therefore, the colours generally originate from stars and emission from the cold dust of star formation. 

\section{Conclusions}

We compiled and presented the largest (to date) sample of 74 restarting radio sources, using  optical, radio, and infrared data to determine their physical properties and compare them with a sample of typical FRII radio galaxies with single activity. We came to the following conclusions:
\\
(i) The black hole masses of radio sources with recurrent activity are similar to those observed in FRII radio galaxies. 
\\
(ii) Recurrent and typical radio sources show different compositions of stellar populations. The hosts of restarting radio sources contain a larger amount of young stars. 
\\
(iii) The total mass of stars in the host galaxies of the recurrent activity sources is, on average, less than that in the FRII hosts. The concentration index for the restarting radio sources tends to be slightly lower than that of the FRII sources, which indicates that they can have hosts with more disturbed morphologies. Both the facts could be the evidence that the restarting radio sources are more common in galaxies after mergers.
\\
(iv) Emission lines are visible in almost all (89\%) the available spectra of recurrent jet activity hosts. Basing on the diagnostic diagram, the emission lines are excited by an AGN-like driven process. 
\\
(v) There is a significant correlation between the luminosity of the emission line and the radio luminosity of the inner and outer radio lobes. The correlation is on the same level for the H$_\alpha$ as well as for [OIII] line. 
\\
(vi) There is a strong correlation between the radio luminosity of the inner and outer lobes.
\\
(vii) In the case of 13 restarting radio galaxies, the inner lobes are more luminous than the outer ones. This can be seen not only in very compact ($<$1kpc) inner doubles, but in much more extended ones as well, with linear sizes of few hundred kiloparsecs.
\\
(viii) Almost half of restarting radio sources have brighter/larger inner and outer lobes on the opposite sides of host galaxy. 
\\
(ix) The infrared WISE colour-colour diagram shows that the 67\% hosts of recurrent jet activity radio galaxies reside in the region typical for spiral galaxies or other dusty, late-type galaxies with some ongoing star formation, while only 41\% of FRII radio sources are found in that region.  
    
\section*{Acknowledgements}
We thank the anonymous referee for her/his very valuable comments. The work is supported by Polish NSC grant DEC-2013/09/B/ST9/00599.






\begin{onecolumn}
\begin{landscape}
\begin{longtable}{l l l l c c c c c c c c c c l}
\caption{Radio sources with evidence of recurrent activity.} \\                                                                                                                                                                                                                  
\hline
Source       &IAU	&	REC$\rm_{J2000.0}$     & DEC$\rm_{J2000.0}$& Opt.& Red- & $l_{in}$&$l_{in}$& $l_{out}$&$l_{out}$&$S_{in}$&$S_{out}$&$logM_{BH}$	&Class	&Ref.  \\
Name         &Name	&$\rm^{h}$ $\rm^{m}$ $\rm^{s}$ & $\degr$ ' ''      & Id. & shift& kpc     &  arcsec&  kpc     &  arcsec &mJy   	& mJy	   &M$\odot$	&A$/$B	&Cmt.\\
    (1)      & (2)	&	(3)	               & (4)               & (5) &  (6) &  (7)    &  (8)   &  (9)     &   (10)  &  (11)  &  (12)   &(13)	&(14)	& (15) \\
\hline
\endfirsthead
\multicolumn{15}{l}{{\it Table \ref{ddrs} continued}}\\
\hline
Source       &IAU	&	REC	    & DEC         & Opt.& Red- & $l_{in}$&$l_{in}$& $l_{out}$&$l_{out}$&$S_{in}$&$S_{out}$&$logM_{BH}$	&Class	&Ref.  \\
Name         &Name	&	J2000	    & J2000       & Id. & shift& kpc     &  arcsec&  kpc     &  arcsec &mJy   	& mJy     &M$\odot$	&A$/$B	&Comment \\
    (1)      & (2)	&	(3)	    & (4)         & (5) &  (6) &  (7)    &  (8)   &  (9)     &   (10)  &  (11)  &  (12)   &(13)		&(14)	& (15) \\
\hline
\endhead
\endfoot
\endlastfoot
J0009$+$1244 		& 4C12.03  & 00 09 52.60 & $+$12 44 04.64 &  G	& 0.156		& 114.9	& 43	& 553.1		& 207		& 100		& 907		& --	& A	&1,f\\
J0037$+$1319	 	& 3C16     & 00 37 44.57 & $+$13 19 55.00 &  G  & 0.405		& 96.91	& 18	& 419.9  	& 78      	& 35     	& 1765    	& --	& A	&1,2	\\
J0041$+$3224$^\star$ &  	   & 00 41 46.12 & $+$32 24 52.65 &  G	& (0.45)	& 172	& 30	& 974		& 170		& 525		& 409		& --	& A	&3	\\
{\bf J0042$-$0613} 	&          & 00 42 46.85 & $-$06 13 52.92 &  G	& 0.123		& 518.1	& 237.1 & 753.6		& 344.9		& 690		& 1140		& 9.23	& A	&4,a,e \\ 
J0104$-$6609		&	   & 01 04 21.26 & $-$66 09 17.30 &  G  & (1.19)	& 66.7	& 8	& 750		& 90		& 0.37		& 5.22		& --	& A	&5  \\ 
J0111$+$3906$^\star$ & B0108+388& 01 11 37.32 & $+$39 06 28.10 &  G  & 0.6685	& 0.07	& 0.01	& 126.1  	& 18      	& 519    	& 8       	& --	& 	B?	&6,b,c\\
J0116$-$4722 		&	   & 01 16 25.04 & $-$47 22 41.60 &  G	& 0.146		& 455	& 180	& 1441		& 570		& 260		& 2640		& --	& A	&7	\\
J0301$+$3512$^\star$ & 4C 34.09 & 03 01 42.37 & $+$35 12 20.68 &  G  & 0.0165	& 0.66	& 2	& 233.7  	& 706     	& 1800   	& 119     	& 8.26	& 	B	&2,i	\\ 
J0301$+$3550$^\star$ & 4C 35.06 & 03 01 51.50 & $+$35 50 30.00 &  G  & 0.0463	& 32.29	& 36	& 403.7  	& 450     	& 492    	& 258     	& --	& 	A	&8      \\   
J0303$+$1626 		& 3C76.1   & 03 03 15.02 & $+$16 26 19.06 &  G	& 0.0325	& 21.8	& 34.1	& 107.1		& 167.4		& 450.2		& 2602		& --	& B	&1	\\
J0351$-$2744 		&PKS0349-27& 03 51 35.76 & $-$27 44 34.70 &  G	& 0.0662	& 251.8	& 200.8	& 437.8		& 349.1		& 2636		& 3009		& --	& A	&9	\\         
{\bf J0504$+$3806} 	& 3C134    & 05 04 42.19 & $+$38 06 11.40 &  G	& --		& -- 	& 115.6	& --		& 166.1		& 1485		& 7862		& --	& A	&a	\\
J0709$-$3601 		&PKS0707-35& 07 09 14.09 & $-$36 01 21.80 &  G	& 0.218		& 627.1	& 179.4	& 1720		& 492		& 480		& 1444		& --	& A	&10,11\\
J0741$+$3112$^\star$ &B2 0738+31& 07 41 10.70 & $+$31 12 00.22 &  Q& 0.632		& 0.03	& 0.005	& 478.5  	& 70      	& 2188   	& 38      	& 9.37	& A	&12,b,h	\\       
J0746$+$4526 		&          & 07 46 17.92 & $+$45 26 34.46 &  G	& 0.5502	& 95	& 15.2	& 640		& 100.1		& 24.2		& 191.6		& 9.96	& A	&13	\\
J0804$+$5809 		&          & 08 04 42.79 & $+$58 09 34.94 &  S	& -- 		& --	& 21.6	& -- 		& 106.3		& 58.6		& 192.1		& --	& A	&13	\\
J0821$+$2117$^\star$ &B0818+214 & 08 21 07.50 & $+$21 17 02.87 &  G	& 0.418		& 2.7	& 0.5	& 209.8		& 38.24		& 148		& 46.4		& --	& A	&14	\\
J0840$+$2949$^\star$ & 4C29.30  & 08 40 02.36 & $+$29 49 02.63 &  G	& 0.0647	& 39.3	& 32    & 533.3		& 434.6		& 446.7		& 216.9		& 8.32	& B	&15	\\
J0847$+$3147 		& IC2402   & 08 47 59.04 & $+$31 47 08.37 &  G	& 0.0674	& 203	& 159.3 & 362.1		& 284.2		& 173.9		& 1303		& 9.28	& A	&16	\\   
J0855$+$4204 		&          & 08 55 49.15 & $+$42 04 20.11 &  G	& (0.279)	& 35.3	& 8.4   & 545.9		& 130		& 18.8		& 155.7		& --	& A	&13	\\
J0910$+$0345$^\star$    &          & 09 10 59.10 & $+$03 45 31.68 &  G	& (0.588)	& 42.3	& 6.4   & 218.8		& 33.1		& 50.7		& 53.4		& --	& A	&13	\\
{\bf J0914$+$1006}$^\star$&	   & 09 14 19.53 & $+$10 06 40.59 &  G  & 0.308		& 216.2	& 48	& 1709		& 379.7		& 252.5		& 129.1		& 8.56 	& A	&a \\ 
J0921$+$4538 		& 3C219	   & 09 21 08.61 & $+$45 38 57.36 &  G	& 0.174		& 70.1	& 24	& 438.5		& 150		& 90		& 8046		& 8.43	& A	&17,18,19\\
{\bf J0924$+$0602} 	&          & 09 24 49.04 & $+$06 02 42.80 &  G	& 0.231		& 80.8	& 22.1	& 424		& 116		& 4.8		& 90		& 8.85	& A	&a\\
J0927$+$2932 	&          & 09 27 44.88 & $+$29 32 32.30 &  S  & --		& --	& 24	& --        	& 115     	& 19     	& 17      	& --	& 	A	&8	\\      
J0927$+$3510 		&  	   & 09 27 50.59 & $+$35 10 50.73 &  G	& (0.55)	& 575.5	& 90	& 2206		& 345		& ~3		& 96		& --	& A	&20	\\
J0929$+$4146 	&   	   & 09 29 10.66 & $+$41 46 45.59 &  G	& 0.365		& 655.6	& 130	& 1876		& 372		& 64		& 99		& --	& A	&21,d,g\\
J0935$+$0204 		& 4C02.27  & 09 35 18.19 & $+$02 04 15.54 &  Q	& 0.6491	& 69.9	& 10.1	& 498		& 71.96		& 230.3		& 547.6		& 9.56	& A	&22	\\ 
J0943$-$0819$^\star$ & B0941-080& 09 43 36.94 & $-$08 19 30.81 &  G  & 0.228 	& 0.18	& 0.05	& 72.34   	& 20     	& 3232   	& 26      	& --	& B?	&23,h\\             
{\bf J1004$+$5434}&		   & 10 04 51.83 & $+$54 34 04.29 &  G   & 0.047	& 55.2	& 60.6	& 694.2		& 762.9		& 90		& 110		& 8.72	& A 	&24,a \\ 
J1006$+$3454 		& 3C236    & 10 06 01.73 & $+$34 54 10.52 &  G	& 0.101		& 1.8	& 1	& 4248		& 2310		& 2500		& 3300		& 8.70	& A	&25,26,27\\
{\bf J1021$+$1216}$^\star$ &          & 10 21 24.21 & $+$12 17 05.44 &  G	& 0.129		& 876.4	& 385	& 1865		& 819.4		& 67.3		& 55	& 8.56	& A	&4,a\\ 
J1039$+$0536 		&          & 10 39 28.21 & $+$05 36 13.62 &  G	& 0.35		& 81.5	& 16.6  & 488.7		& 99.6		& 54.9		& 594.9		& --	& A	&13	\\
J1103$+$0636 		&          & 11 03 13.29 & $+$06 36 16.00 &  G	& 0.4406	& 66.3	& 11.7  & 548.8		& 96.9		& 13.2		& 79.9		& 8.24	& A	&13	\\
J1158$+$2621 		& 4C26.35  & 11 58 20.13 & $+$26 21 12.07 &  G	& 0.112		& 139	& 69	& 483.8		& 240		& 67		& 962		& 7.96	& A	&13,28\\
J1159$+$5820 		&	   & 11 59 05.68 & $+$58 20 35.57 &  G	& 0.054		& 23.2	& 22.4  & 348.2		& 335.8		& 5.3		& 319.1		& 8.60	& A	&29	\\
J1208$+$0821 		&          & 12 08 56.78 & $+$08 21 38.57 &  G	& 0.5841	& 111.3	& 16.9  & 648.9		& 98.5		& 2		& 51.9		& 9.27	& A	&13	\\
J1238$+$1602 		&          & 12 38 21.20 & $+$16 02 41.42 &  S	& -- 		& -- 	& 40.8  & --		& 115.6		& 8.9		& 56.5		& --	& A	&13	\\
J1242$+$3838 		&  	   & 12 42 36.82 & $+$38 38 06.15 &  G	& 0.408		& 308.3	& 57	& 735.5		& 136		& 8		& 24		& 8.72	& A	&30	\\
J1247$+$6723$^\star$ &  	   & 12 47 33.31 & $+$67 23 16.46 &  G	& 0.107		& 0.019	& 0.01	& 1196		& 618		& 260		& 126		& 8.63	& A	&31,32\\
J1325$-$4301 		& Cen A    & 13 25 27.62 & $-$43 01 08.81 &  G	& 0.0018	& 21.3	& 67    & 266.4$^j$	& 1800$^j$	& 28$\cdot10^4$	& 52$\cdot10^4$	& --	& A	&33,34\\
  			&      	   &   		 &  		  &   	&  		&  	&       & 533		& 14400		& 		& 96$\cdot10^4$	& 	&  &\\
J1326$+$1924 		&          & 13 26 13.67 & $+$19 24 23.75 &  G	& 0.1762	& 26	& 8.8	& 150.6		& 51		& 6		& 81.8		& 8.93	& A	&13	\\
J1328$+$2752 		&          & 13 28 48.45 & $+$27 52 27.81 &  G	& 0.0911	& 97.3	& 58	& 220.8		& 131.7		& 27.9		& 219.9		& 7.82	& A	&13	\\
J1344$-$0030 		&          & 13 44 46.92 & $-$00 30 09.31 &  G	& 0.5801	& 85.4	& 13    & 631		& 96.1		& 20.1		& 48.7		& 8.73	& A	&13	\\
J1352$+$3126 		& 3C293    & 13 52 17.88 & $+$31 26 46.49 &  G	& 0.045		& 1.1	& 1.2   & 179.5		& 204.2		& 3703		& 1209		& 8.15	& A	&35	\\      
J1407$+$5132 		& 4C51.31  & 14 07 18.48 & $+$51 32 04.63 &  G	& (0.324)	& 84.8	& 18.2  & 707.5		& 151.8		& 9.2		& 646.2		& --	& A	&13	\\
J1409$-$0302 		&          & 14 09 48.85 & $-$03 02 32.53 &  G  & 0.1378	& 52.9	& 22	& 308		& 128	        & 4		& 48		& 8.61	& A	&36,d\\ 
  			&      	   &   		 &  		  &   	&  		&  	&       & 1373		& 570		& 		& 112		& 	&  &\\
J1443$+$5201 		& 3C303    & 14 43 02.75 & $+$52 01 37.23 &  G	& 0.1412	& 81.9	& 33.3  & 90.9		& 37		& 1500		& 935		& 8.07	& B	&1	\\
J1453$+$3308 		& 4C33.33  & 14 53 02.86 & $+$33 08 42.40 &  G	& 0.249		& 158.5	& 41	& 1299		& 336		& 34		& 426		& 9.02	& A	&30,37 \\
J1500$+$1542 		&          & 15 00 55.18 & $+$15 42 40.56 &  G	& (0.456)	& 124.2	& 21.5  & 480.7		& 83.2		& 11.1		& 17.8		& --	& A	&13	\\
J1504$+$2600 		& 3C310    & 15 04 57.12 & $+$26 00 58.46 &  G	& 0.0538	& 152.1	& 147.1	& 255.4		& 247		& 1547		& 5846		& 8.29	& B	&38	\\
J1513$+$2607 		& 3C315    & 15 13 40.05 & $+$26 07 30.46 &  G	& 0.1083	& 5.9	& 3	& 262		& 134		& 350.2		& 3967		& 9.98	& A	&39,f\\      
J1516$+$0701 	& 3C317    & 15 16 44.48 & $+$07 01 17.83 &  G	& 0.0345	& 0.05	& 0.08	& 50.85   	& 75      	& 353    	& 5191    	& 8.33	& B?		&40,41	\\
{\bf J1520$-$0546} 		&          & 15 20 13.29 & $-$05 46 27.01 &  G	& 0.0601	& 118.8	& 103.7 & 1513		& 1320		& 16.5		& 132.8	& 9.14	& A	&4,a	\\   
J1521$+$5214 		&          & 15 21 05.90 & $+$52 14 39.91 &  G	& (0.537)	& 61.9	& 9.8	& 396		& 62		& 8.5		& 18.5		& --	& A	&13	\\
{\bf J1528$+$0544} 		&          & 15 28 04.95 & $+$05 44 28.18 &  G	& 0.0401	& 15.0	& 19.2  & 645.2		& 824		& 32.2		& 379.3	& 7.53	& A	&4,a\\ 
J1534$+$1016 		&          & 15 34 18.63 & $+$10 16 47.54 &  G	& 0.1333	& 164.6	& 70.3  & 507.1		& 216.6		& 43		& 280		& --	& A	&28	\\
J1538$-$0242 		&          & 15 38 41.31 & $-$02 42 05.51 &  G	& (0.575)	& 58.2	& 8.9   & 517.1		& 79.1		& 8.3		& 75.2		& --	& A	&13	\\
J1545$+$5047 		&          & 15 45 17.20 & $+$50 47 53.94 &  G	& 0.4309	& 61.5	& 11    & 361.1		& 64.6		& 14.3		& 96.9		& 8.39	& A	&13	\\
J1548$-$3216 		&	   & 15 48 58.05 & $-$32 16 57.60 &  G	& 0.108		& 312.2	& 160	& 961.8		& 493		& 78		& 1722		& --	& A	&42,43\\
J1604$+$3438 		&	   & 16 04 45.89 & $+$34 38 16.53 &  G	& 0.282		& 211.6	& 50	& 846.4		& 200		& ~5		& 146		& --	& A	&20	\\
J1605$+$0711 		&          & 16 05 13.74 & $+$07 11 52.56 &  G	& 0.3112	& 344.1	& 75.9	& 538.6		& 118.8		& 55.3		& 212.6		& 8.36	& A	&13	\\
J1627$+$2906 		&          & 16 27 54.63 & $+$29 06 20.01 &  G	& (0.722)	& 73.8	& 10.2	& 697.4		& 96.4		& 10.4		& 112.9		& --	& A	&13	\\
J1628$+$3933 	& 3C338    & 16 28 38.24 & $+$39 33 04.55 &  G	& 0.0304	& 7.8	& 13	& 48      	& 80      	& 224    	& 3380    	& 9.01	&	 B	&7,44	\\
J1649$+$4133 		&          & 16 49 28.32 & $+$41 33 41.61 &  S	& -- 		& --	& 6.7	& --		& 39.7		& 2.4		& 39		& --	& A	&13	\\
J1651$+$0459 		& 3C348	   & 16 51 08.15 & $+$04 59 33.32 &  G	& 0.1550	& 212.6	& 80	& 319.5		& 120.2		&		&	& 9.07	& B?	&45	\\	
J1705$+$3940 		&          & 17 05 17.83 & $+$39 40 29.25 &  G	& (0.701)	& 305.3	& 42.7	& 592.7		& 82.9		& 35.2		& 70.7		& --	& A	&13	\\
J1706$+$4340 		&          & 17 06 25.44 & $+$43 40 40.16 &  S	& (0.525) 	& 194	& 31.8	& 687.2		& 110.1		& 79.3		& 69.3		& --	& A	&13,46	\\
J1835$+$6204 		&  	   & 18 35 10.92 & $+$62 04 08.14 &  G	& 0.519		& 372.3	& 60	& 1378		& 222		& 200		& 604		& --	& A	&30	\\
J1844$+$4533$^\star$ & 3C388   & 18 44 02.40 & $+$45 33 29.70 &  G & 0.0917	& 67.48	& 40	& 84.35   	& 50      	& 3362   	& 2205    	& 9.47	&	 B	&47,48	\\
J2048$+$0701 		& 3C424    & 20 48 12.03 & $+$07 01 17.48 &  G	& 0.127		& 22.5	& 10	& 49.2 		& 21.9 		& 625		& 1816		& --	& A	&49	\\
J2107$+$2331 	& 4C 23.56 & 21 07 15.08 & $+$23 31 43.71 &  G  & 2.483 	& 457	& 55.7	& 492.2 	& 60      	&        	& 397     	& --	& 	A	&50\\ 
J2223$-$0206 		& 3C445	   & 22 23 49.54 & $-$02 06 12.90 &  G	& 0.056		& 130	& 121	& 612		& 570		& 55		& 5260		& 8.12	& A	&51,52\\
J2345$-$0449 		&          & 23 45 32.71 & $-$04 49 25.32 &  G	& 0.0757	& 338.9	& 238.9	& 1454		& 1025		& 29		& 148.4	& --	& A	&4,53\\
\hline \\
\label{ddrs}
\end{longtable}

\vspace{-0.9cm}

\hspace{-0.63cm}{\bf Column designation:} (1) -- source name, (2) -- other name, (3), (4) -- right ascension and declination (J2000), (5) -- optical identification (G -- galaxy, Q -- quasar, S -- unidentified source), (6) -- redshift, (7) and (9) -- projected linear size of the inner and outer lobes, respectively, (8) and (10) -- apparent angular size of the inner and outer lobes respectively, (11) and (12) -- flux densities at 1.4\,GHz of the inner and outer lobes, respectively, 
(13) -- mass of the central black hole, (14) -- class of radio structure (A or B), (15) -- references. The redshift given in brackets corresponds to the Sloan Digital Sky Survey (SDSS) photometric redshift. For two objects, J1409$-$0302 and J1325$-$4301, we wrote two values of l$\rm_{out}$ and S$\rm_{out}$, which correspond to the length and flux density of the middle and outer lobes, respectively. The bolded names of sources correspond to newly recognized restarting radio galaxies.\\
{\bf Notes:}
(a) a new DDRS;
(b) flux density measured at 1.5~GHz; 
(c) This source is counted as radio relic associated with a giga-hertz peaked (GPS) source by \citet{Stanghellini2005} and by \citet{Shulevski2015}. However,
 the optical DSS map (see the attached figure in the Appendix) reveals a weak extended emission in the vicinity of the extended diffuse relic. Therefore, in our 
opinion, the association is questionable. Deep optical observation are necessary.;  
(d) triple-double radio galaxy;
(e) the structure should be confirmed; the flux density of 1120 mJy is for the entire source;
(f) X-shape radio source;
(g) there are two nearby galaxies; identification consistent with FIRST is REC: 09$\rm^h$29$\rm^m$10$\fs$38 DEC: $+$41$\degr$46$'$44$\farcs$5 (J2000.0);
(h) GHz Peaked Spectrum radio source;
(i) Compact Steep Spectrum radio source;
(j) Size of one side middle lobe;
($^\star$) -- restarting radio galaxies with inner lobes more luminous than the outer ones.
\\
\\
{\bf References:} 
(1) \citet{Leahy1991},
(2) \citet{Shulevski2012},
(3) \citet{Saikia2006},
(4) \citet{Machalski2007},
(5) \citet{Saripalli2012}
(6) \citet{Baum1990},
(7) \citet{Saripalli2002}, 
(8) \citet{Shulevski2015},
(9) \citet{Morganti1993},
(10) \citet{Subrahmanyan1996},
(11) \citet{Saripalli2013}, 
(12) \citet{Siemiginowska2003},
(13) \citet{Nandi2012}, 
(14) \citet{Marecki2009},
(15) \citet{Jamrozy2007}, 
(16) \citet{Giovannini2005},
(17) \citet{Perley1980},
(18) \citet{Bridle1986}, 
(19) \citet{Clarke1992}, 
(20) \citet{Machalski2006},
(21) \citet{Brocksopp2007}, 
(22) \citet{Jamrozy2009},
(23) \citet{Stanghellini2005},
(24) \citet{Sikora2013},
(25) \citet{Willis1974},
(26) \citet{Strom1980},  
(27) \citet{Schilizzi2001},
(28) \citet{Owen1997},
(29) \citet{Koziel2012},
(30) \citet{Schoenmakers2000},
(31) \citet{Marecki2003},   
(32) \citet{Bondi2004},
(33) \citet{Burns1983}, 
(34) \citet{Clarke1992b},
(35) \citet{Bridle1981},
(36) \citet{Hota2011},
(37) \citet{Konar2006},
(38) \citet{vanBreugel1984},
(39) \citet{Saripalli2009},
(40) \citet{Venturi2004},
(41) \citet{Zhao1993},
(42) \citet{Saripalli2003},
(43) \citet{Safouris2008},
(44) \citet{Ge1994},
(45) \citet{Colafrancesco2013},
(46) \citet{Marecki2016},
(47) \citet{Burns1982},
(48) \citet{Roettiger1994},
(49) \citet{Black1992},
(50) \citet{Blundell2011},
(51) \citet{Kronberg1986},
(52) \citet{Leahy1997},
(53) \citet{Bagchi2014}.

\end{landscape}
\end{onecolumn}

%
\begin{onecolumn}
\newpage
\appendix
\section{Radio maps of restarting radio sources -- class A}

This appendix presents radio maps of restarting radio sources, which are of typical double-double radio morphology -- class A sources. Most of the radio maps were obtained with the Very Large Array at 1.4 GHz. Maps of J0041$+$3224, J0351$-$2744, J0927$+$3510, J1648$-$3218, J1604$+$3438, J2107$+$2331 radio sources show radio contours at 4.8 GHz, and of J0037$+$1319, 2048$+$0701 at 8.4 GHz. The map of J0104$-$6609 was taken from \citep{Saripalli2012}. Three radio maps were plotted using data from other radio telescopes. The 330 MHz map of J0116$-$4722 was taken by Giant Metrewave Radio Telescope, the 1.4 GHz map of inner structure of J0821$+$2117 is from the EVN/MERLIN measurement, and the 4.9 GHz inner structure of J1247$+$6723 comes from VLBI. In all the cases the radio contours are overlaid onto the R-band optical image from Digital Sky Survey (DSS).\\
\begin{onecolumn}
\begin{figure}
    \includegraphics[height=5.5cm]{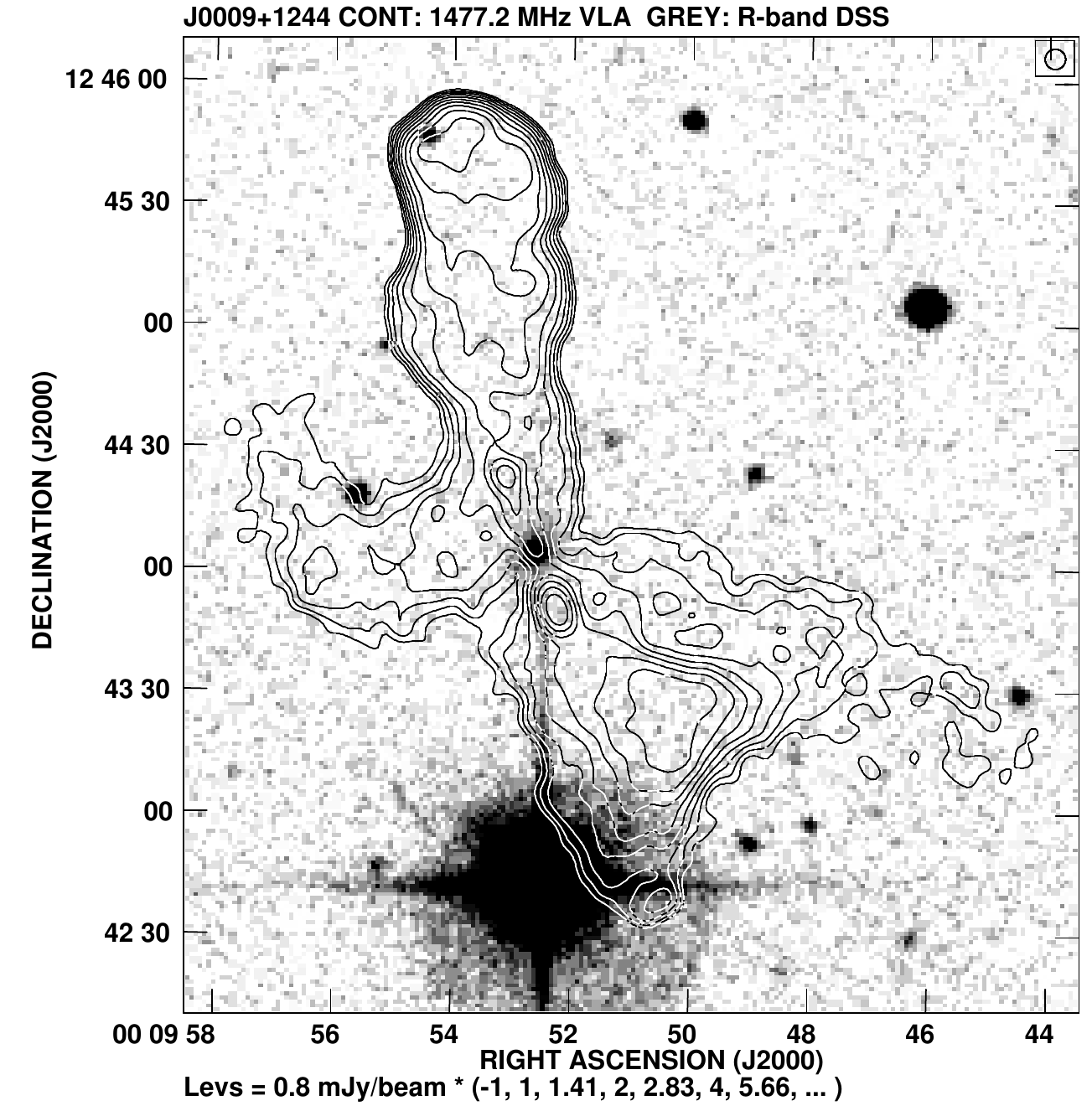} 
    \includegraphics[height=5.4cm]{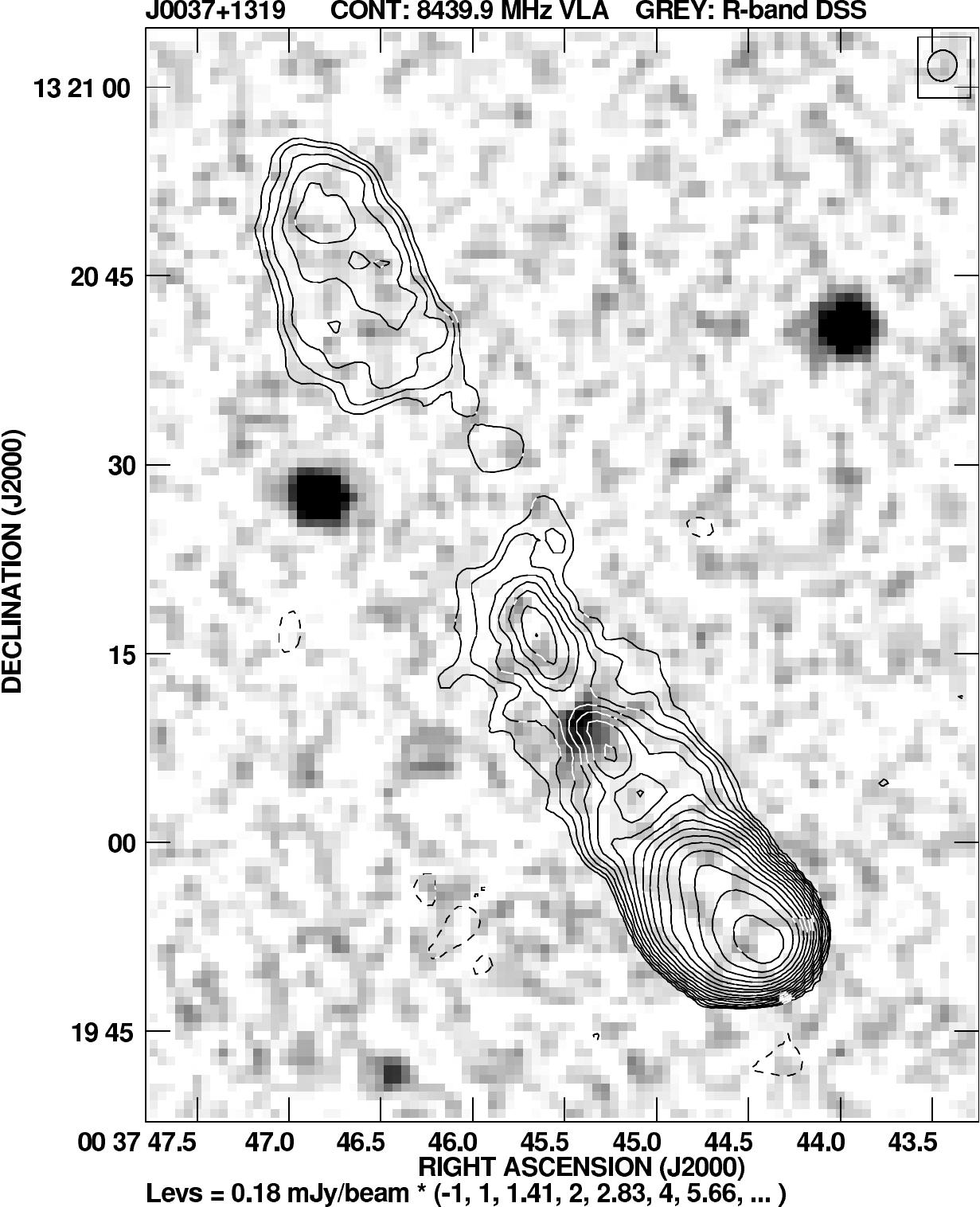} 
    \includegraphics[width=7cm]{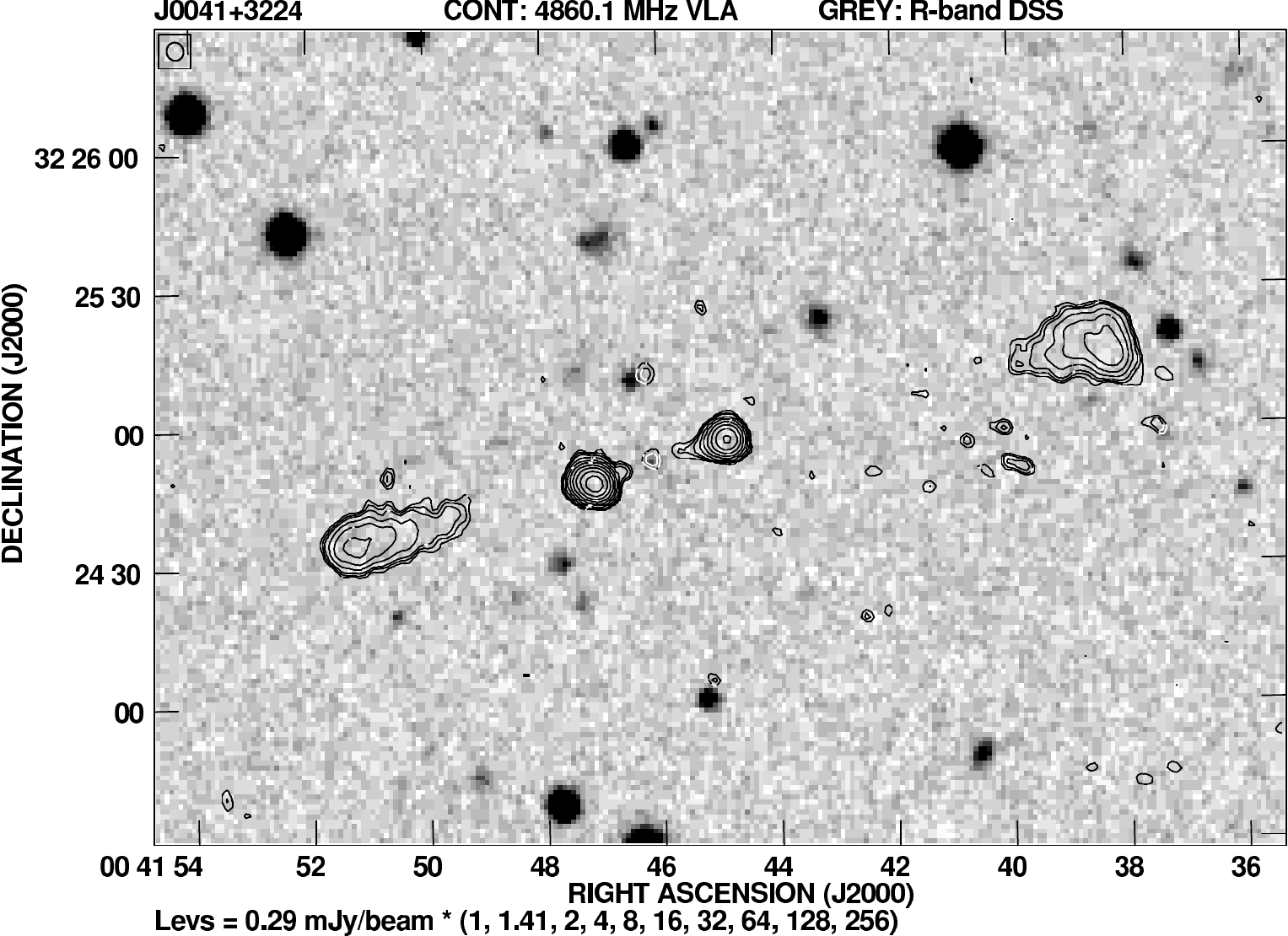}\\
\vspace{0.4cm}
    \hspace{-0.2cm}\includegraphics[height=5.3cm]{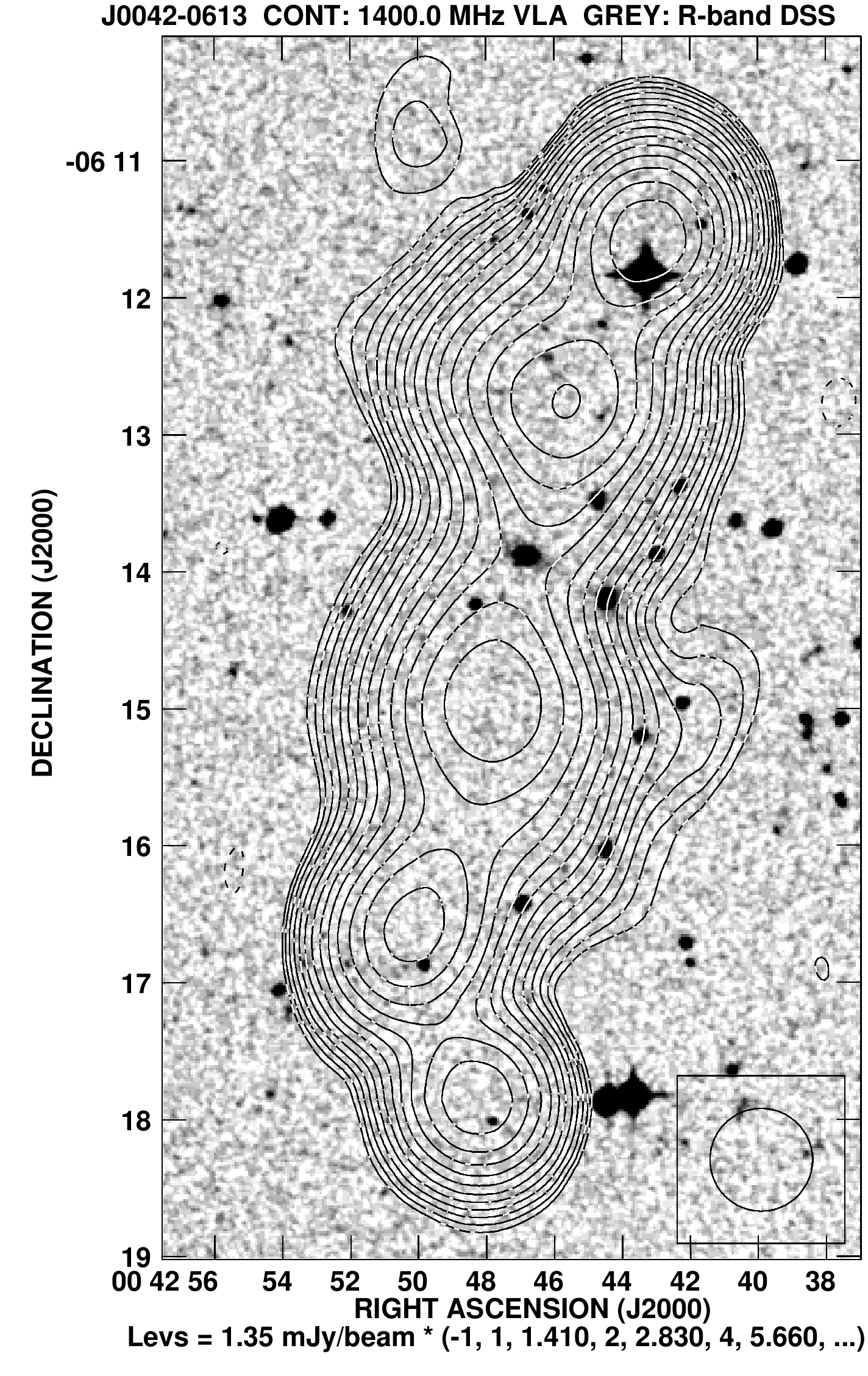} 
    \hspace{-0.2cm}\includegraphics[height=5.5cm]{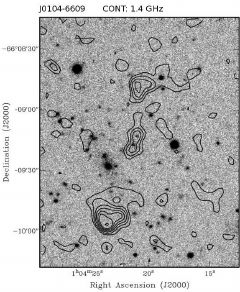}
    \includegraphics[height=5.3cm]{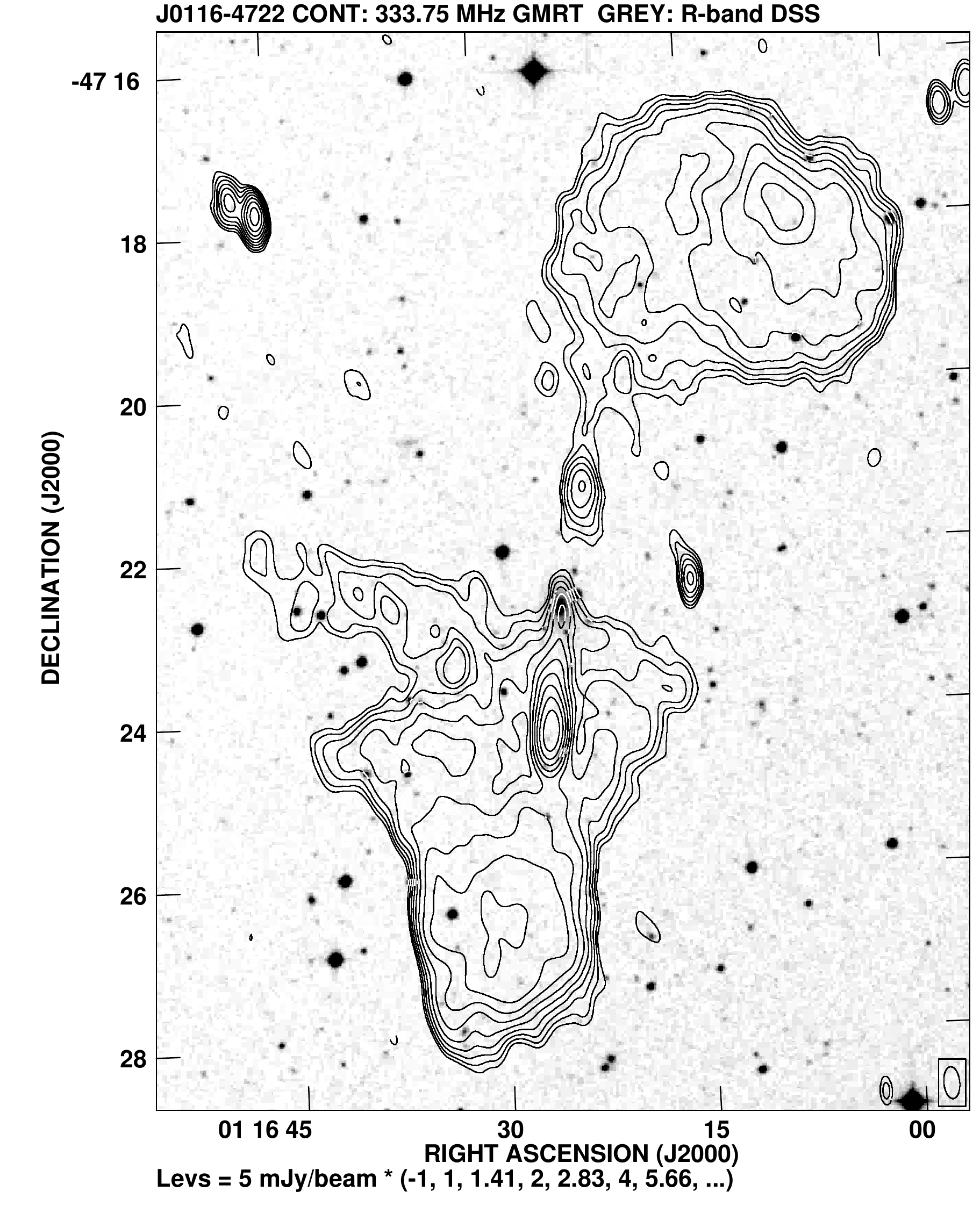}
    \includegraphics[height=5.3cm]{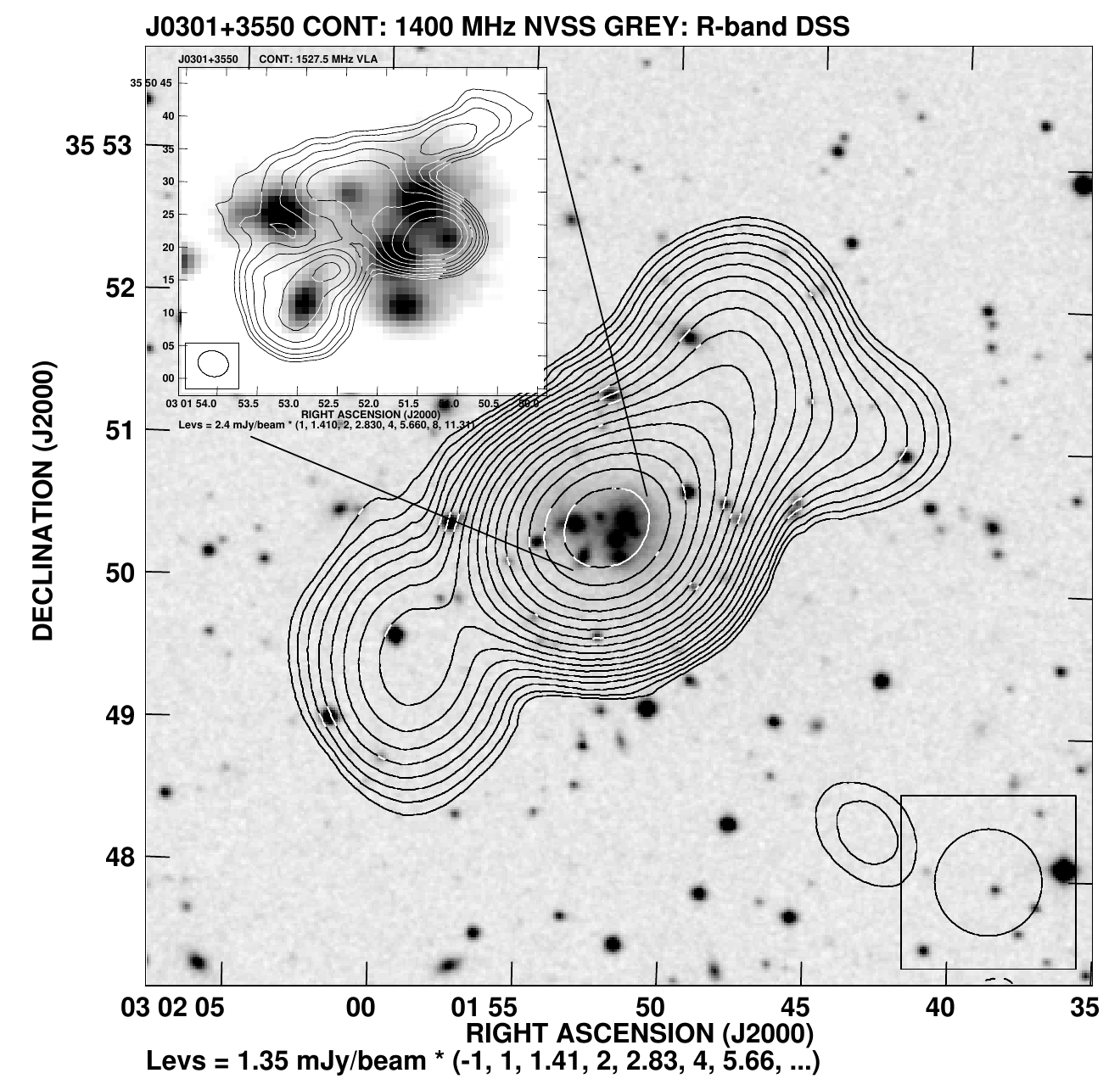}\\
\vspace{0.4cm}
    \includegraphics[height=5.5cm]{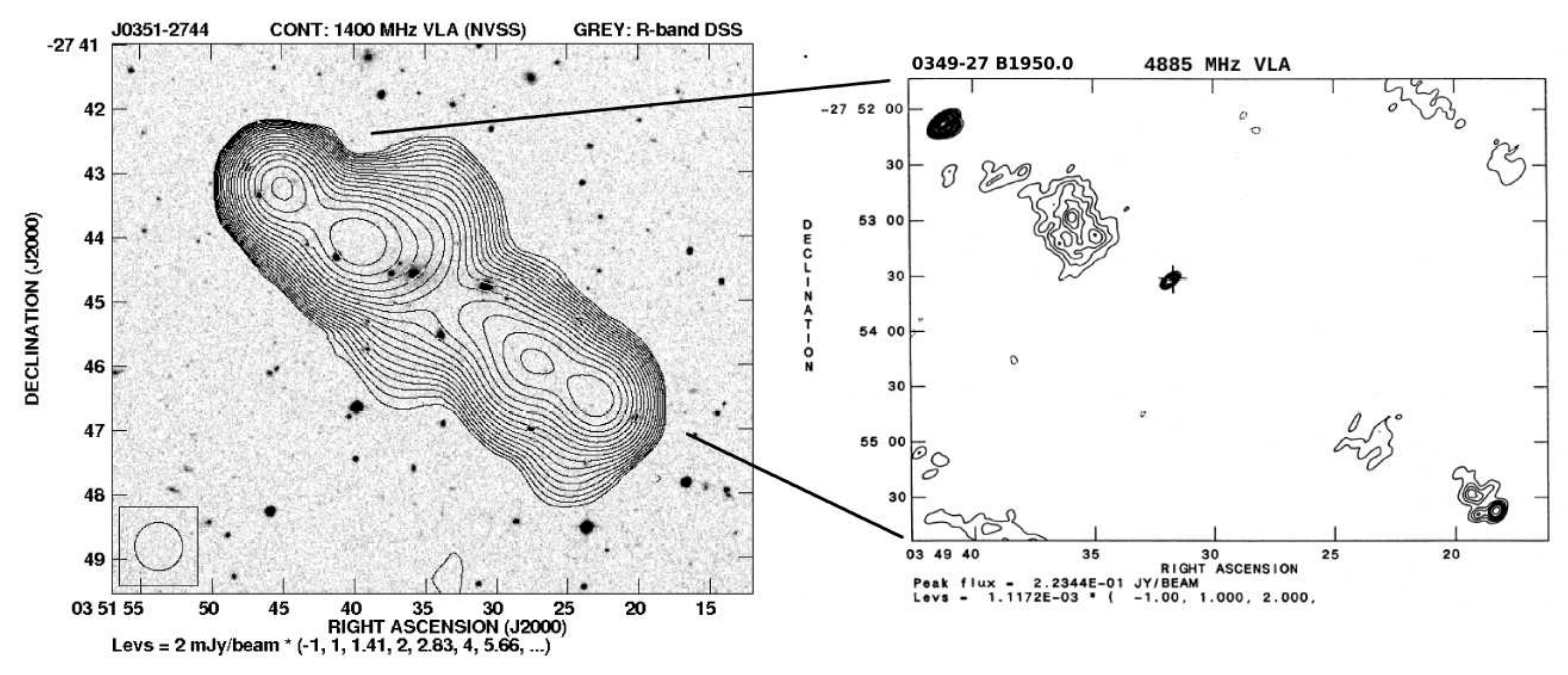}
    \includegraphics[height=5.5cm]{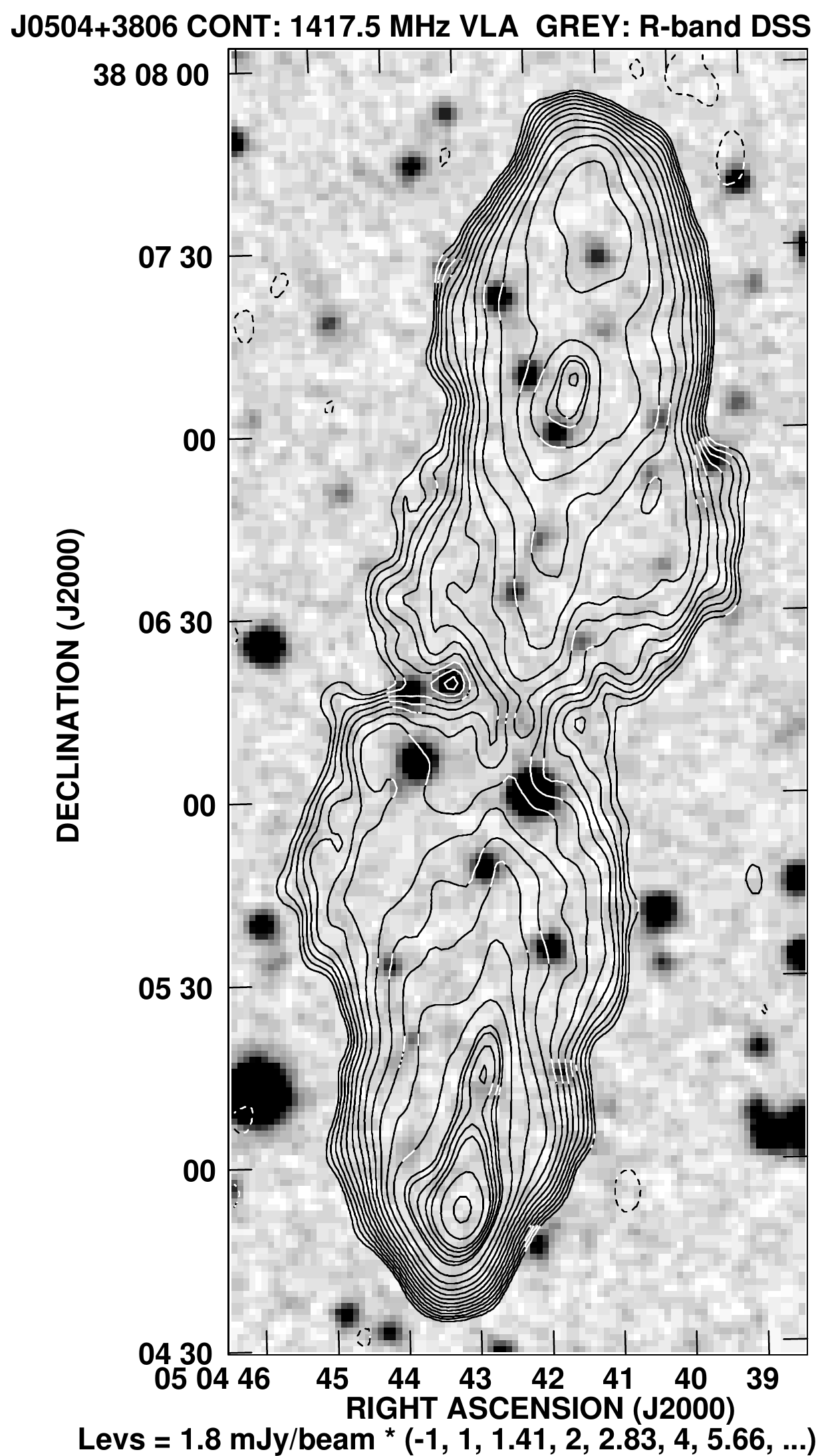}\\
\vspace{0.4cm}
\hspace{2cm}\includegraphics[height=5.5cm]{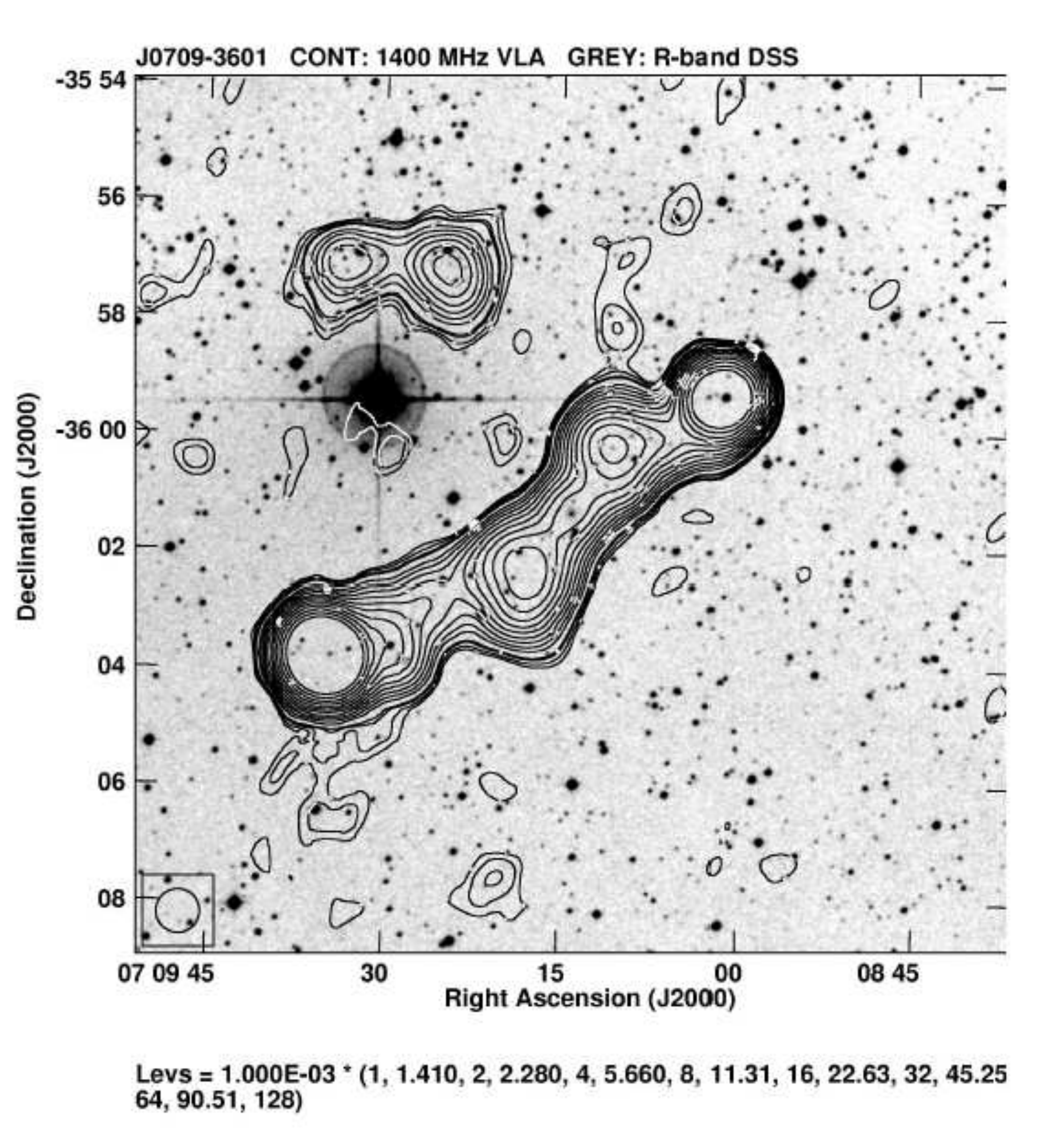} 
    \includegraphics[height=5.5cm]{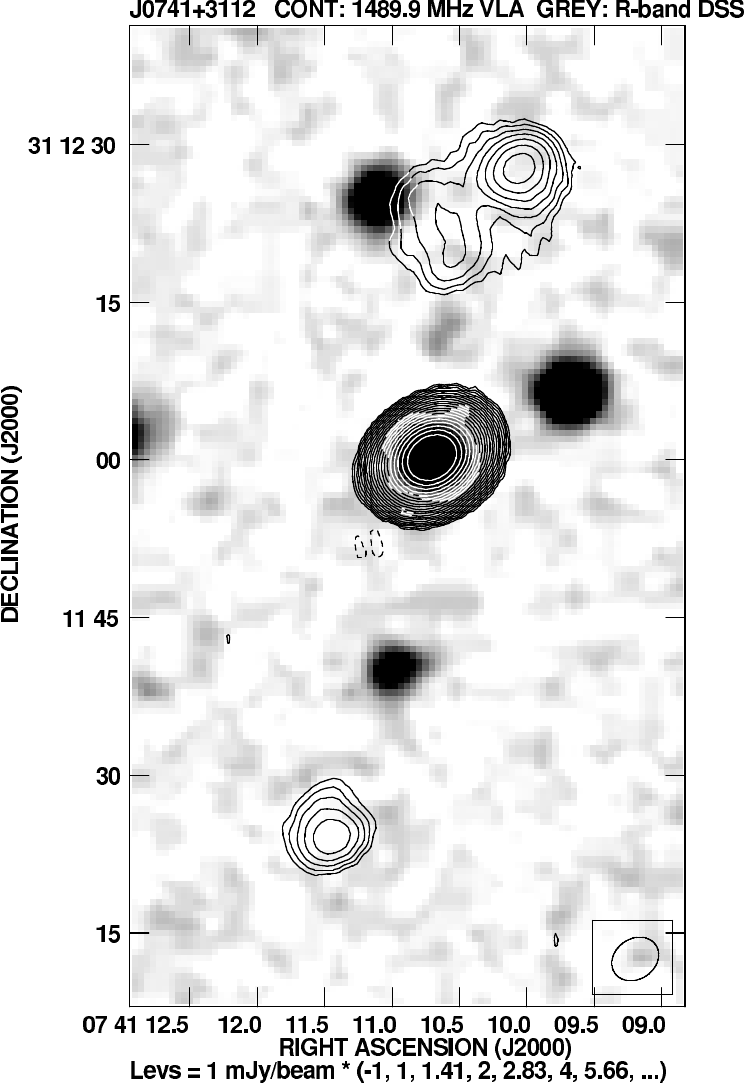}
    \includegraphics[height=5.5cm]{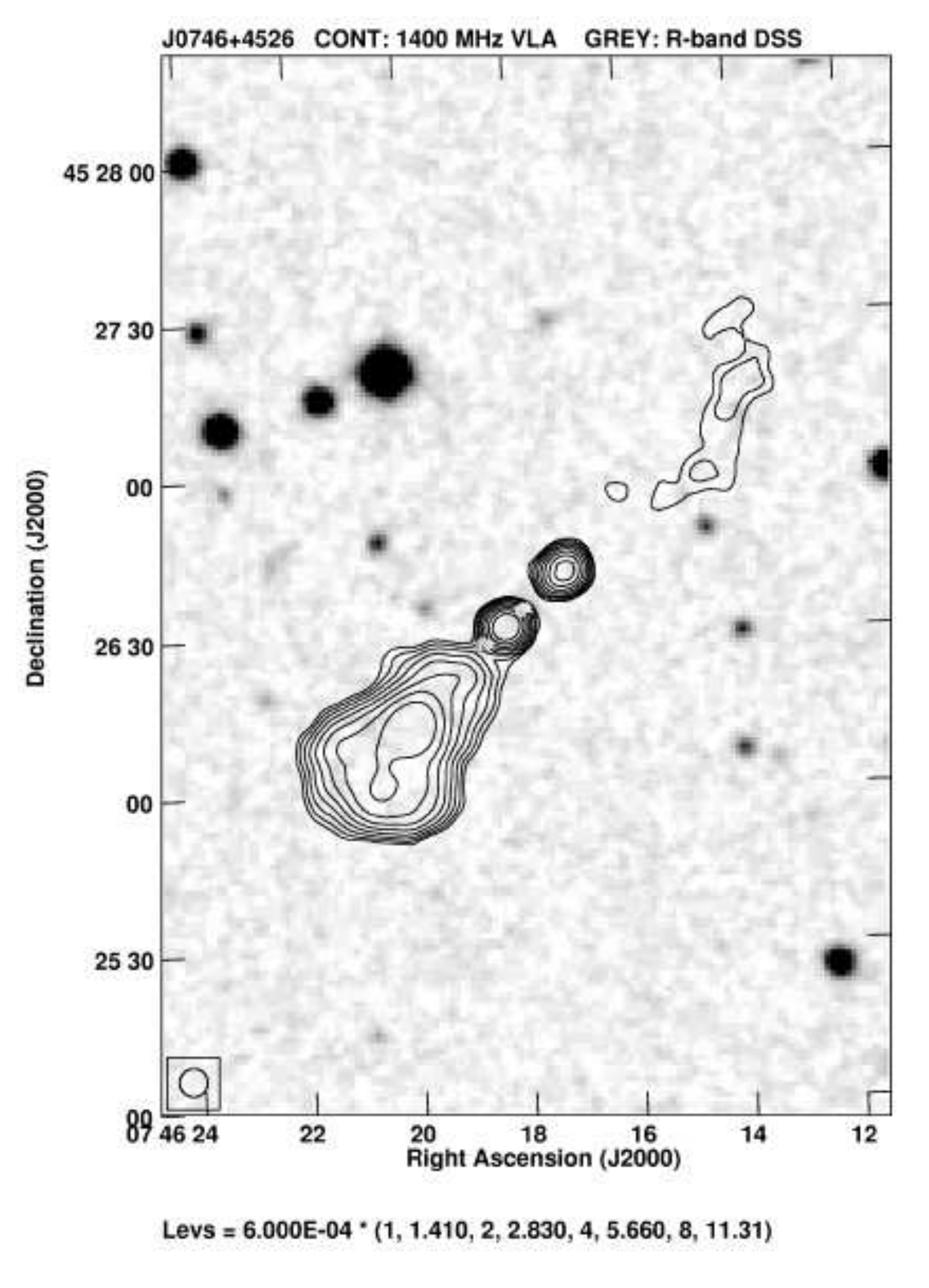}
\end{figure}
\begin{figure}
\centering
    \includegraphics[width=7.5cm]{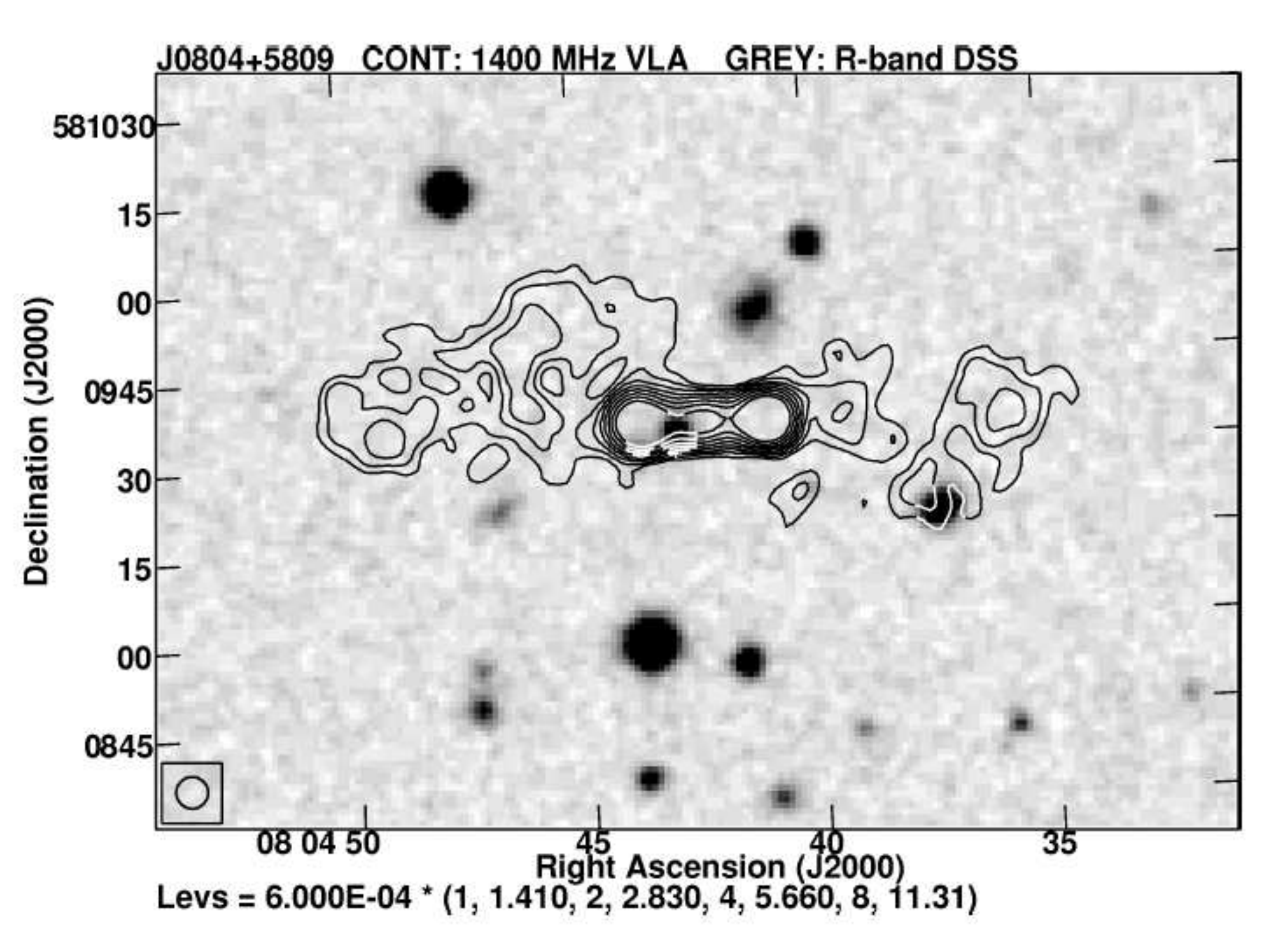}
    \includegraphics[height=5.5cm]{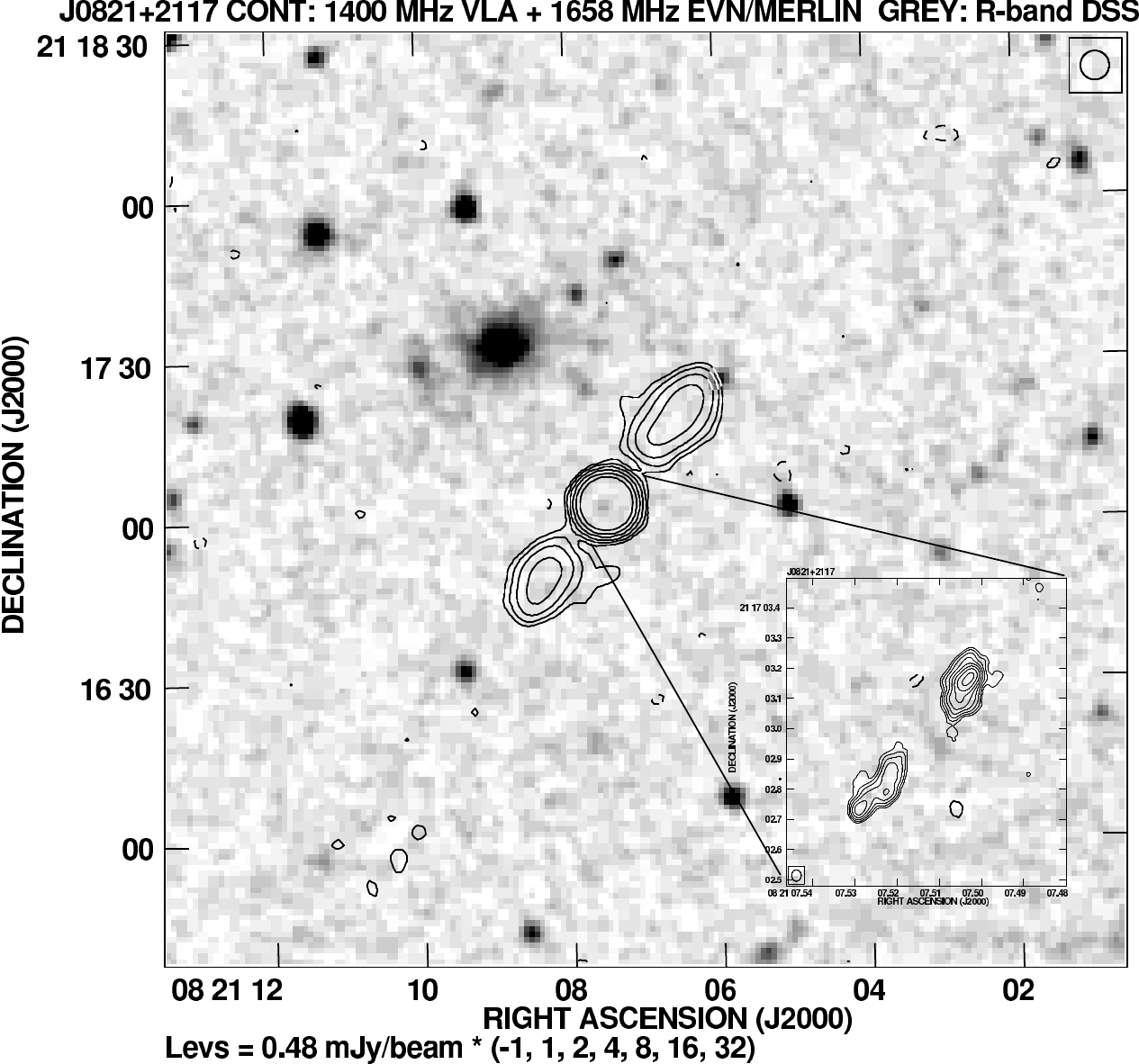}
    \includegraphics[height=5.5cm]{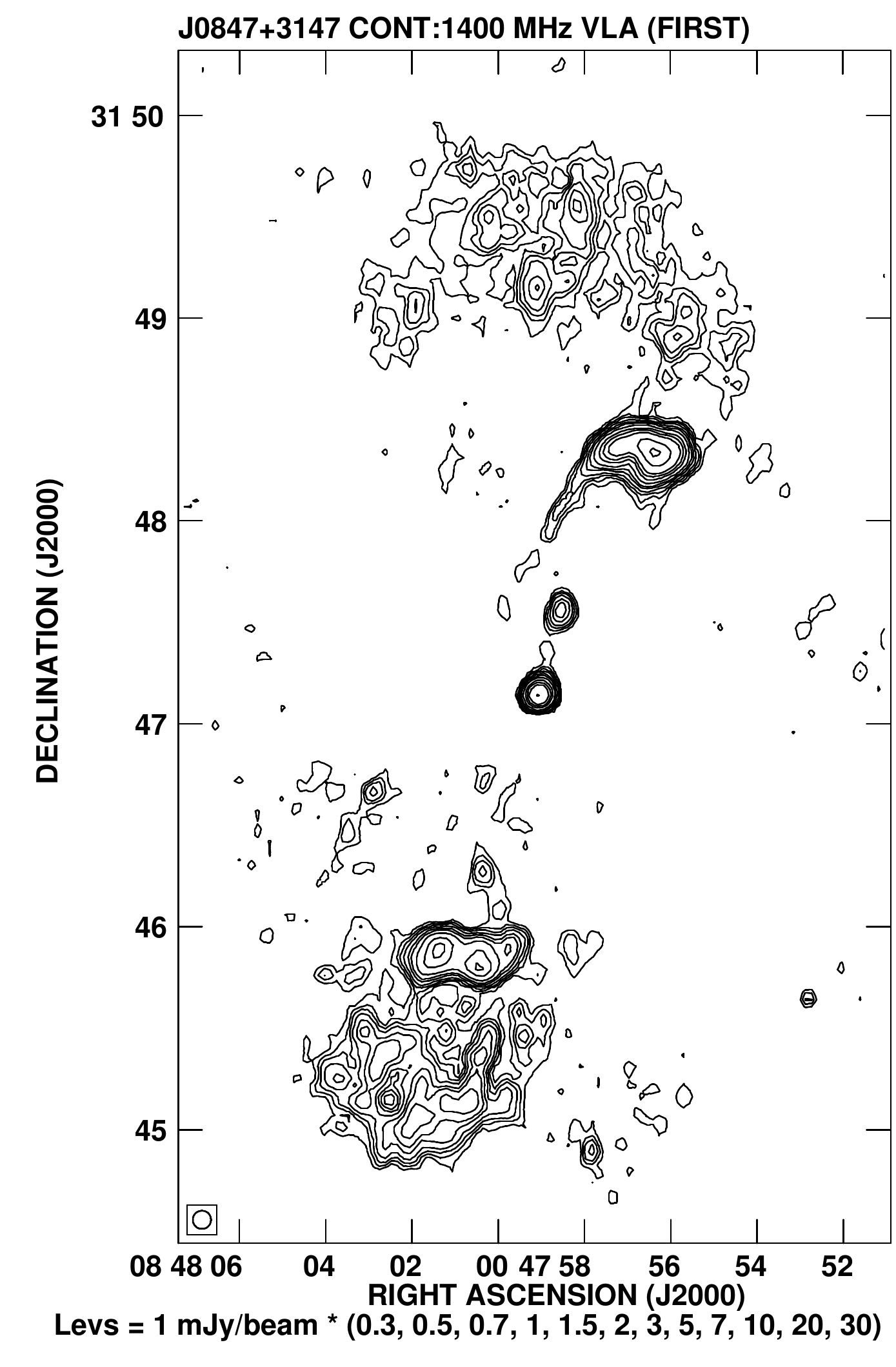} \\
\vspace{0.4cm}
    \includegraphics[height=5.5cm]{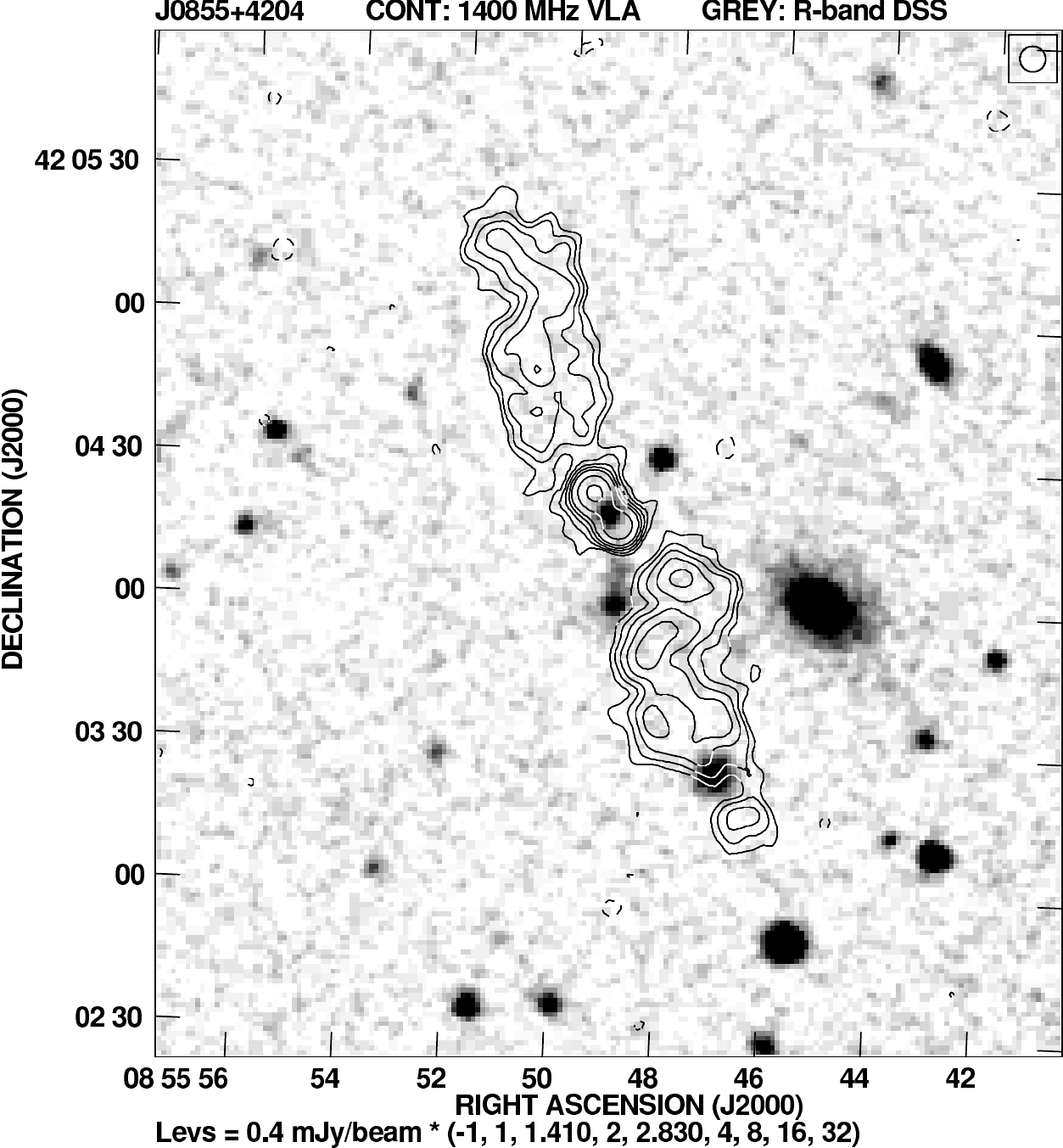} 
    \includegraphics[height=5.5cm]{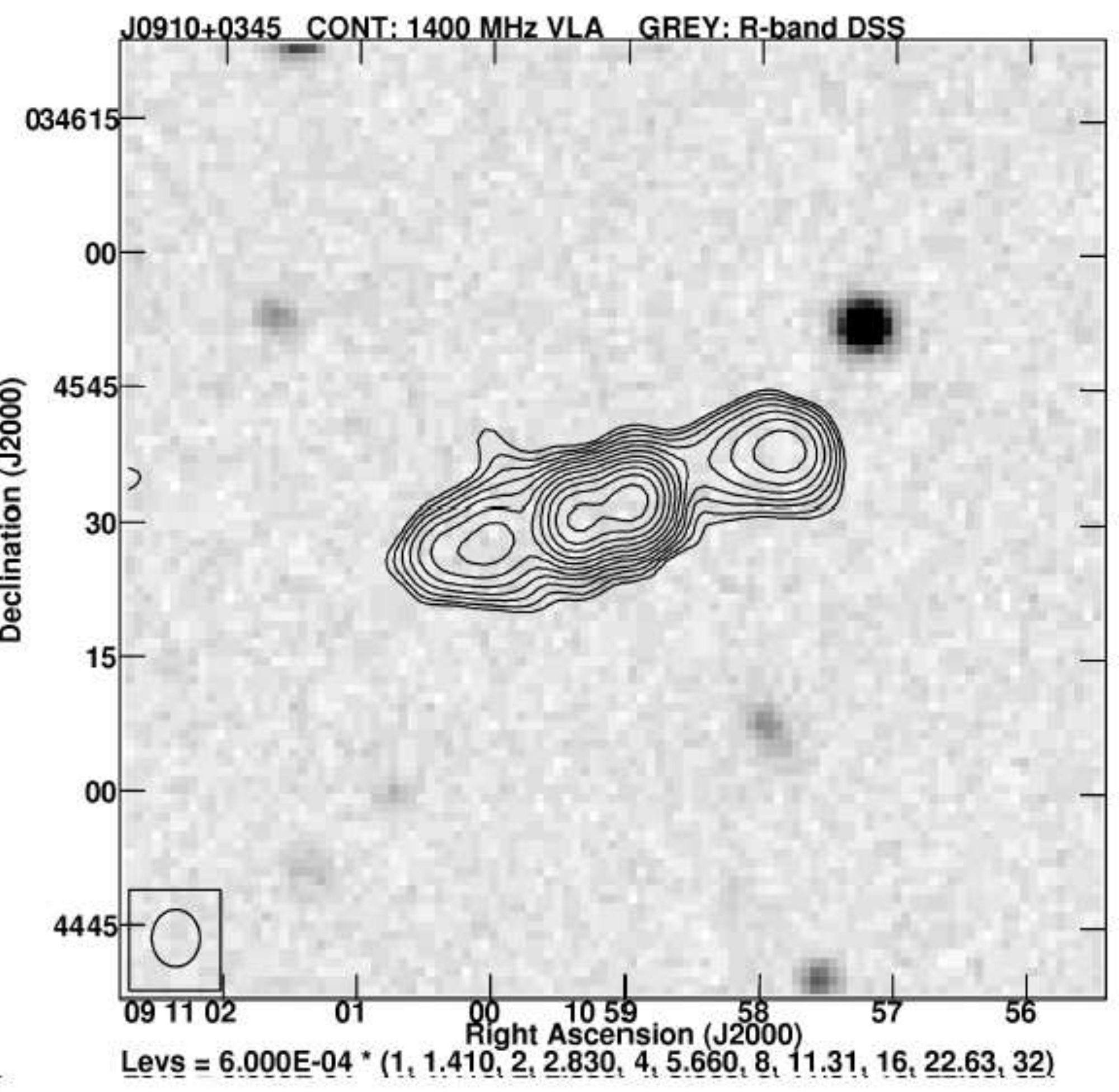}
    \includegraphics[height=5.5cm]{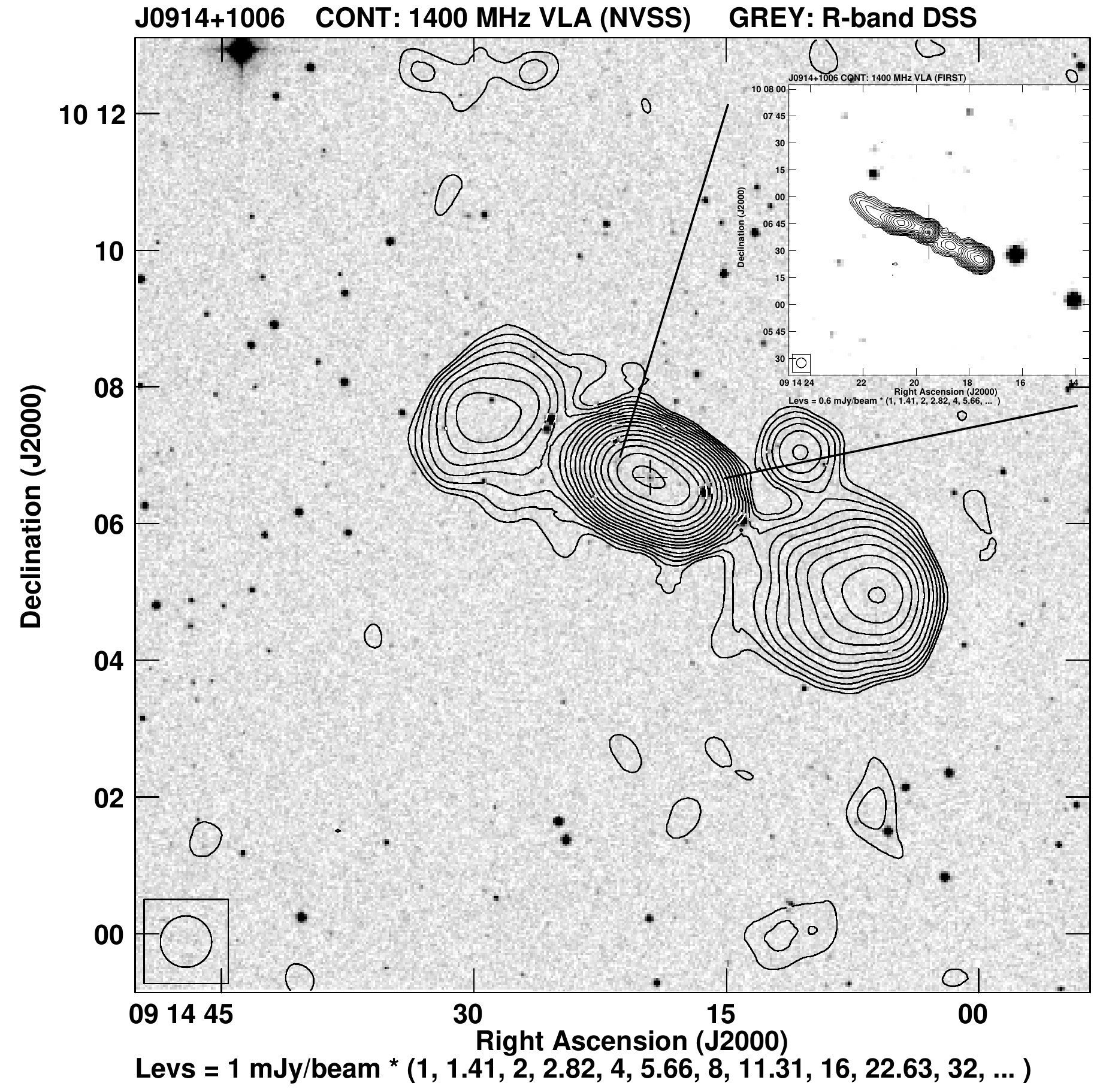}\\
\vspace{0.4cm}
   \includegraphics[height=5.5cm]{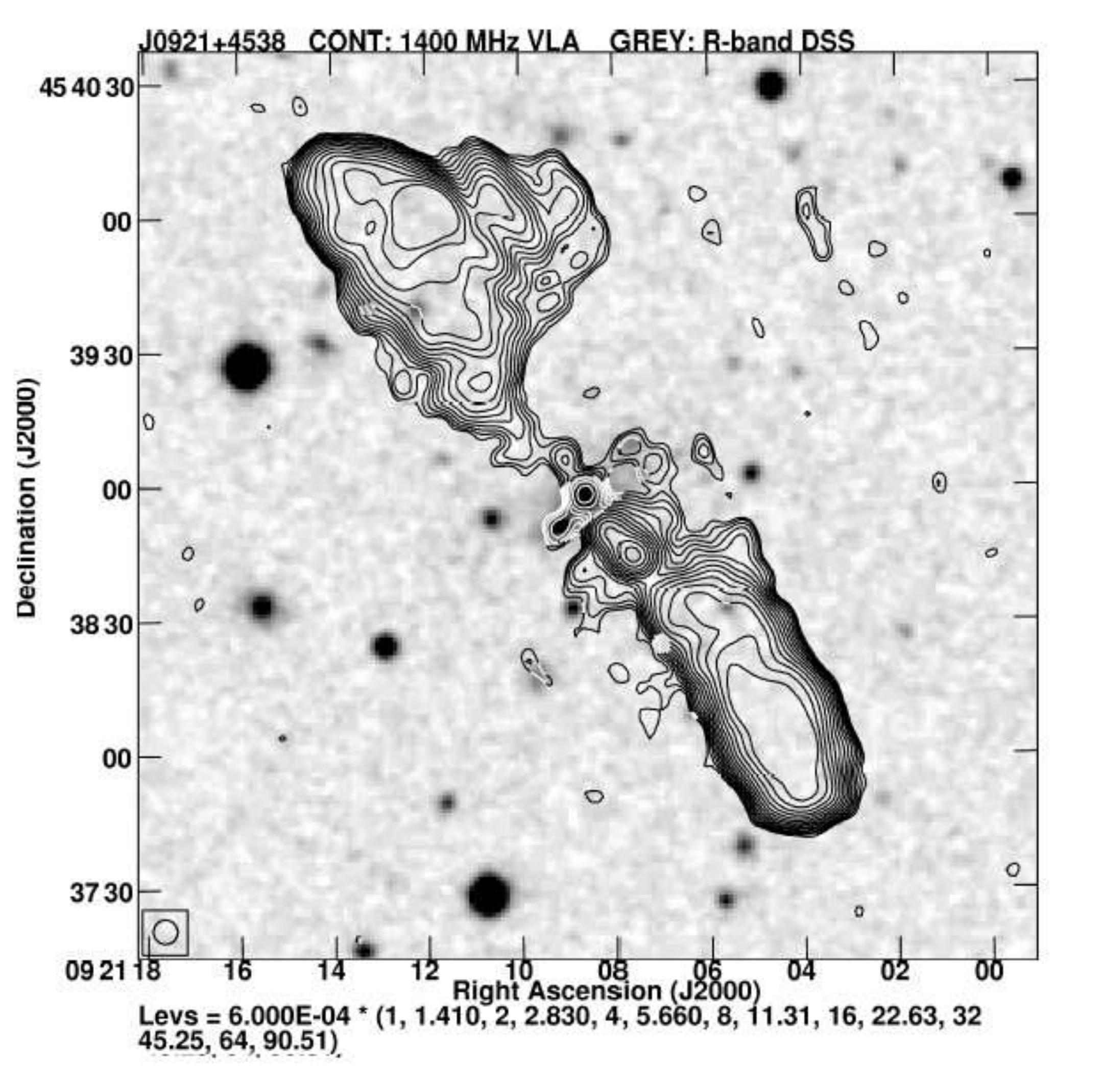}
   \hspace{0.5cm}\includegraphics[width=8.5cm]{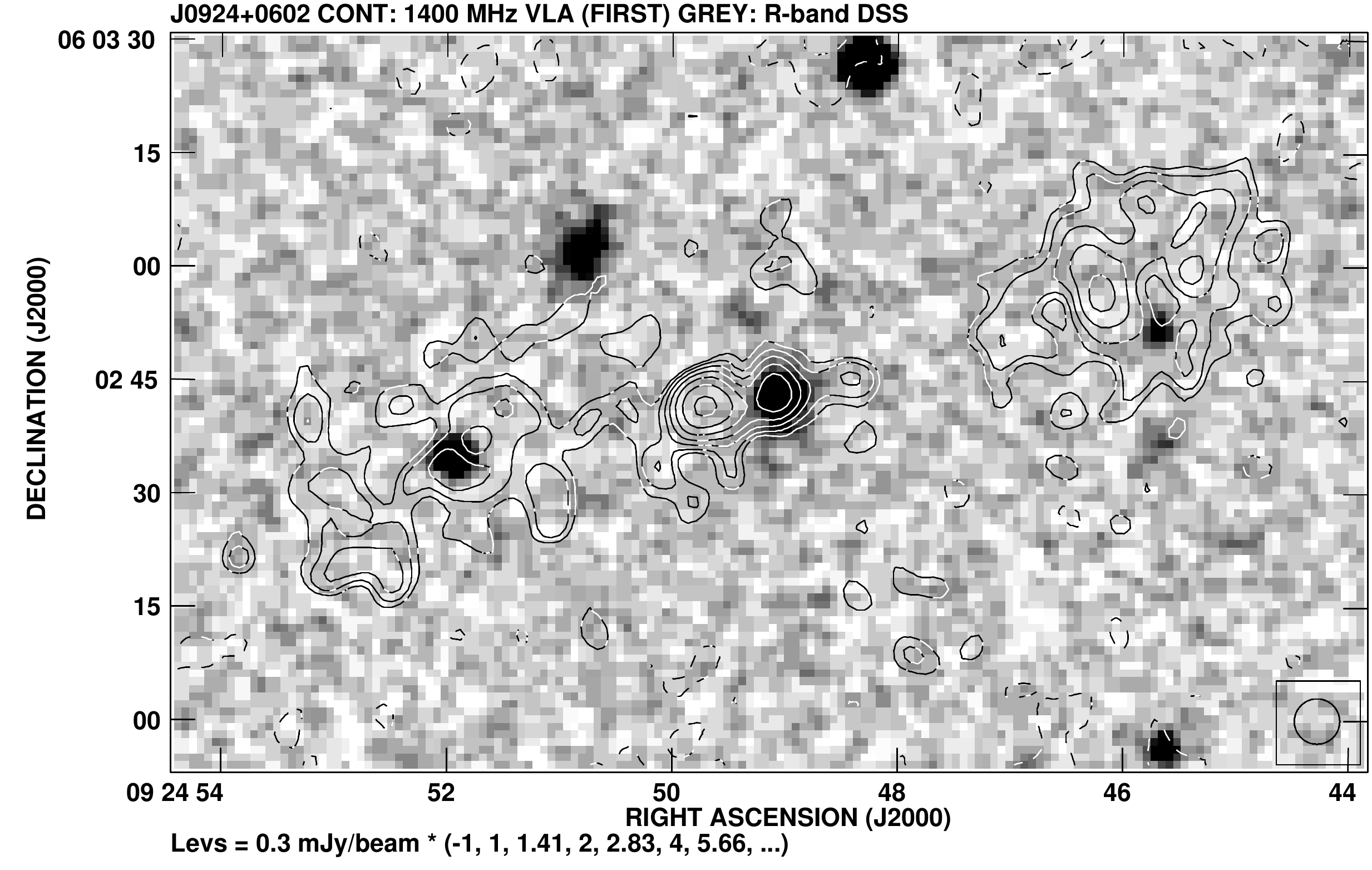} \\
\vspace{0.4cm}
\includegraphics[width=7.5cm]{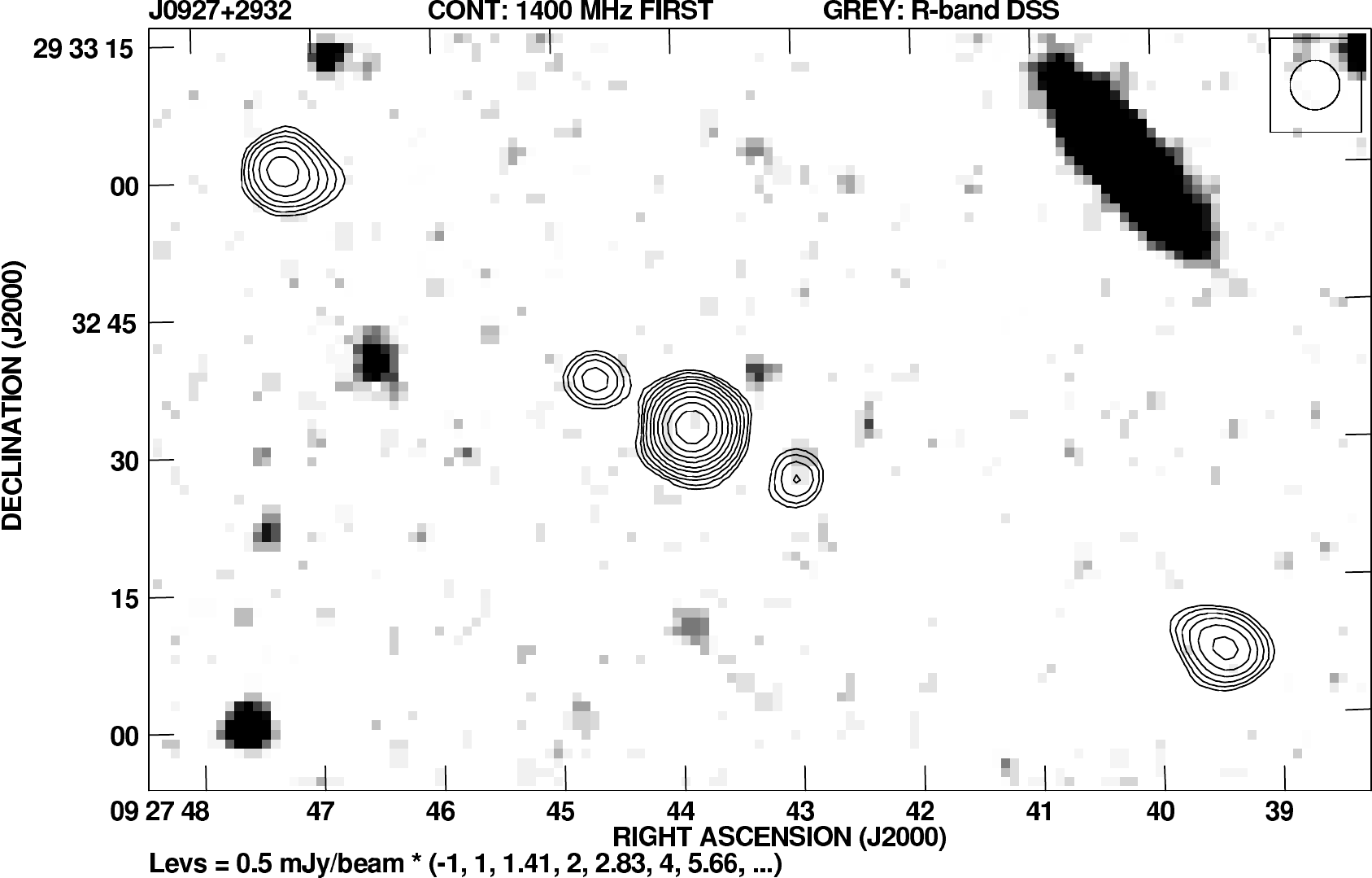}
    \includegraphics[width=7cm]{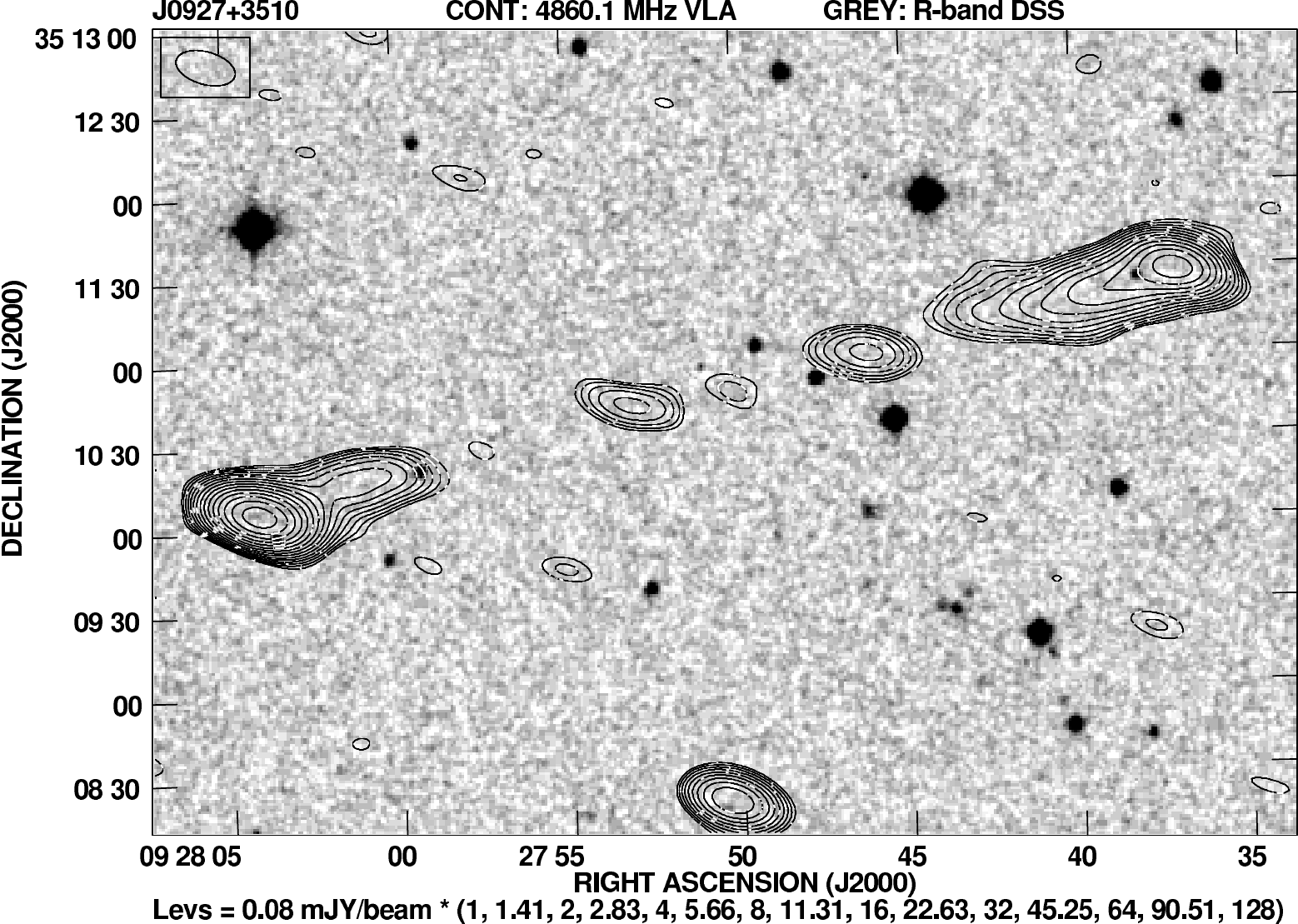}
\end{figure}
\begin{figure}
    \includegraphics[height=5.5cm]{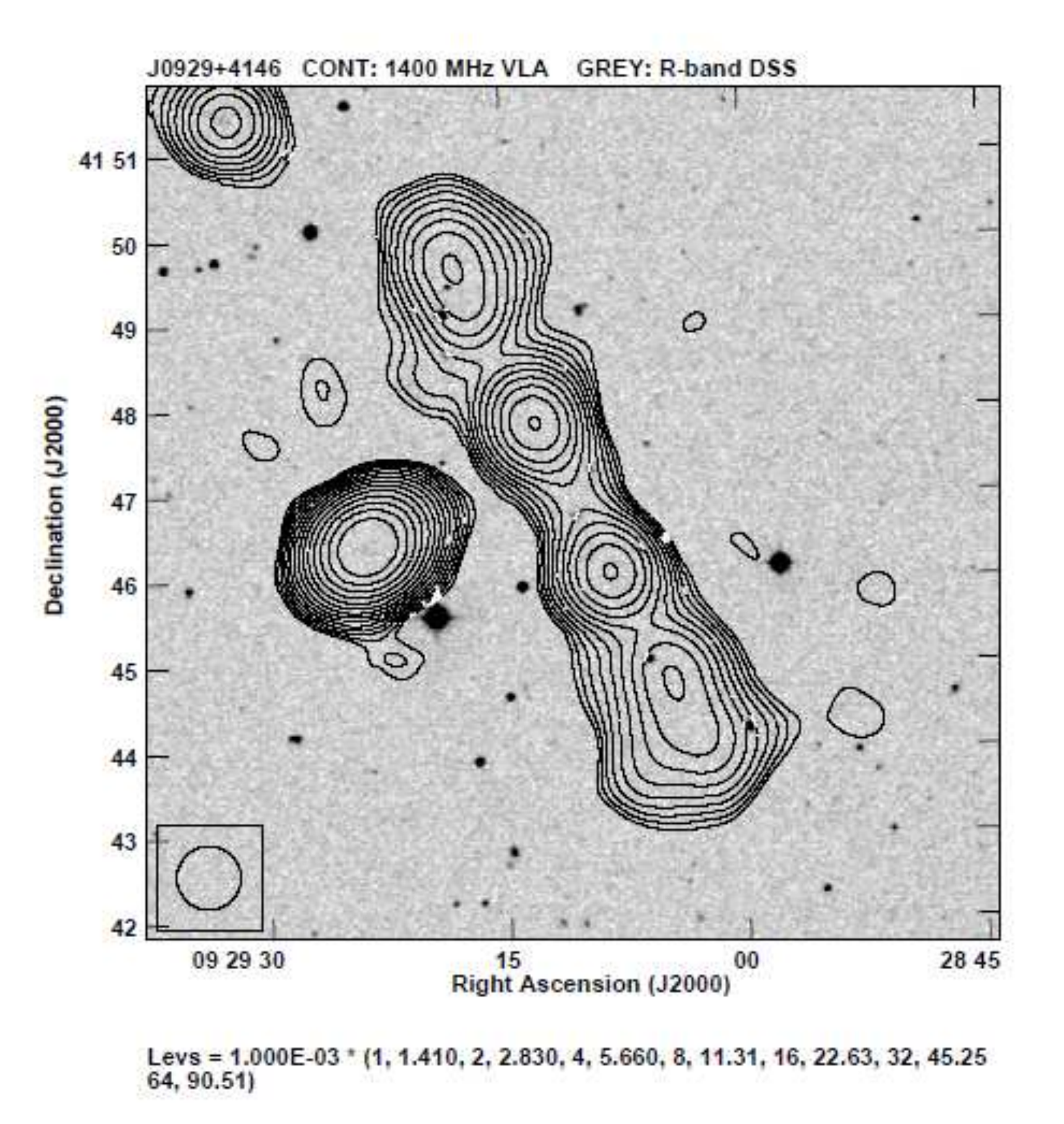}
    \hspace{0.5cm}\includegraphics[height=5.5cm]{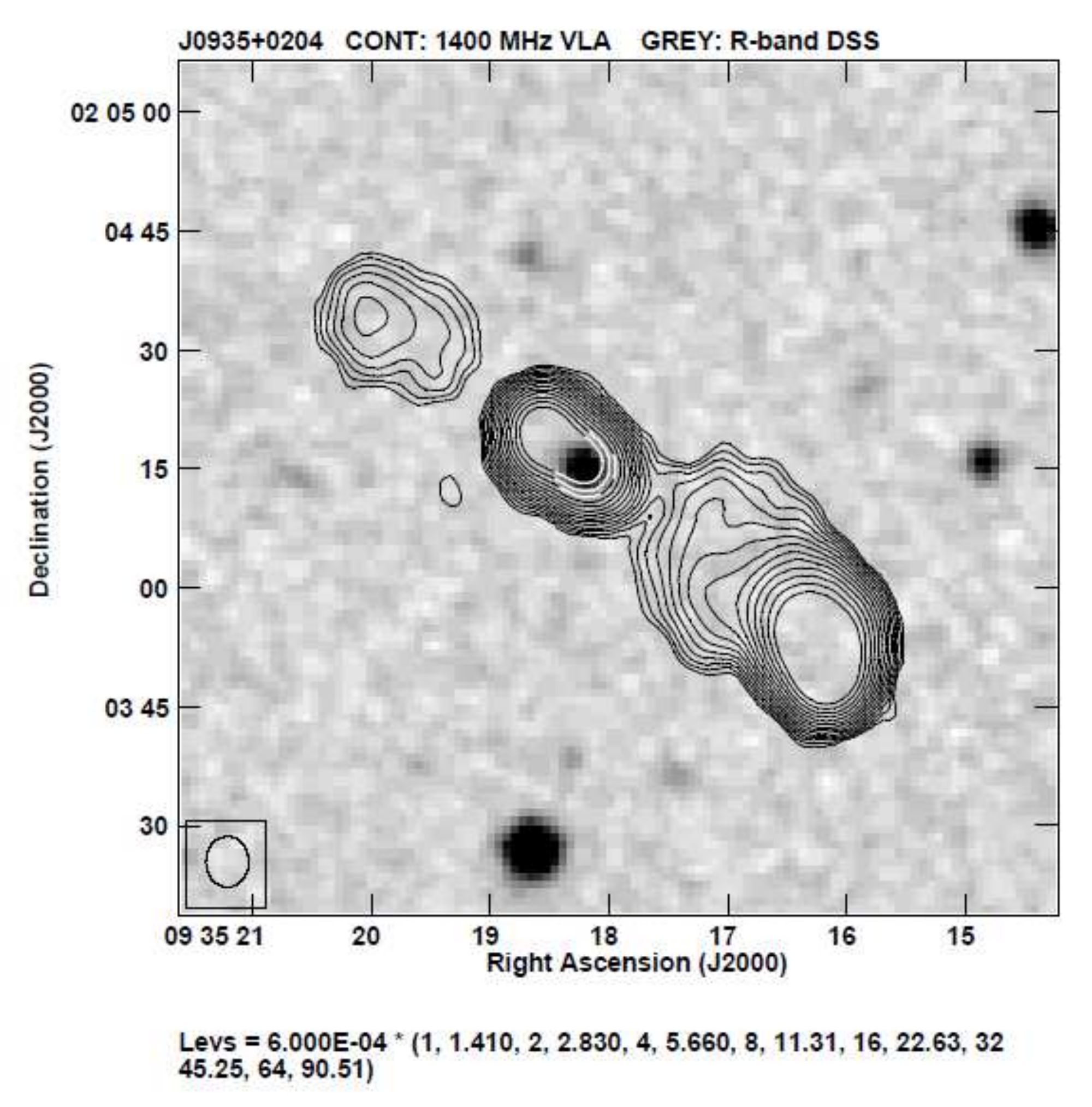}
    \hspace{0.5cm}\includegraphics[height=5cm]{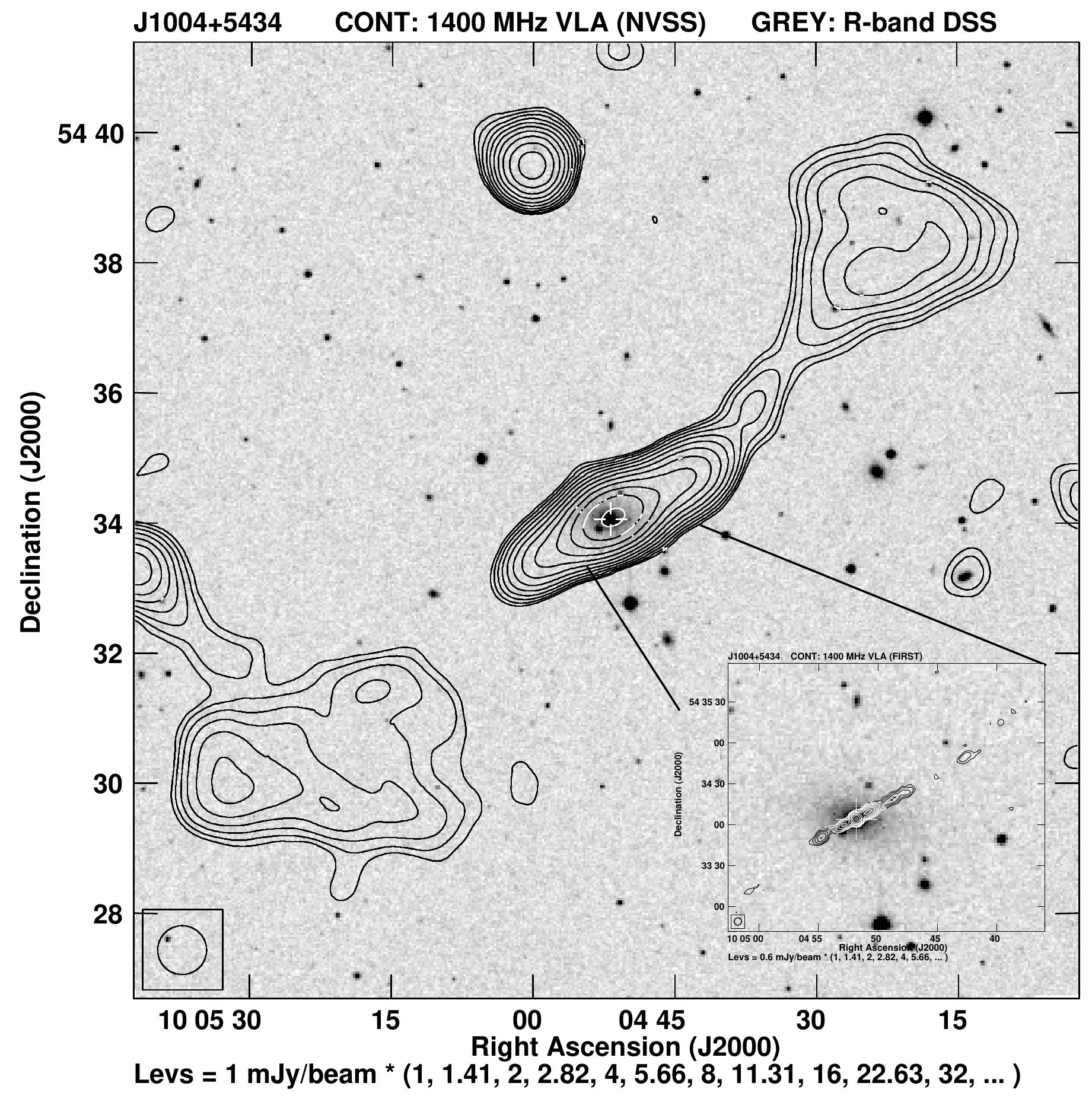}\\
\vspace{0.4cm} 
    \hspace{2cm}\includegraphics[height=5.3cm]{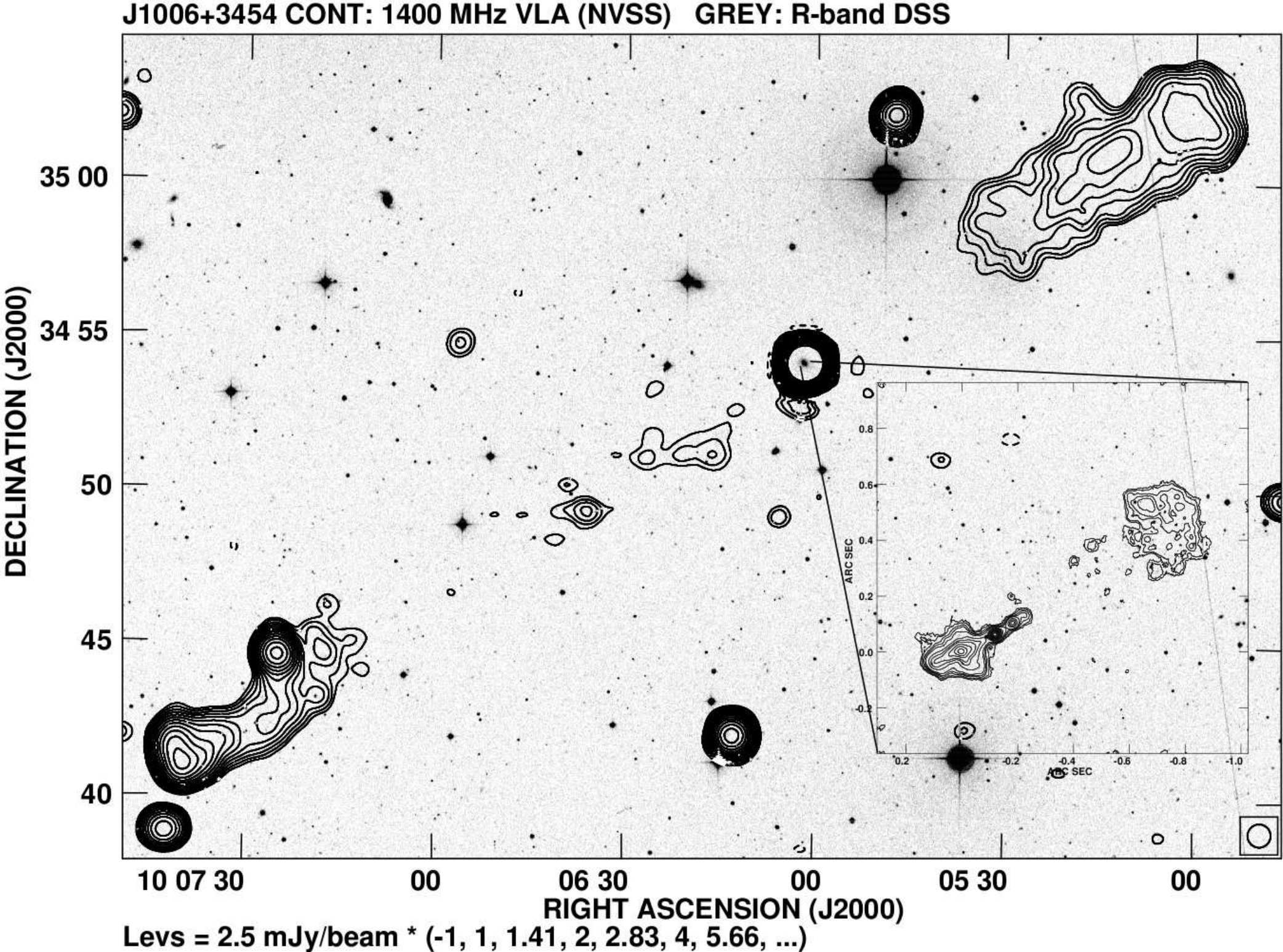} 
    \hspace{1cm}\includegraphics[height=5.5cm]{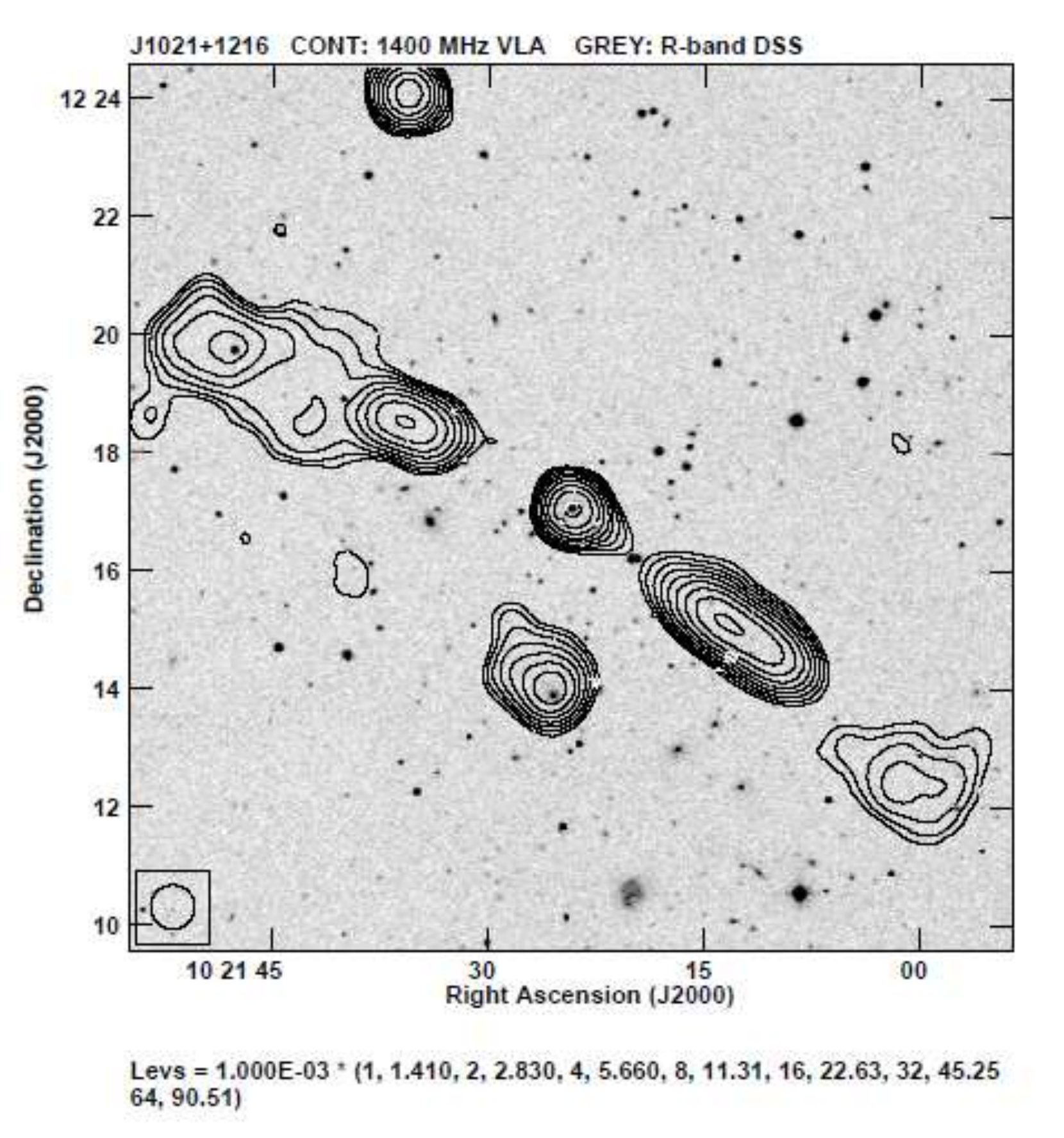}\\
\vspace{0.4cm} 
    \includegraphics[width=8.5cm]{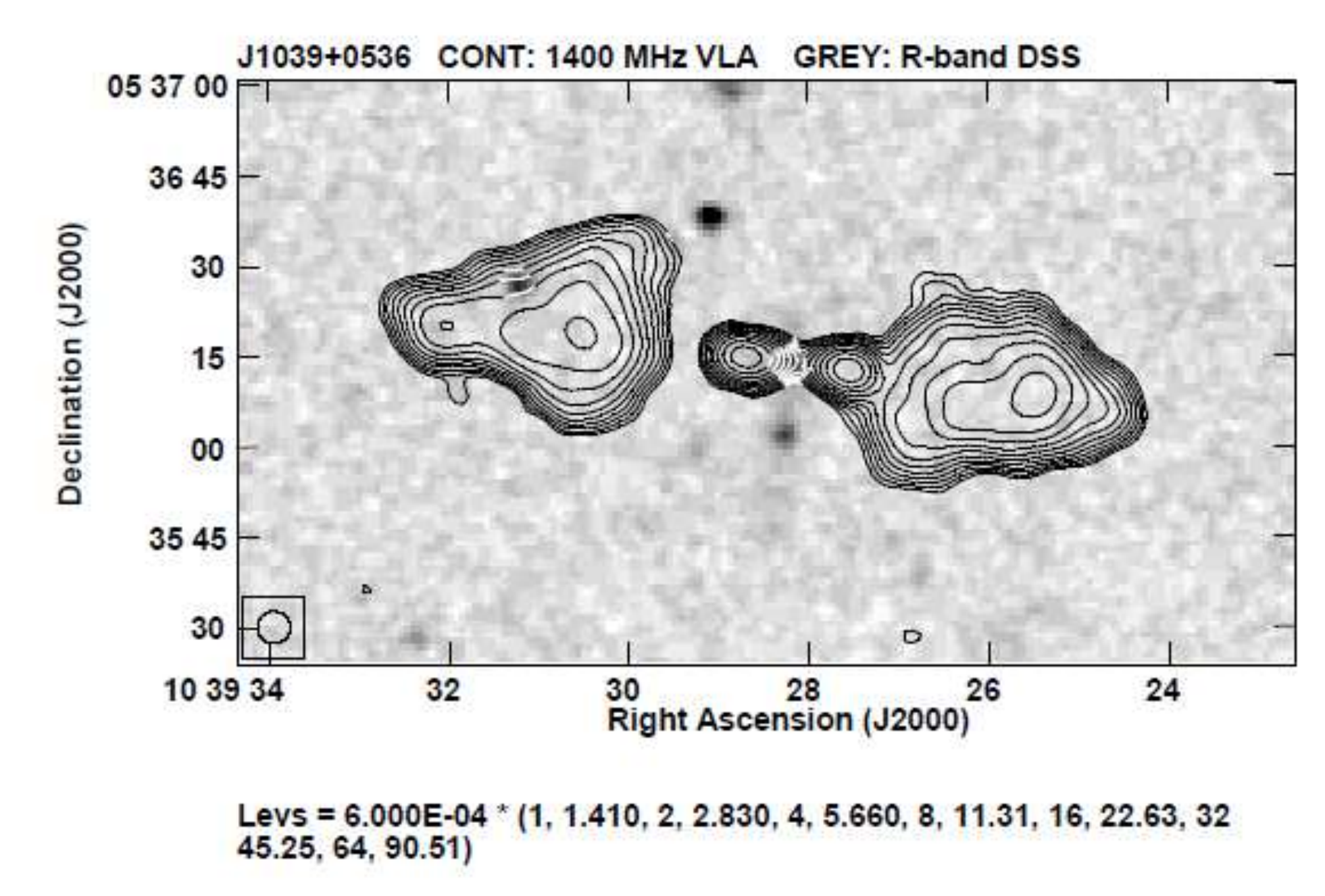} 
    \hspace{1cm}\includegraphics[width=7.5cm]{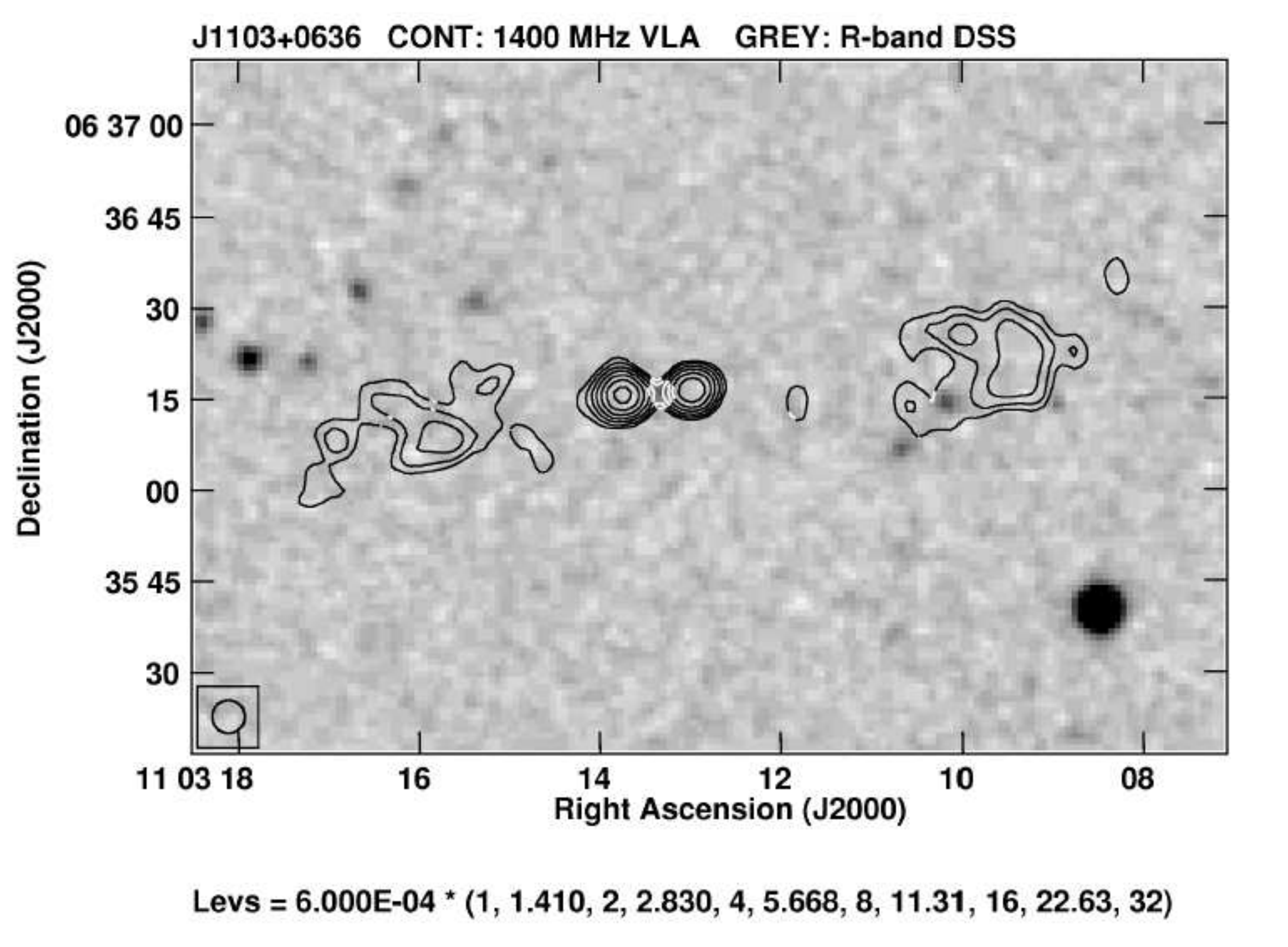} \\
\vspace{0.4cm}
    \includegraphics[height=5.5cm]{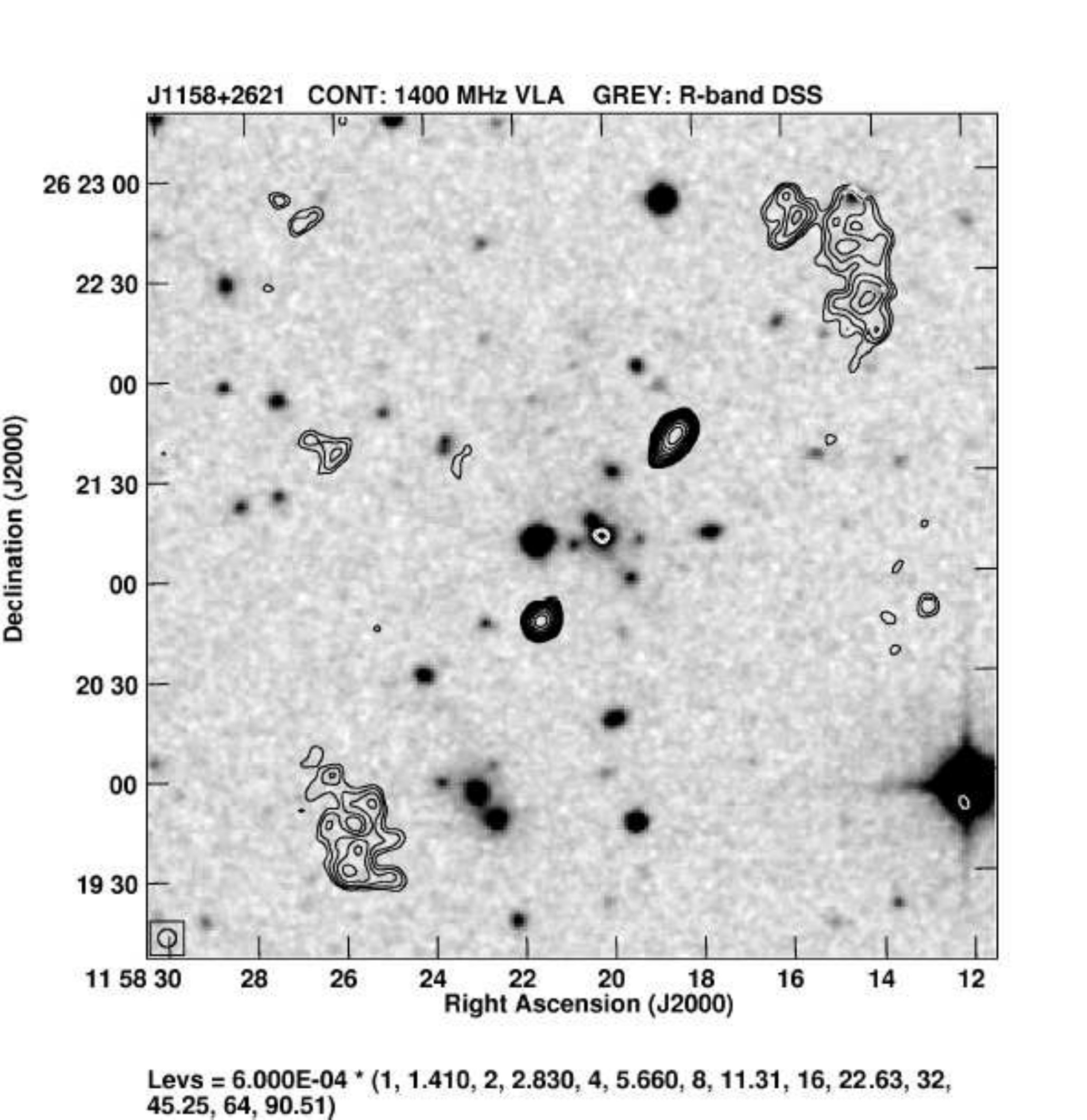}
    \includegraphics[height=5.3cm]{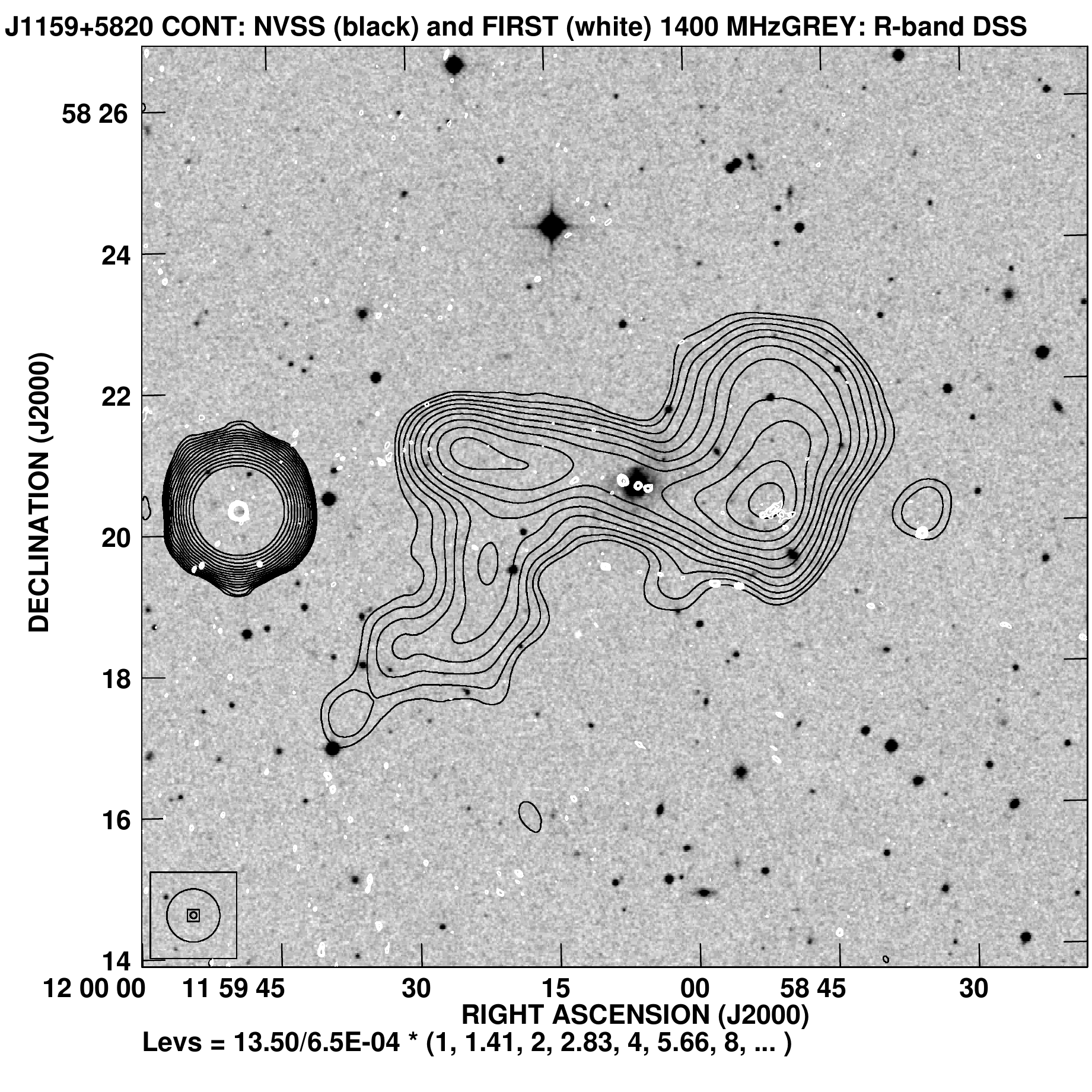} 
    \includegraphics[width=7cm]{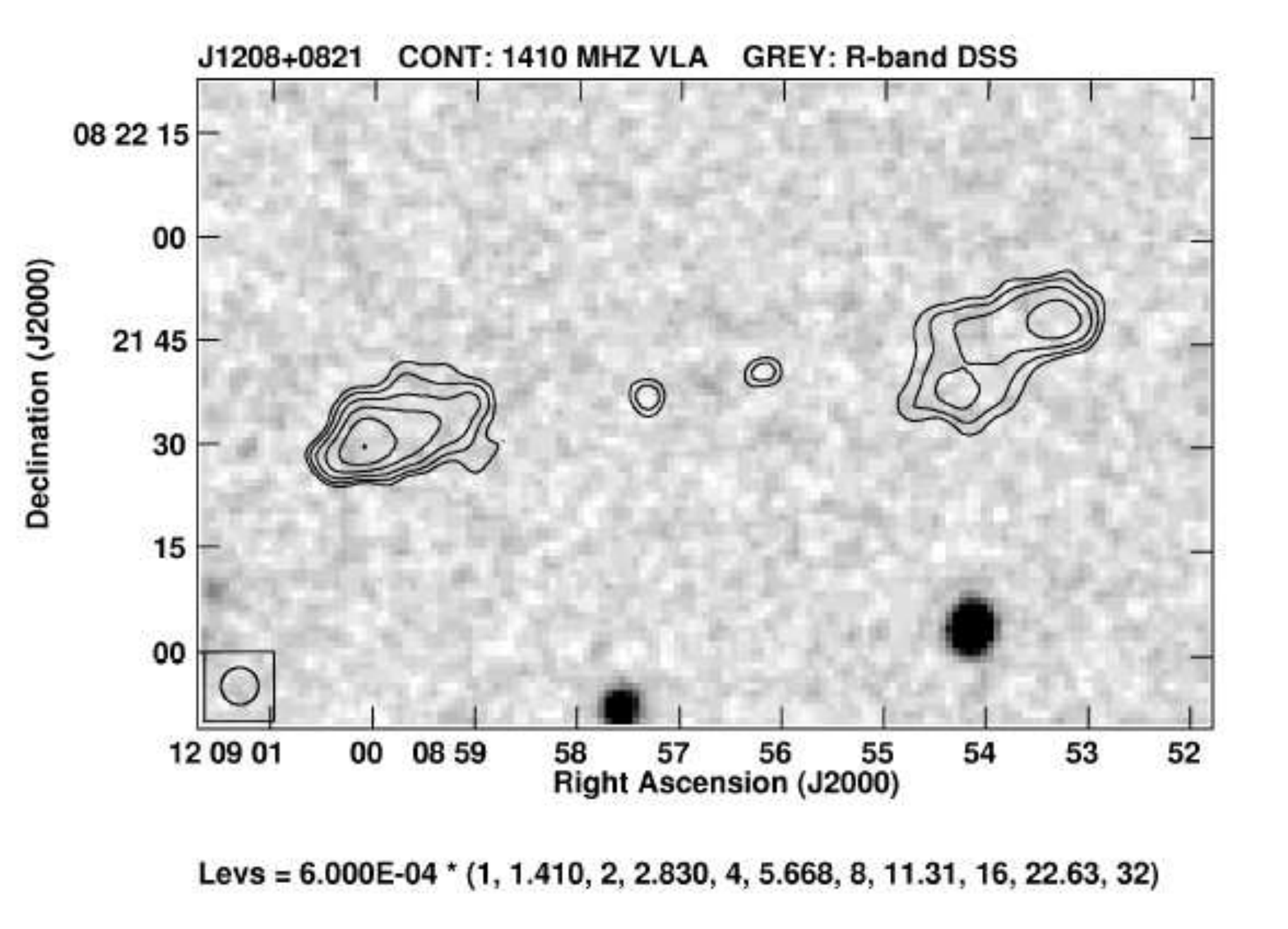}
\end{figure}
\begin{figure}
    \includegraphics[height=5.7cm]{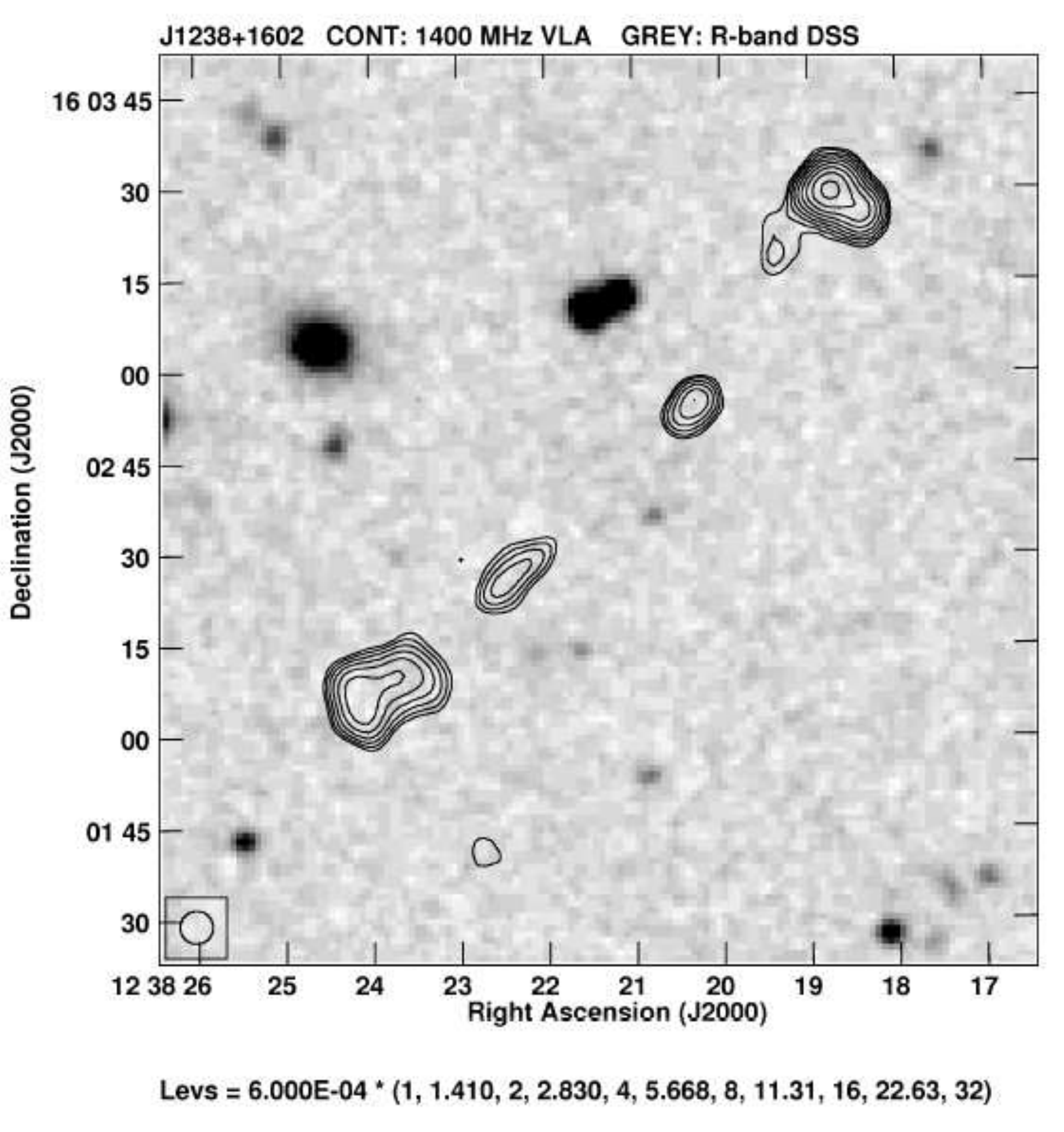} 
    \hspace{0.5cm}\includegraphics[height=5.5cm]{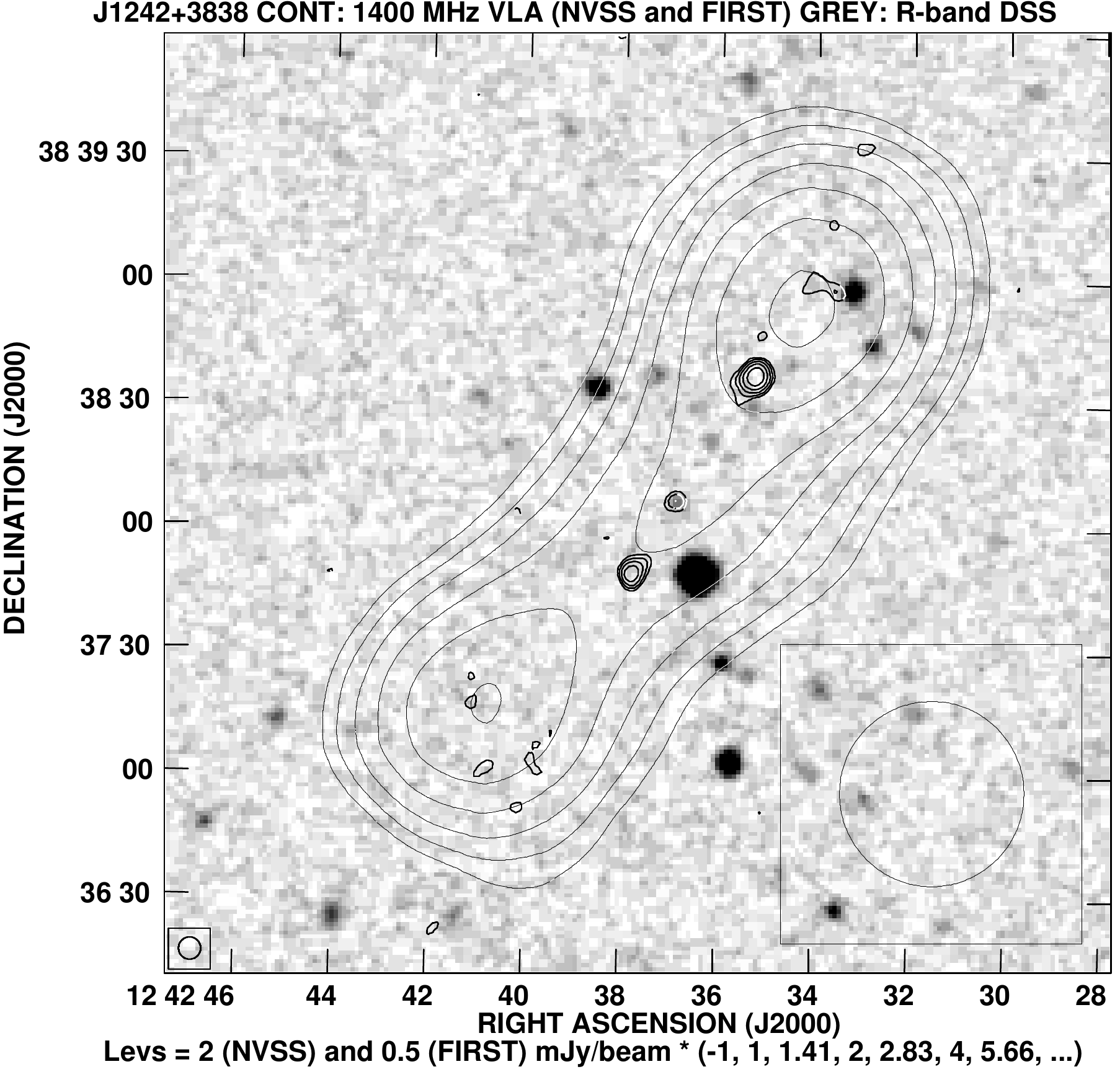} 
    \hspace{0.5cm}\includegraphics[height=5.5cm]{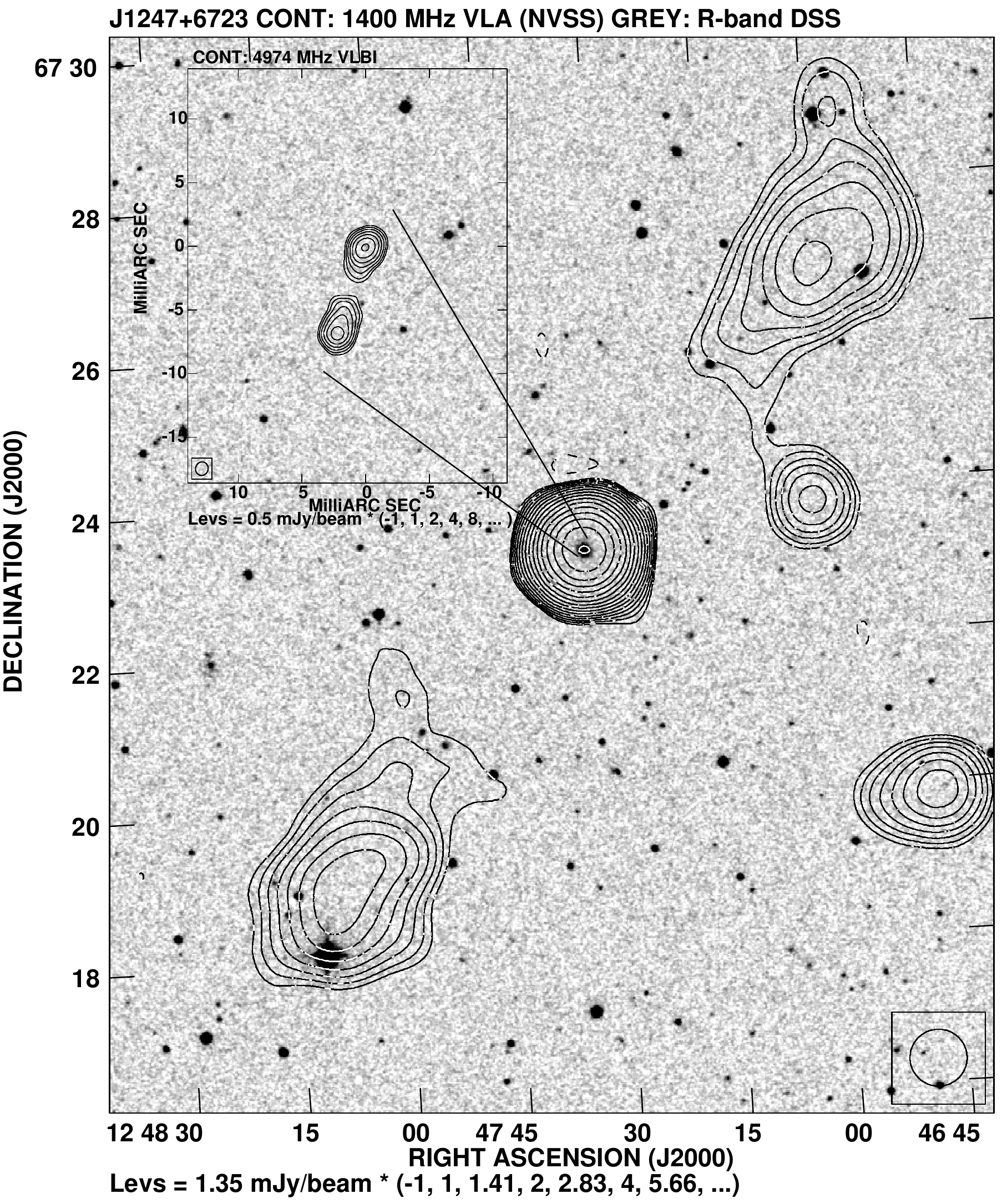}\\
\vspace{0.4cm}
    \includegraphics[height=5cm]{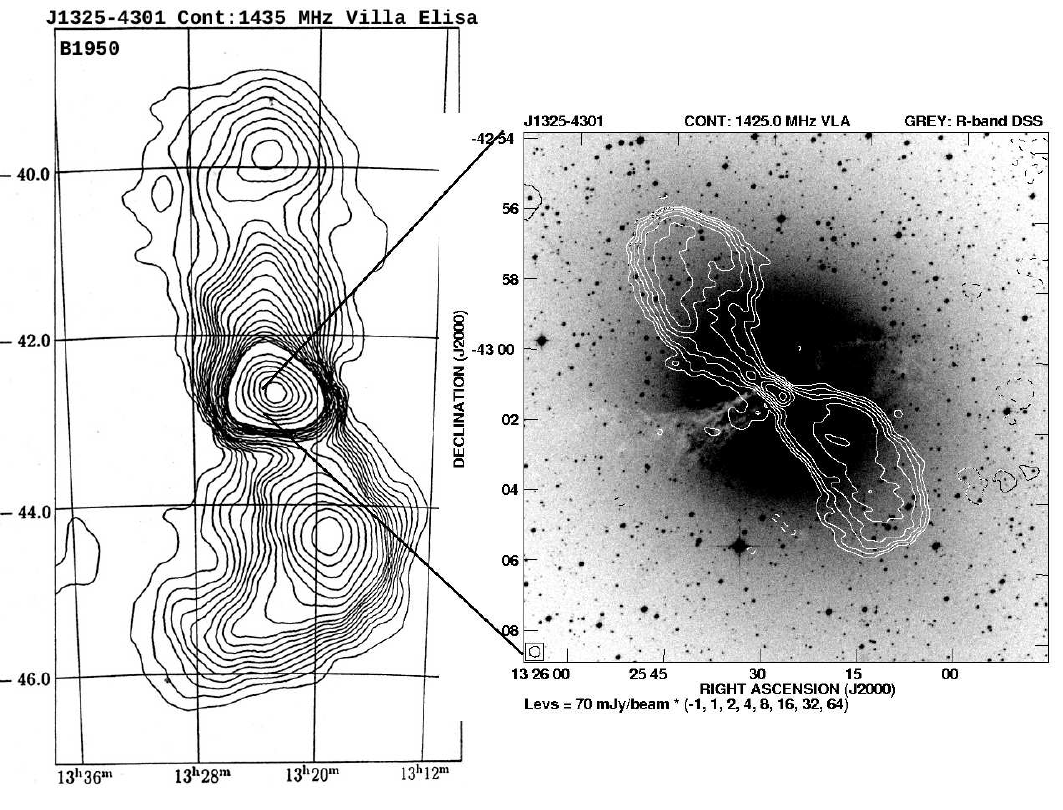} 
    \includegraphics[height=5.5cm]{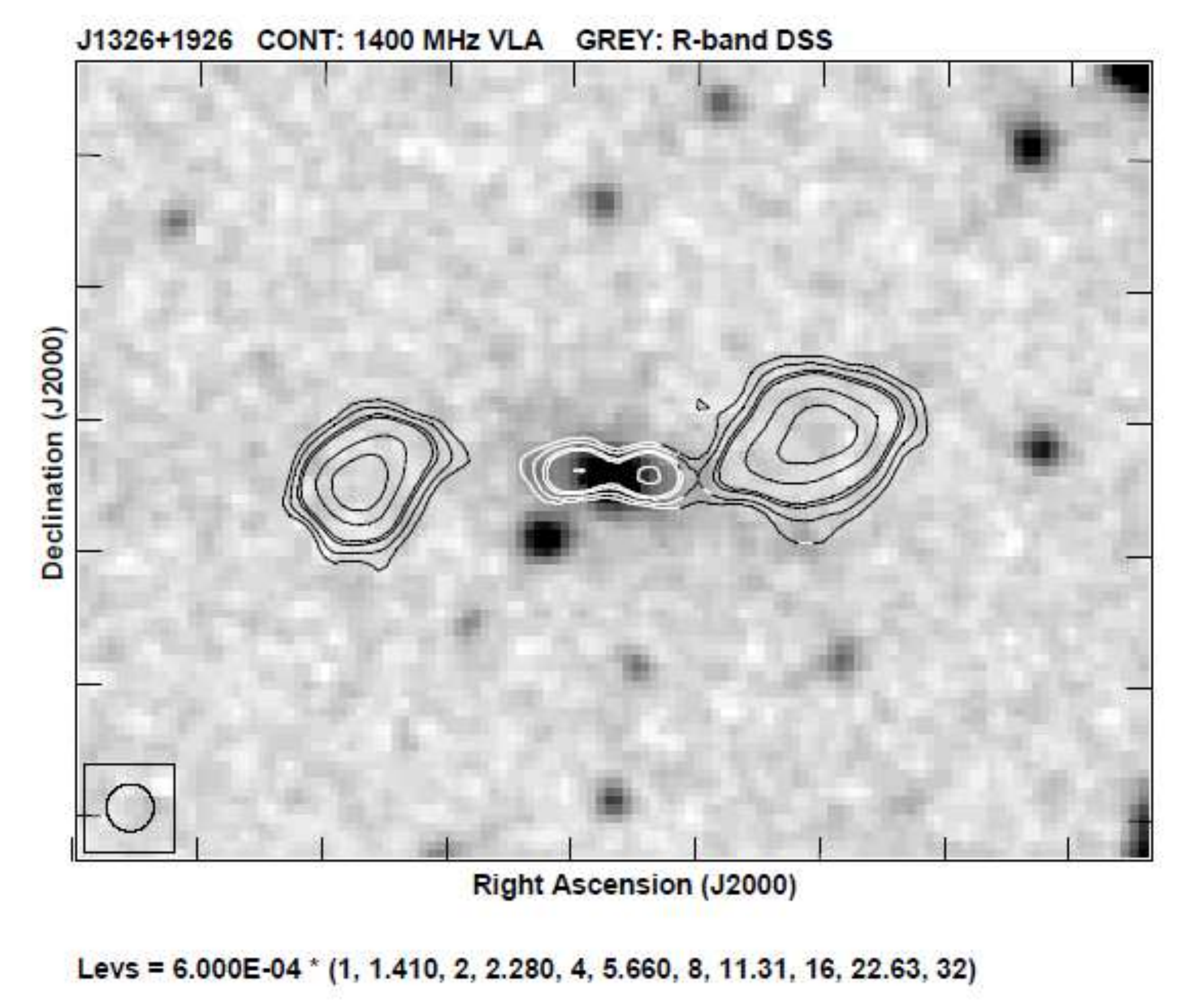}
    \includegraphics[height=5.5cm]{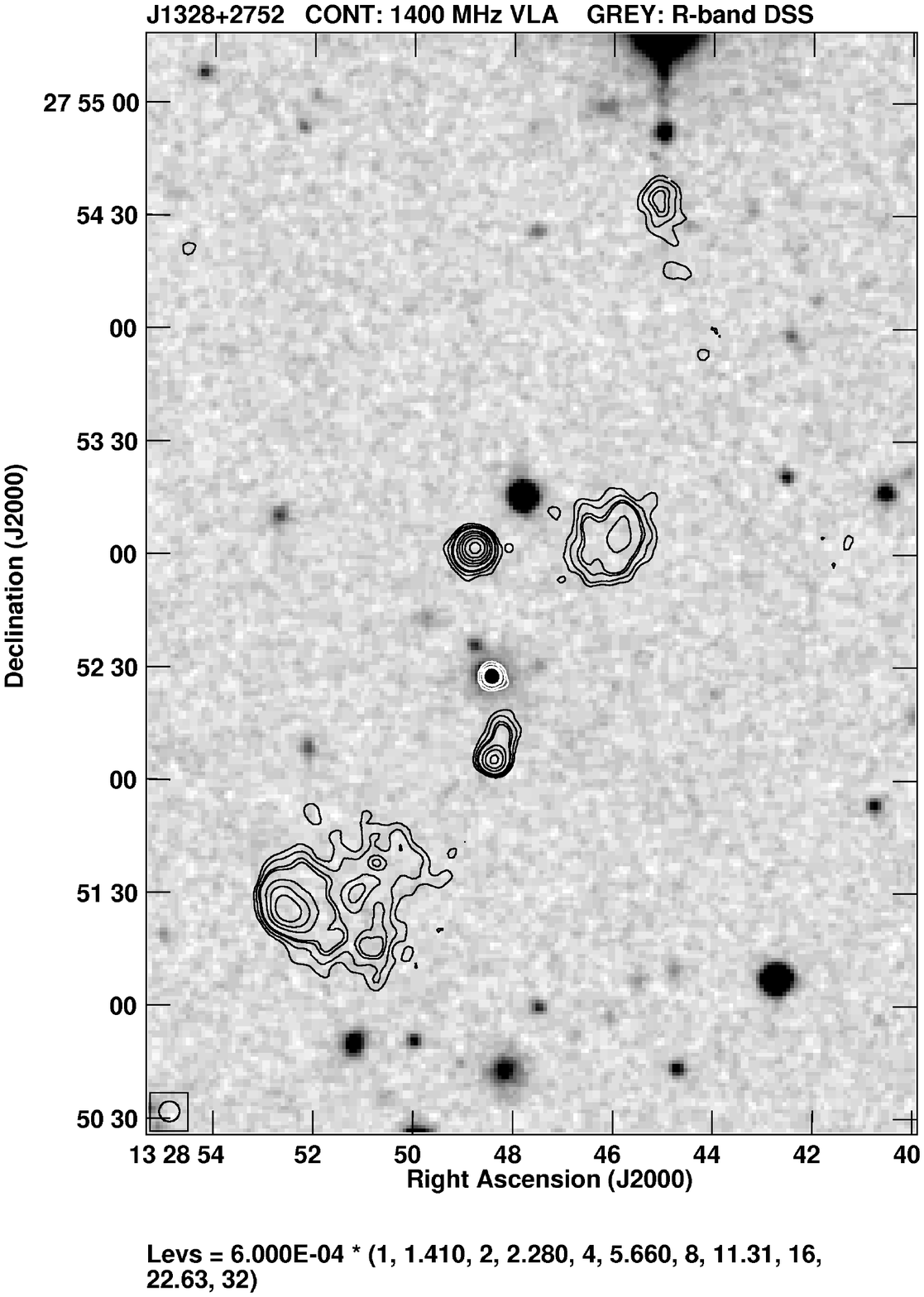} \\
\vspace{0.4cm}
    \includegraphics[height=5.5cm]{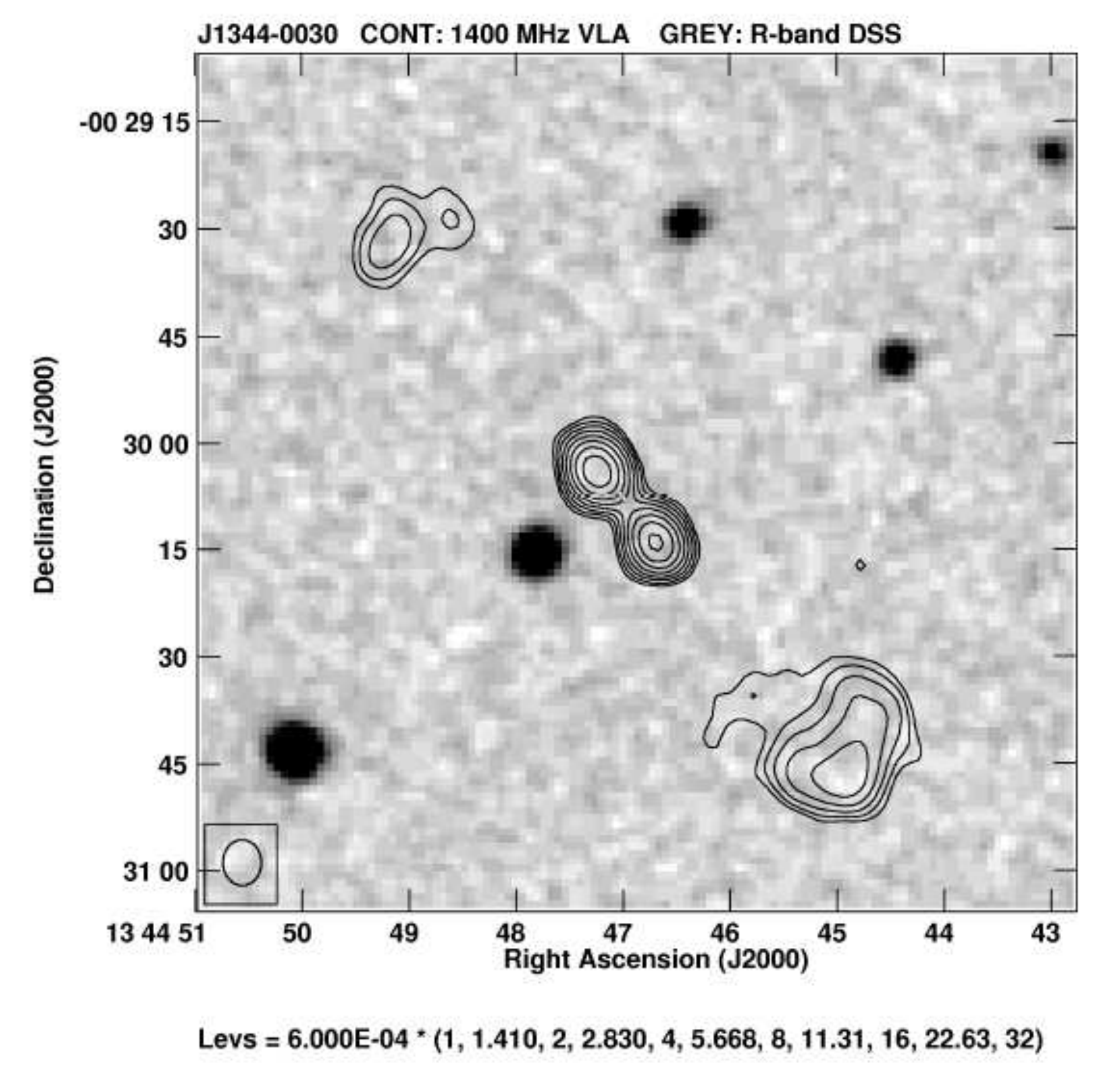} 
    \includegraphics[height=5.5cm]{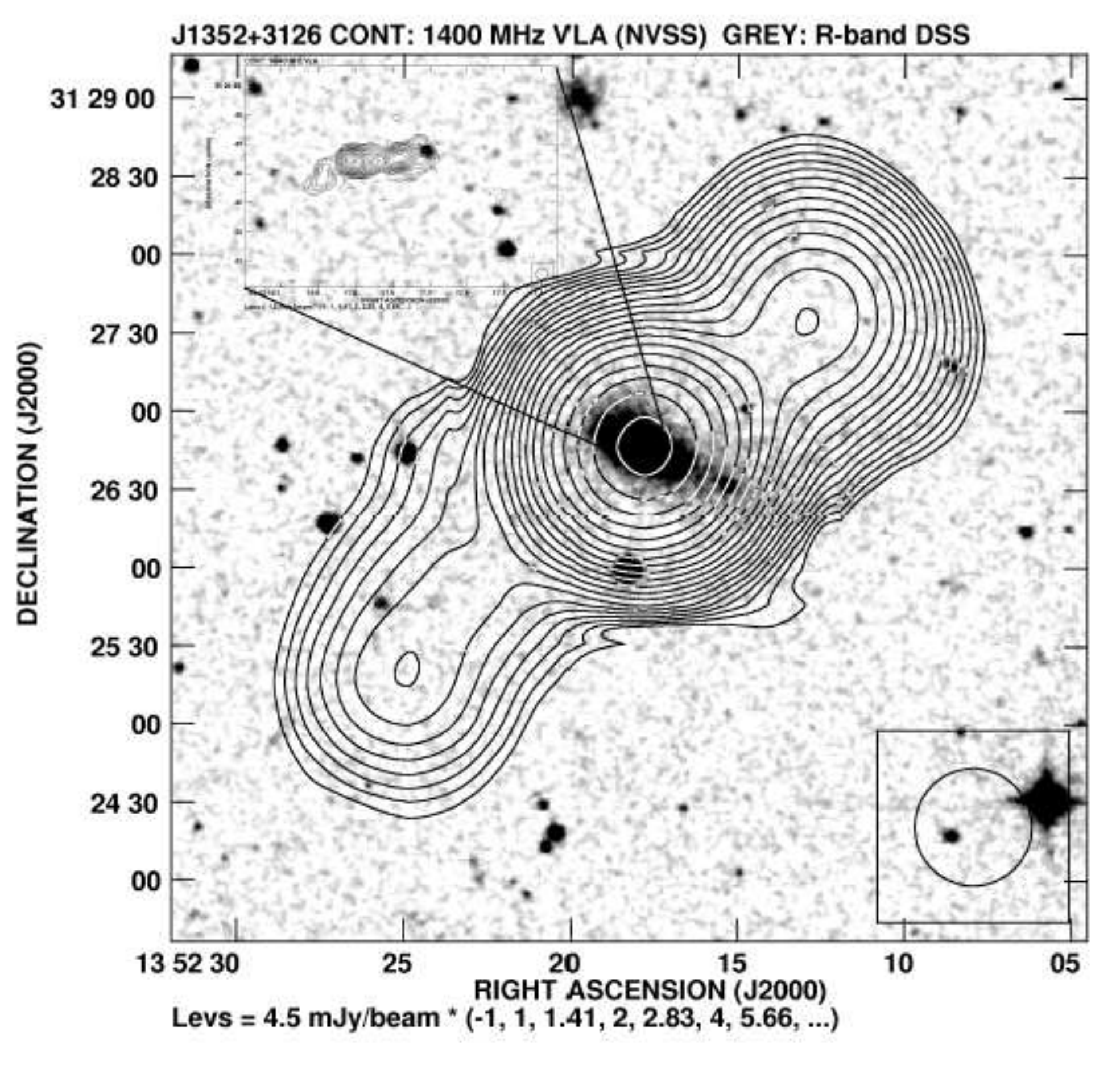} 
    \includegraphics[height=5.5cm]{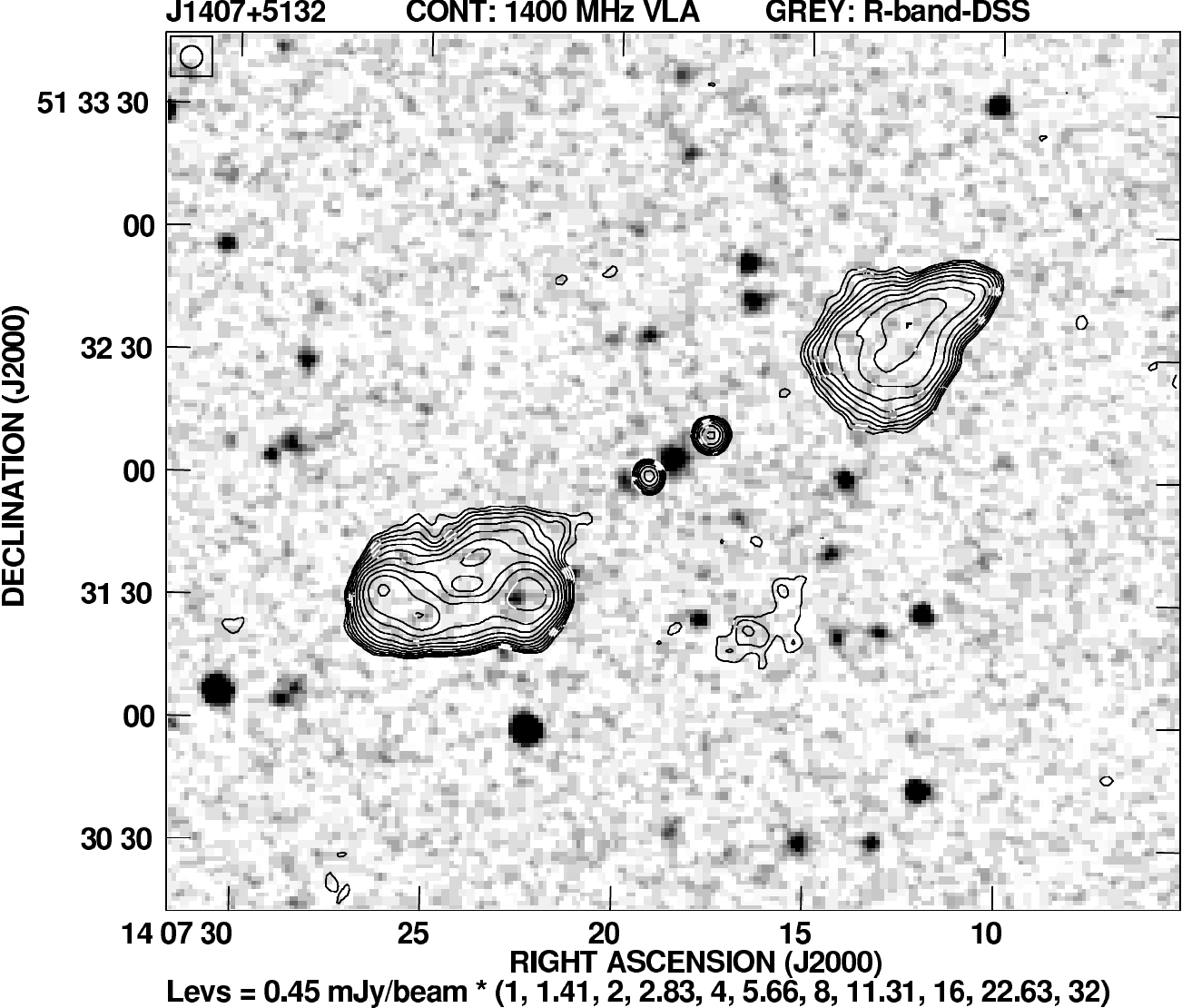}\\
\vspace{0.4cm}
    \includegraphics[height=5.5cm]{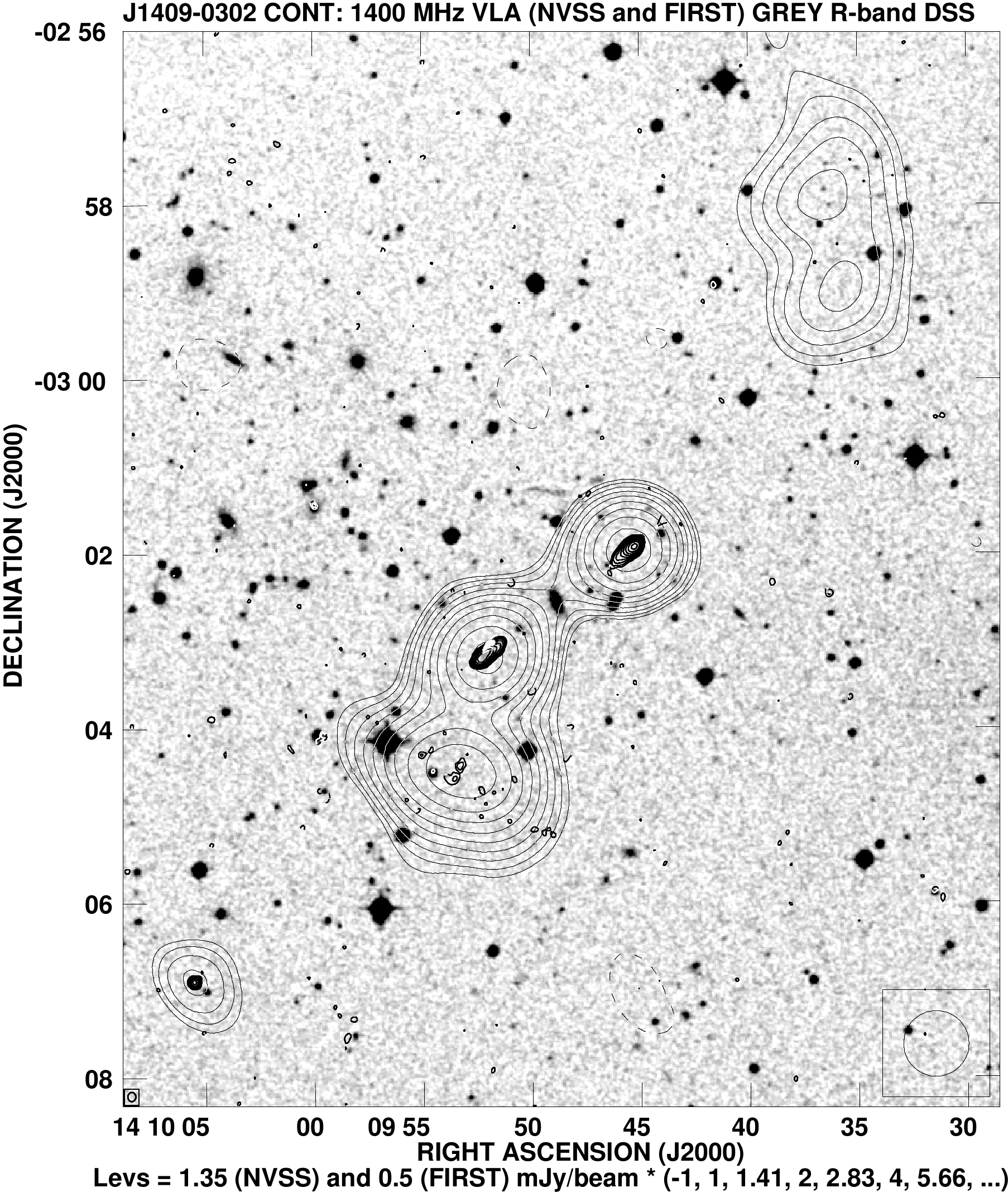}
    \includegraphics[height=5.5cm]{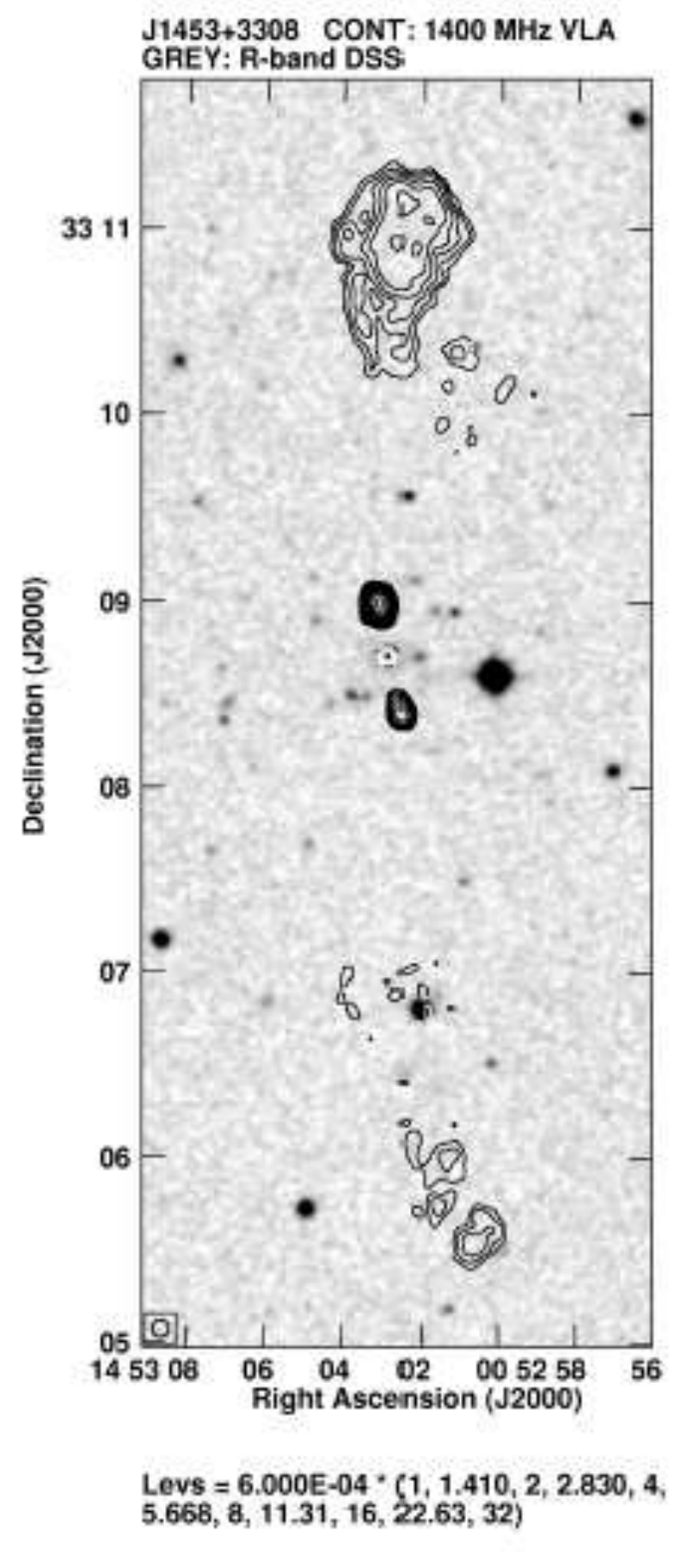} 
    \includegraphics[height=5.5cm]{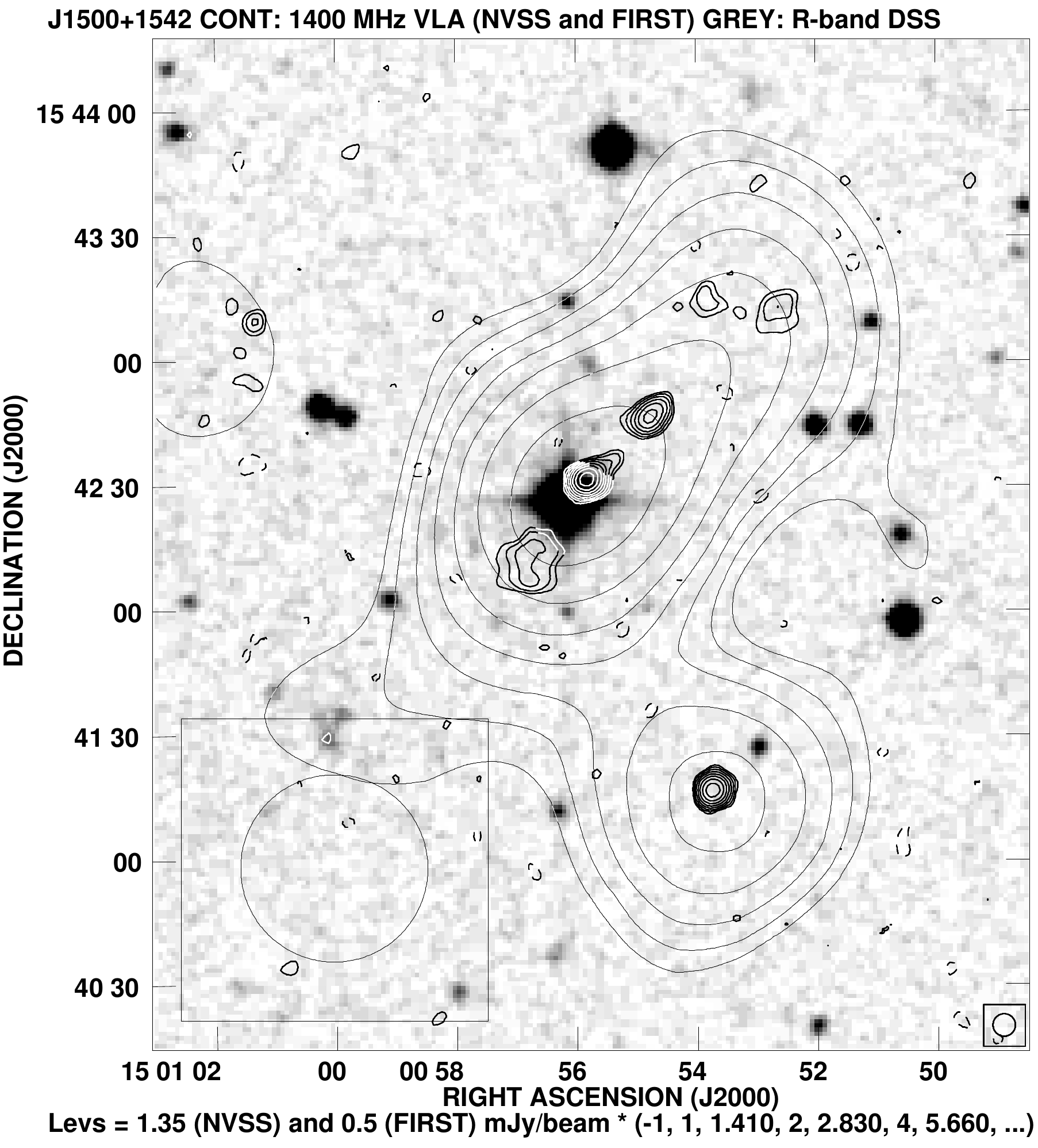} 
    \includegraphics[height=5.5cm]{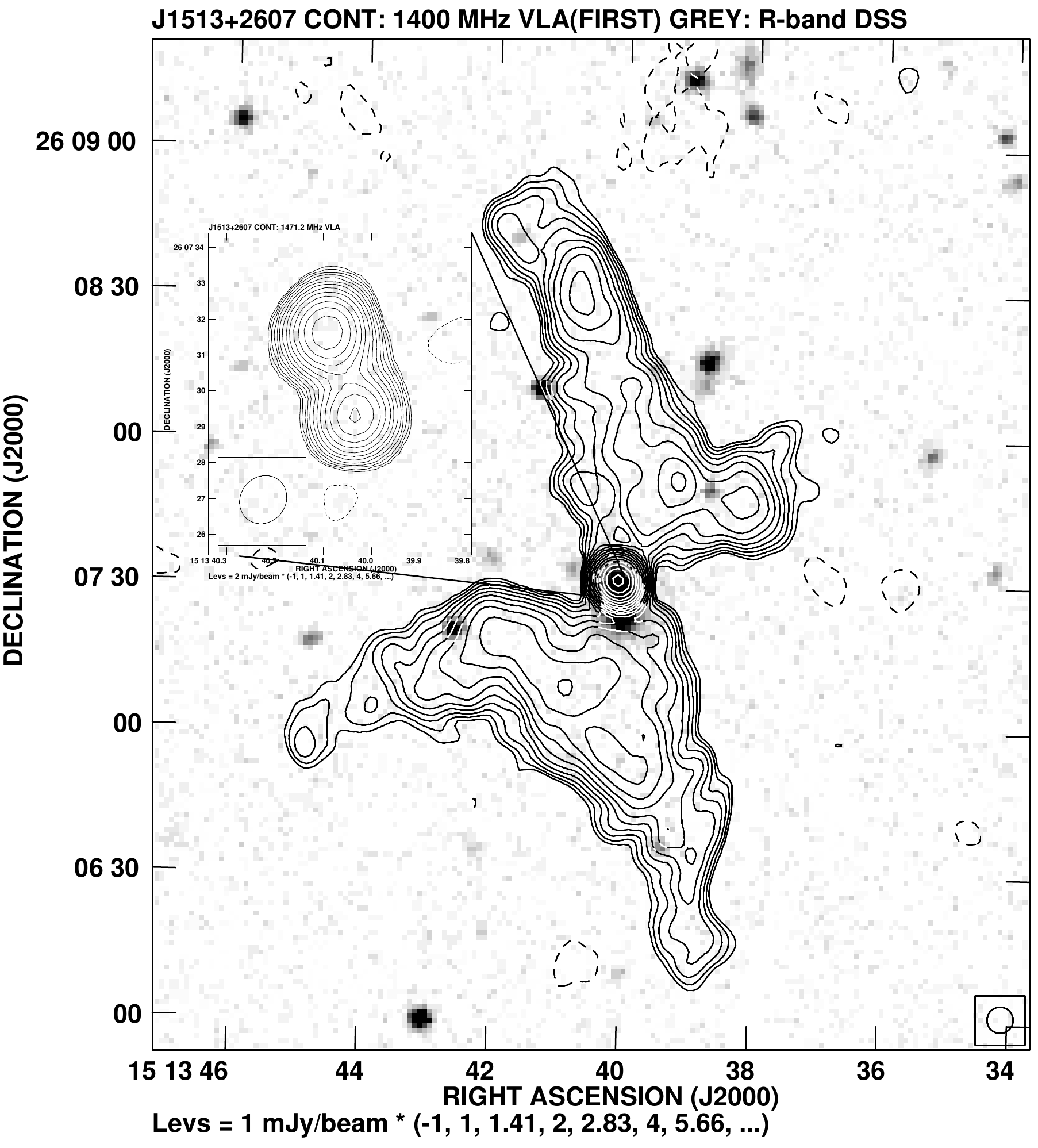}
\end{figure}
\begin{figure}
    \includegraphics[height=5.4cm]{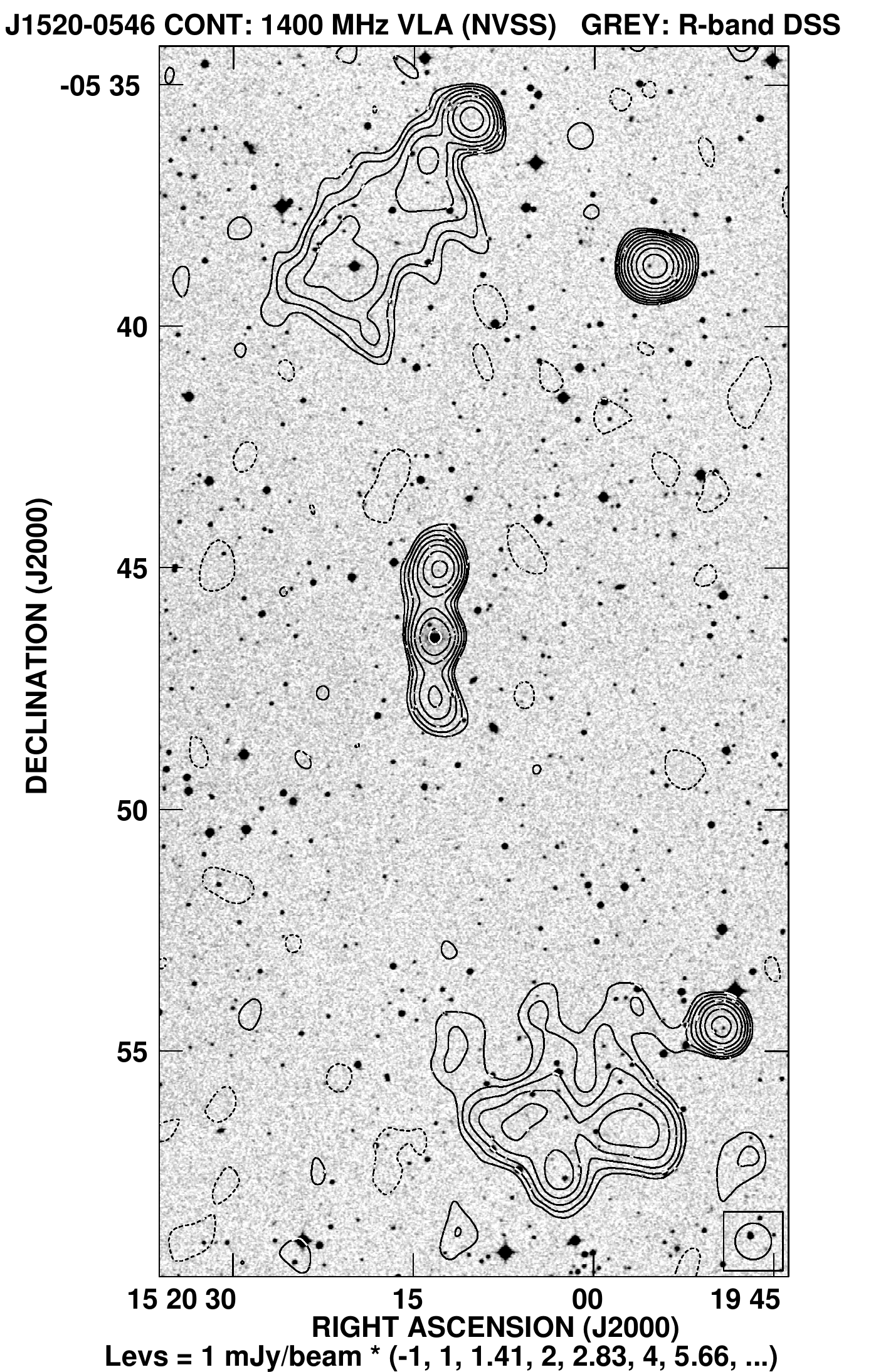}
    \includegraphics[height=5.5cm]{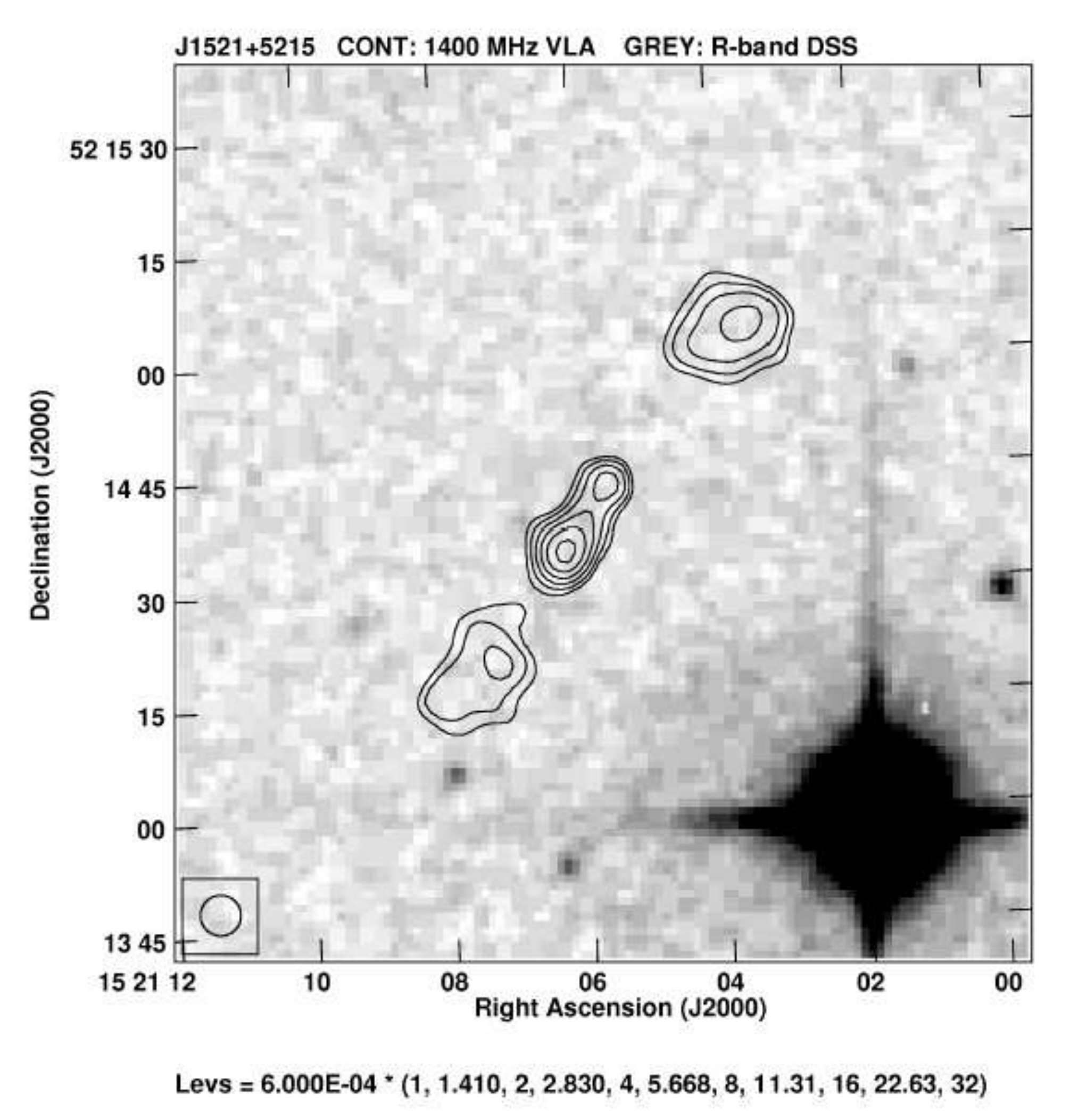}
    \hspace{1cm}\includegraphics[width=8cm]{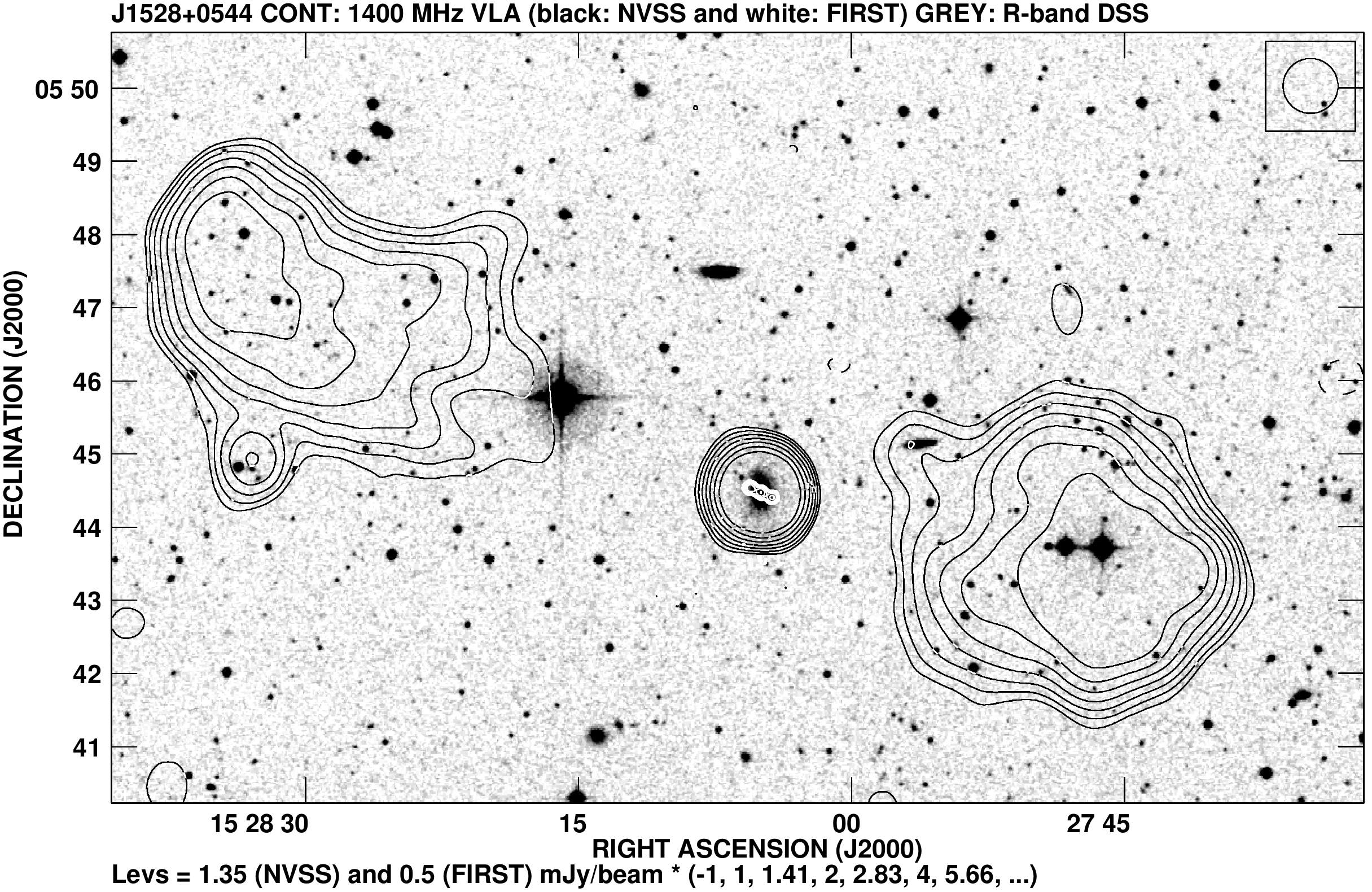}\\
\vspace{0.4cm}
    \includegraphics[height=5.5cm]{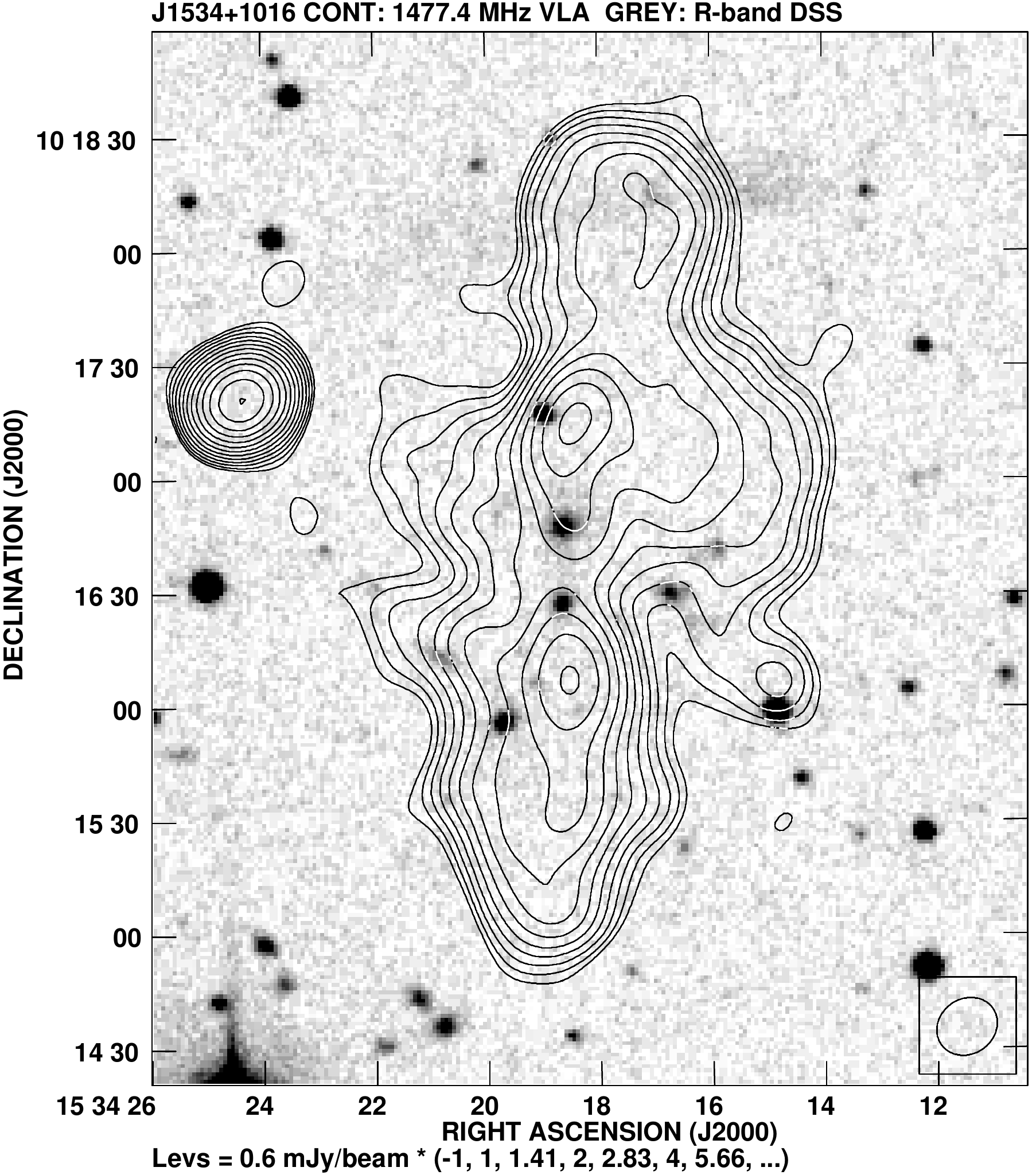}
    \includegraphics[height=6cm]{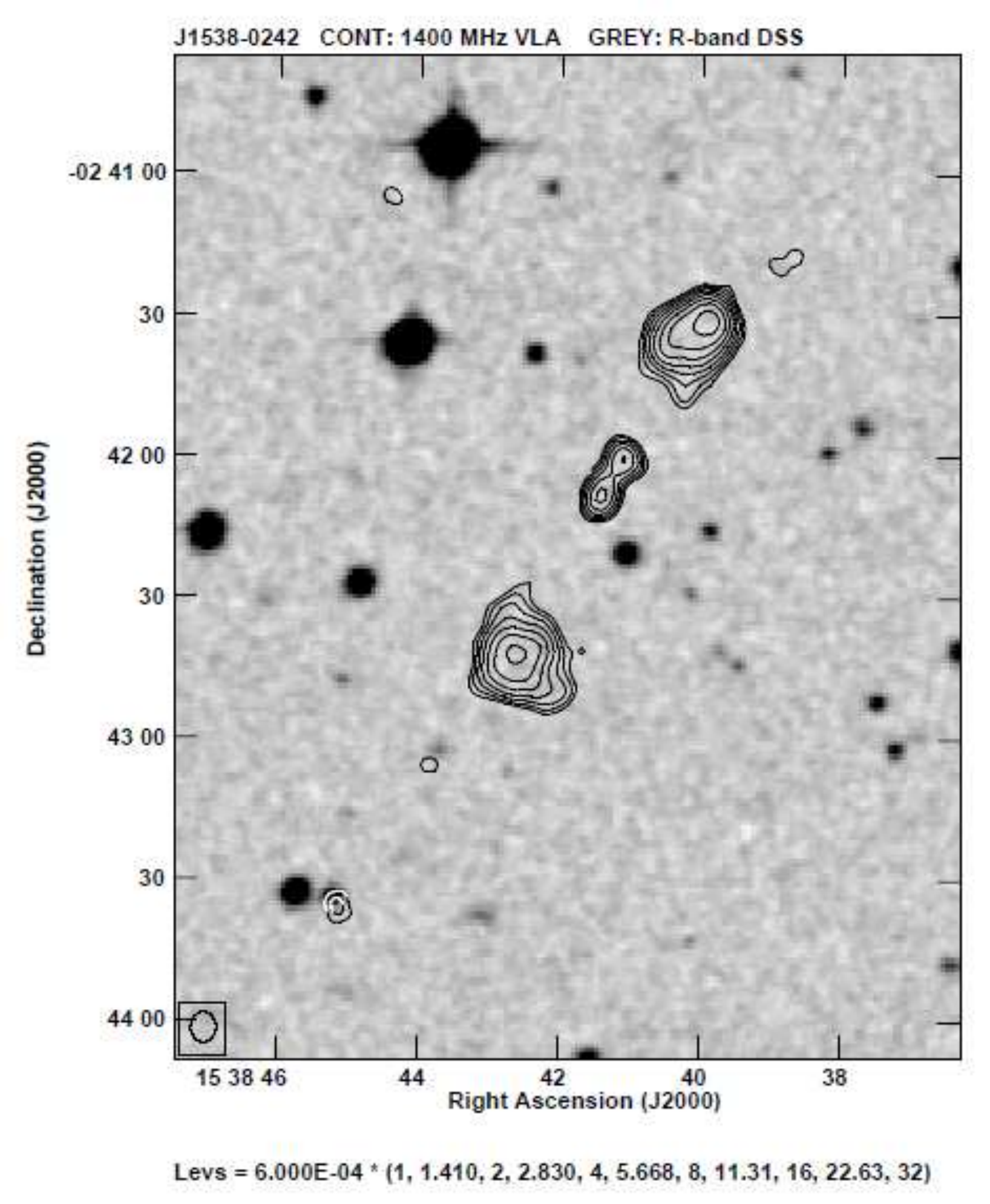}
    \includegraphics[width=8cm]{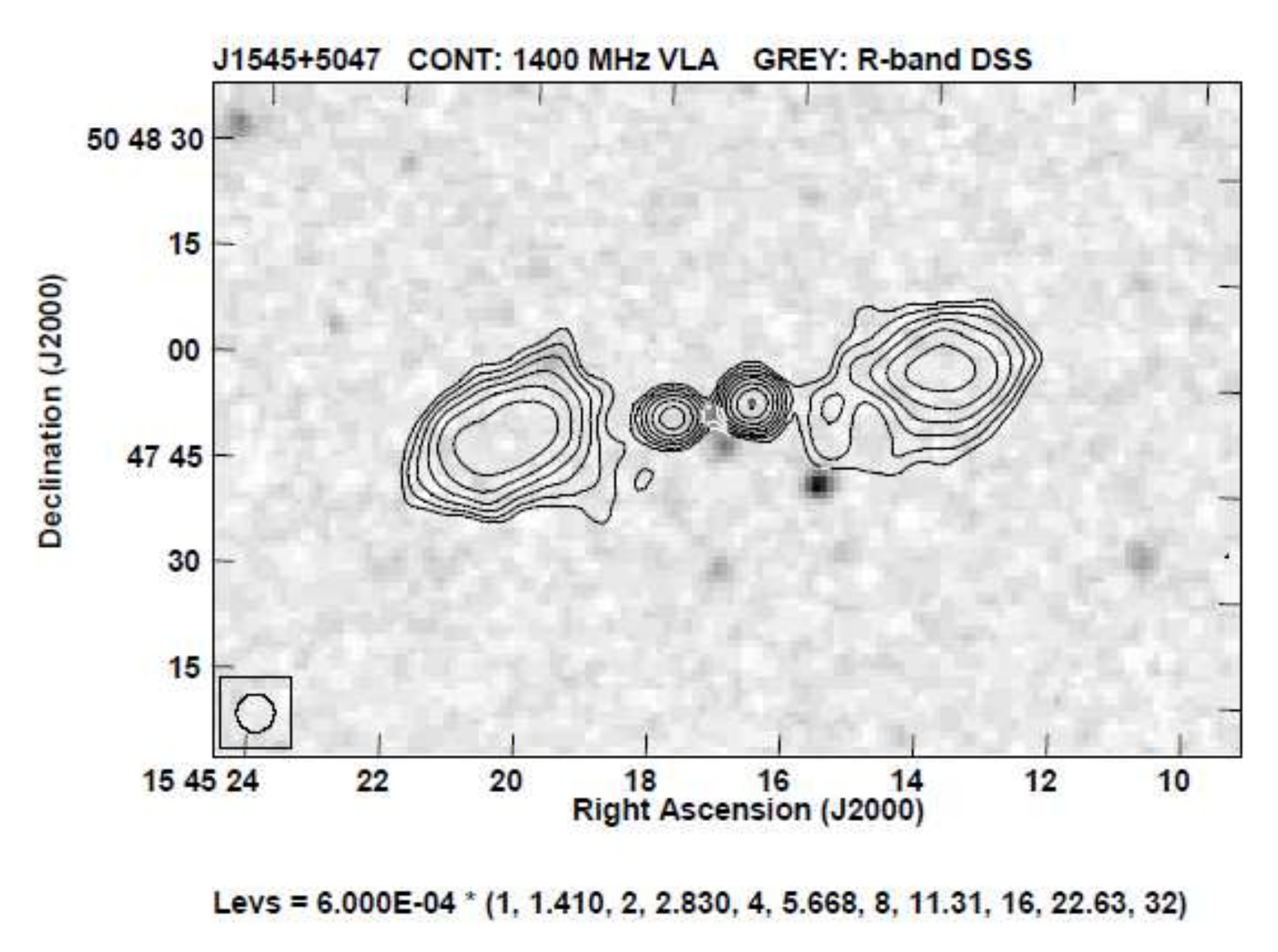}\\
\vspace{0.4cm}
    \includegraphics[height=5.5cm]{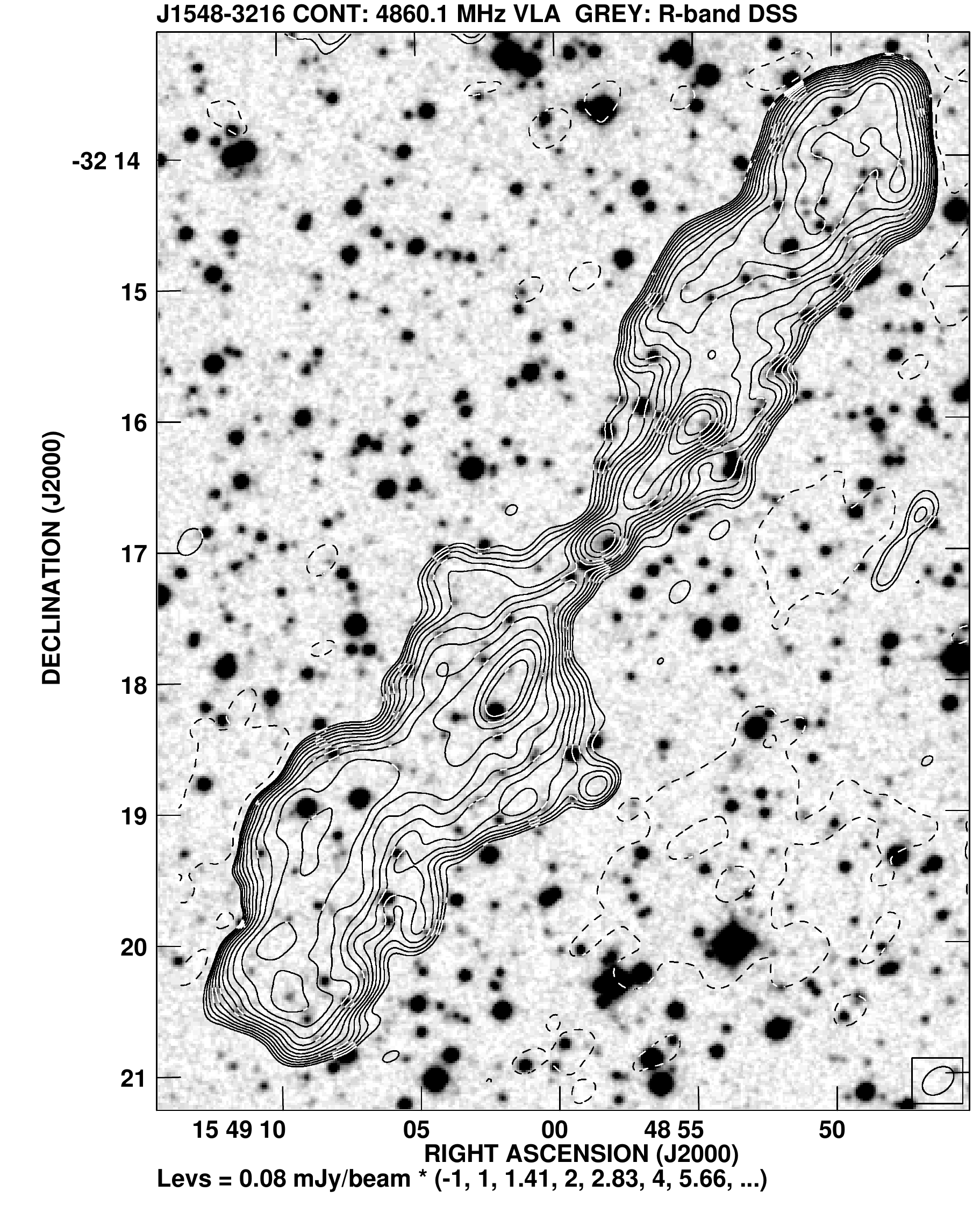}
    \includegraphics[width=8.5cm]{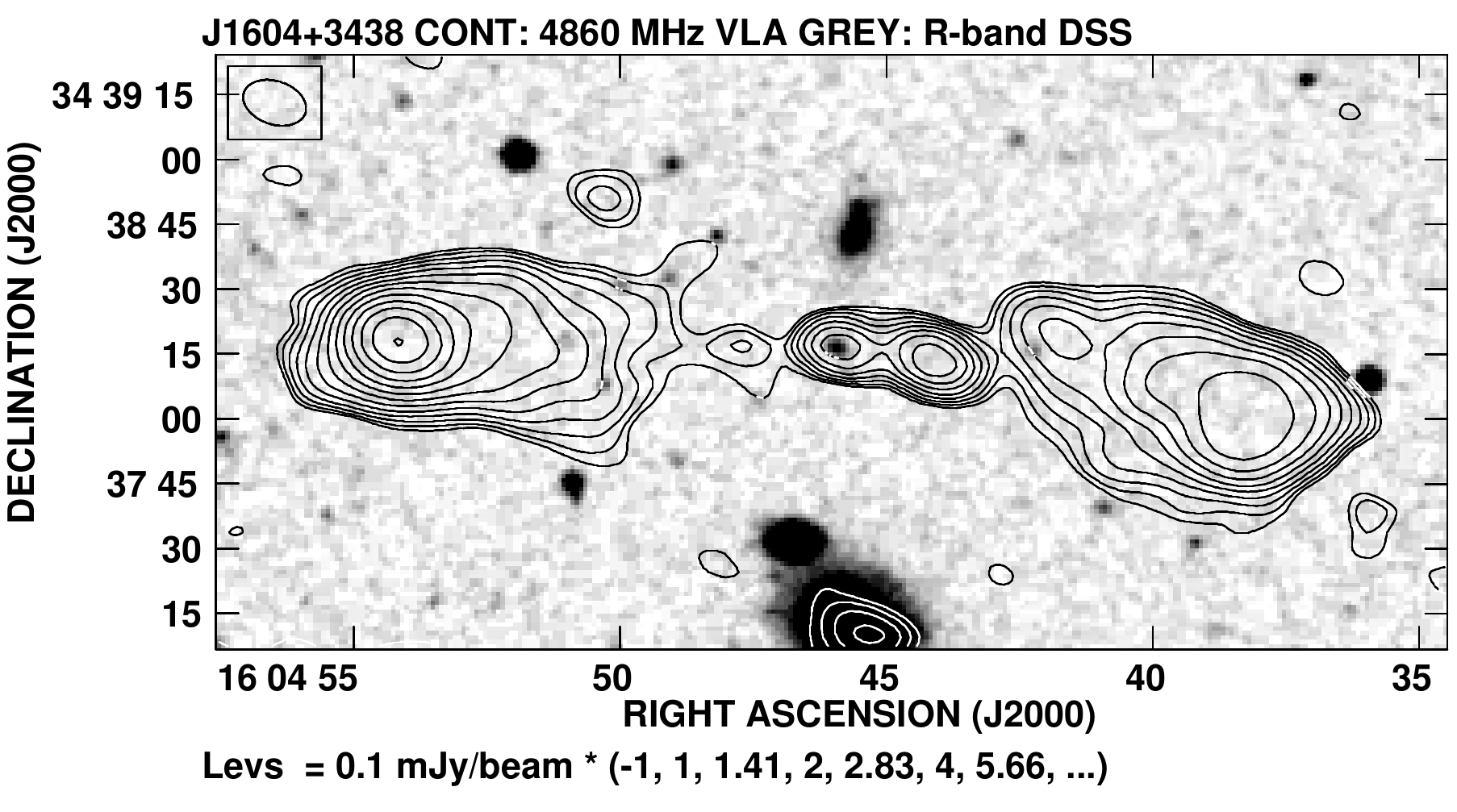} 
    \includegraphics[height=5.5cm]{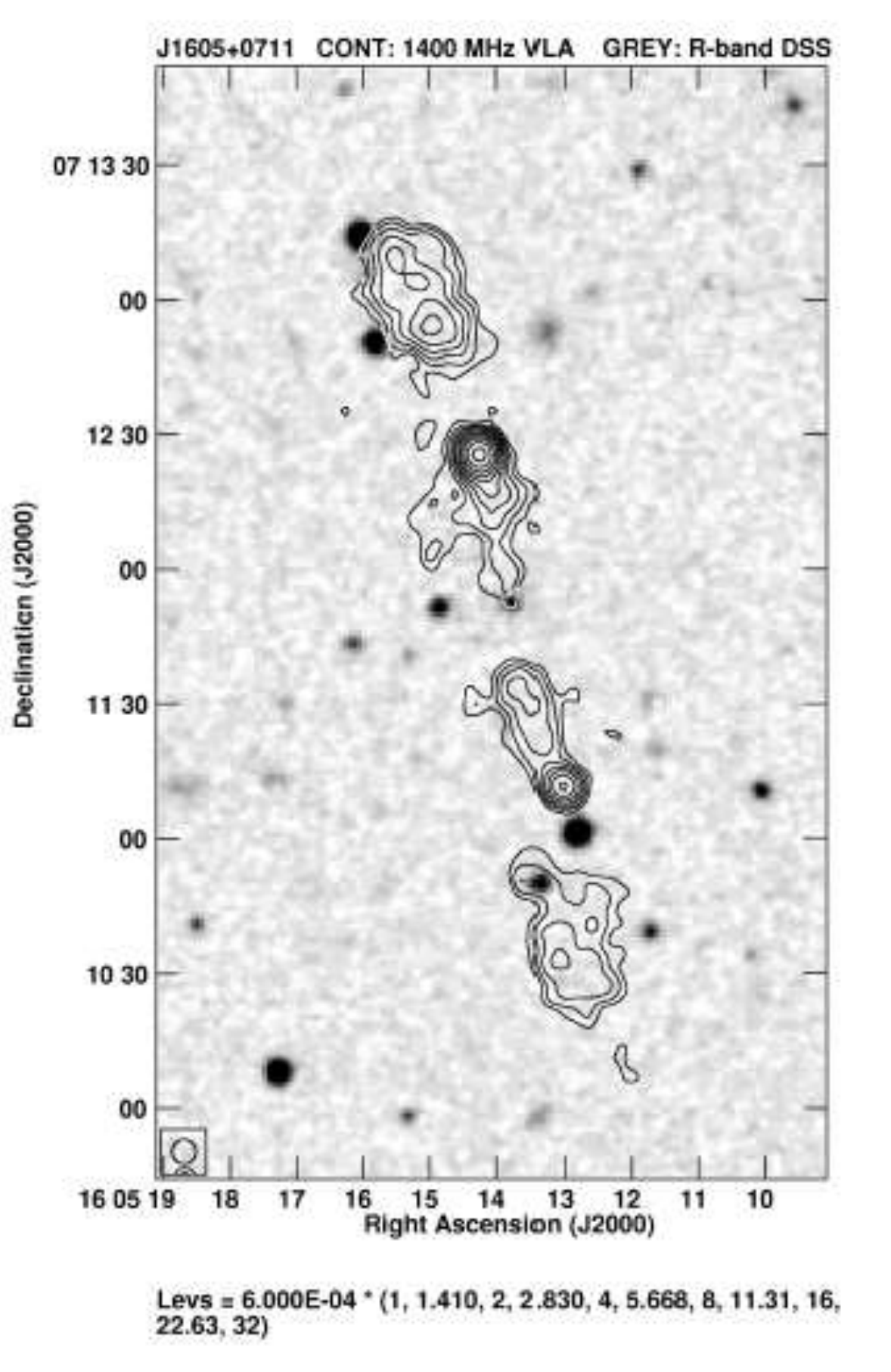} \\
\vspace{0.4cm}
    \includegraphics[height=5.4cm]{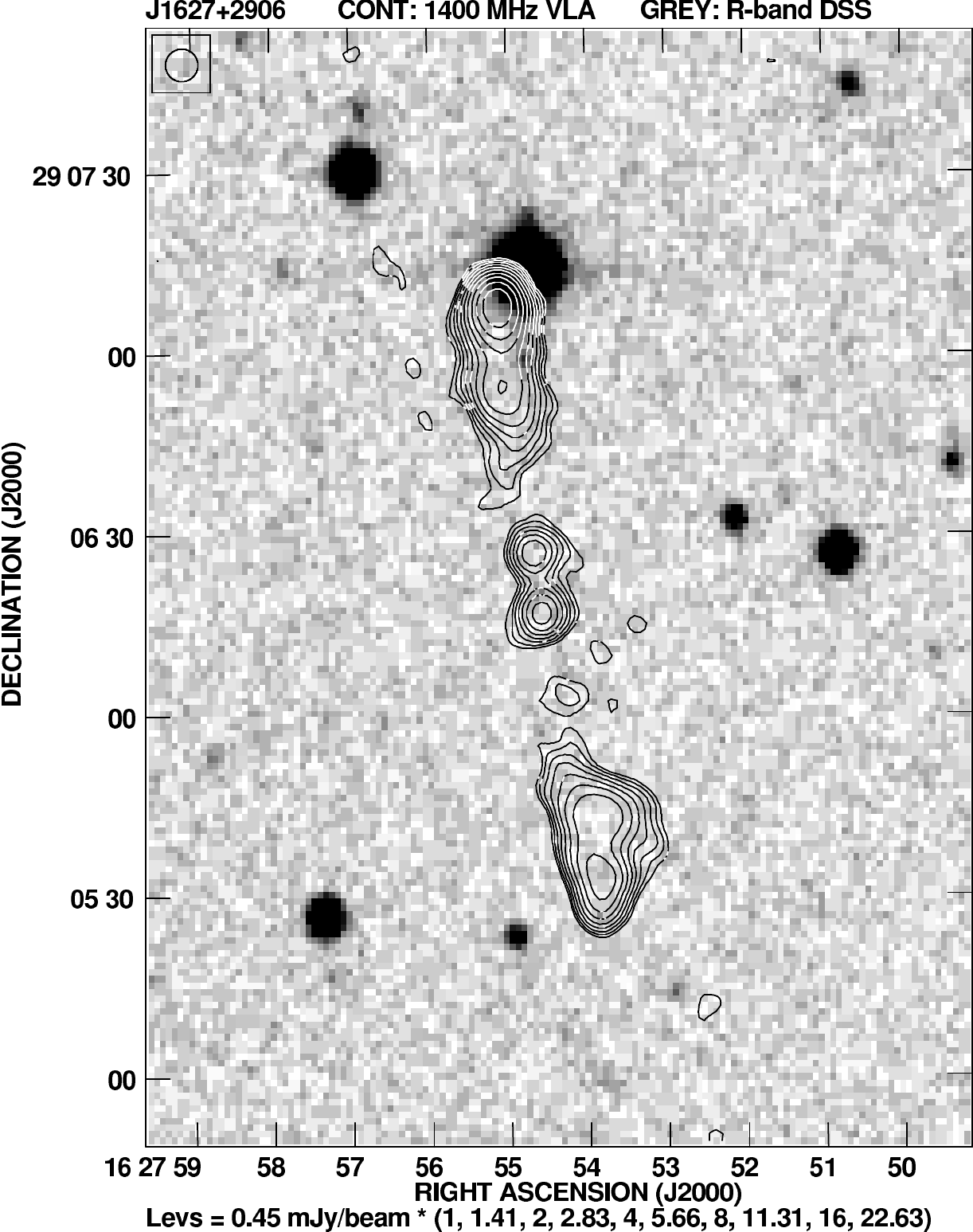}
    \hspace{1cm}\includegraphics[height=5.5cm]{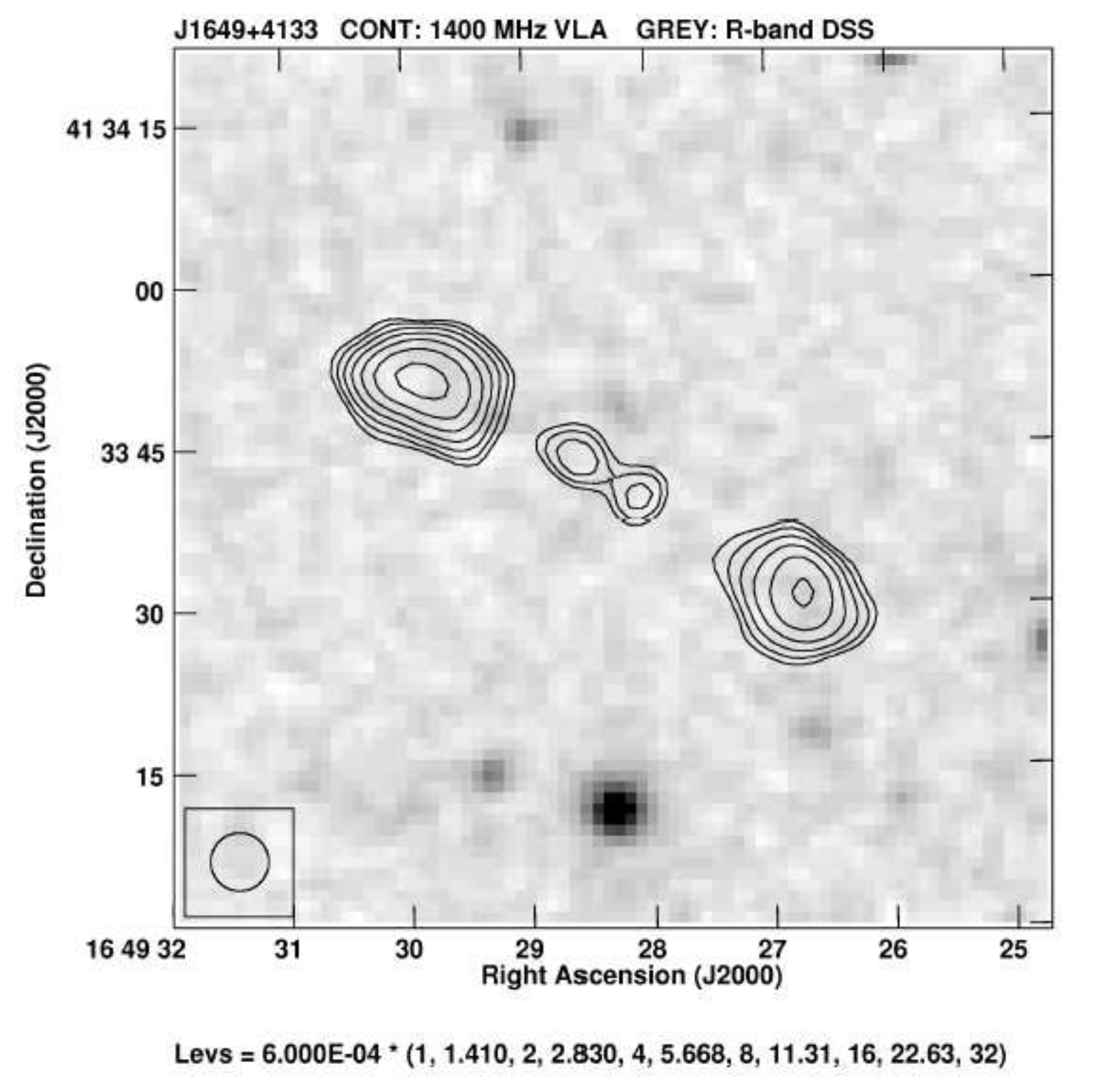} 
    \hspace{1cm}\includegraphics[height=5.5cm]{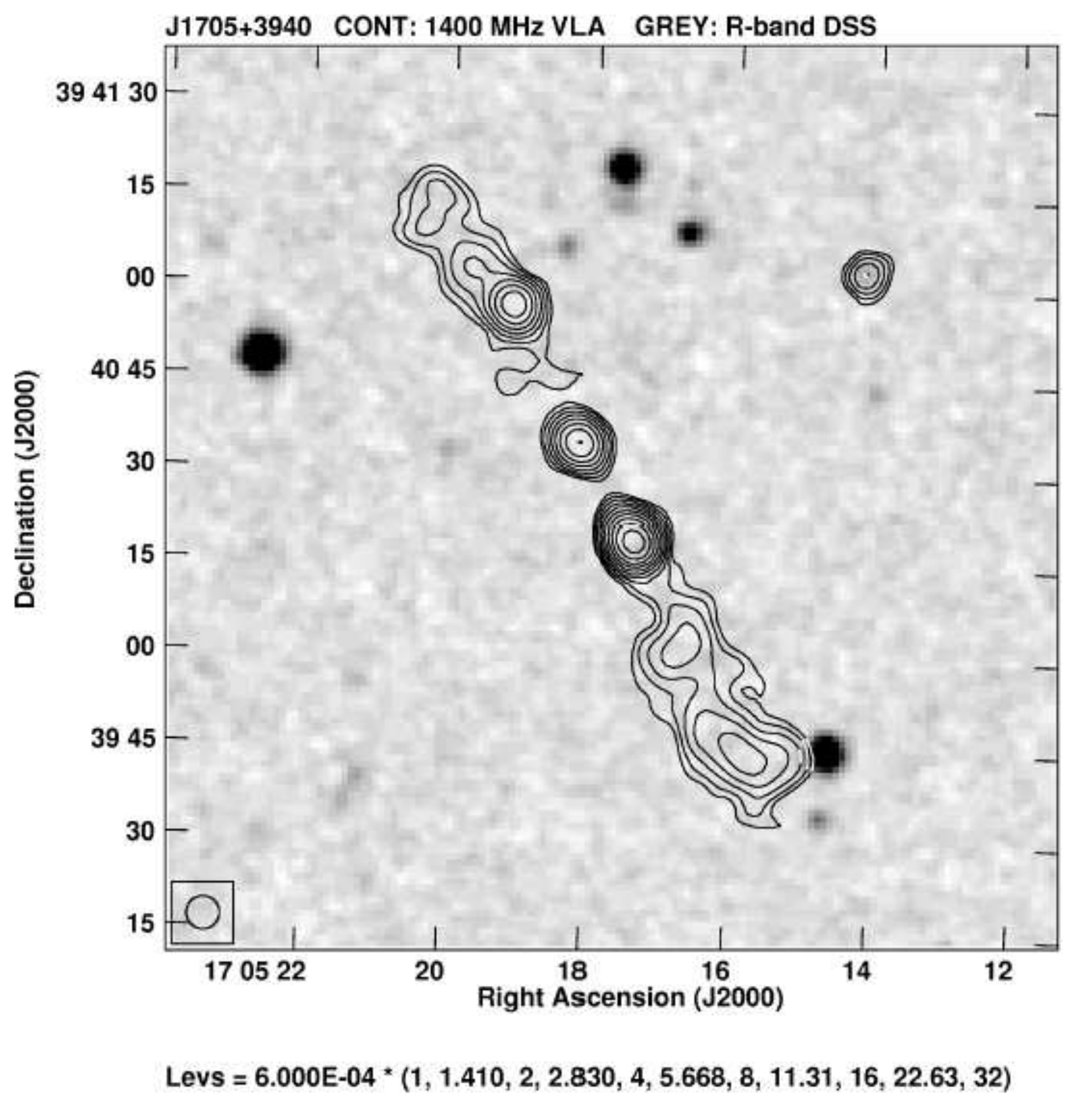}
\end{figure}
\begin{figure}
    \includegraphics[height=5.1cm]{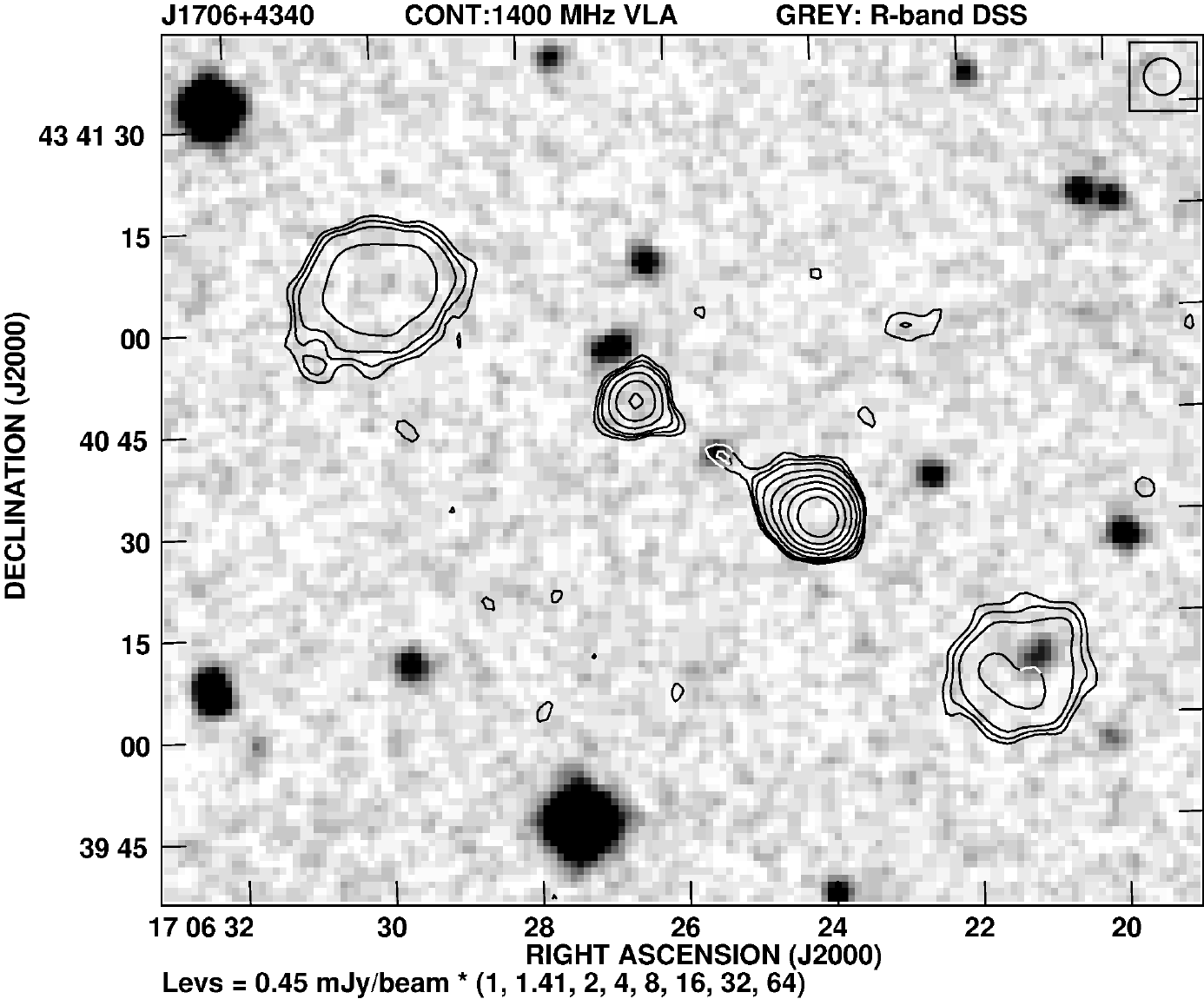} 
    \hspace{1cm}\includegraphics[height=5.1cm]{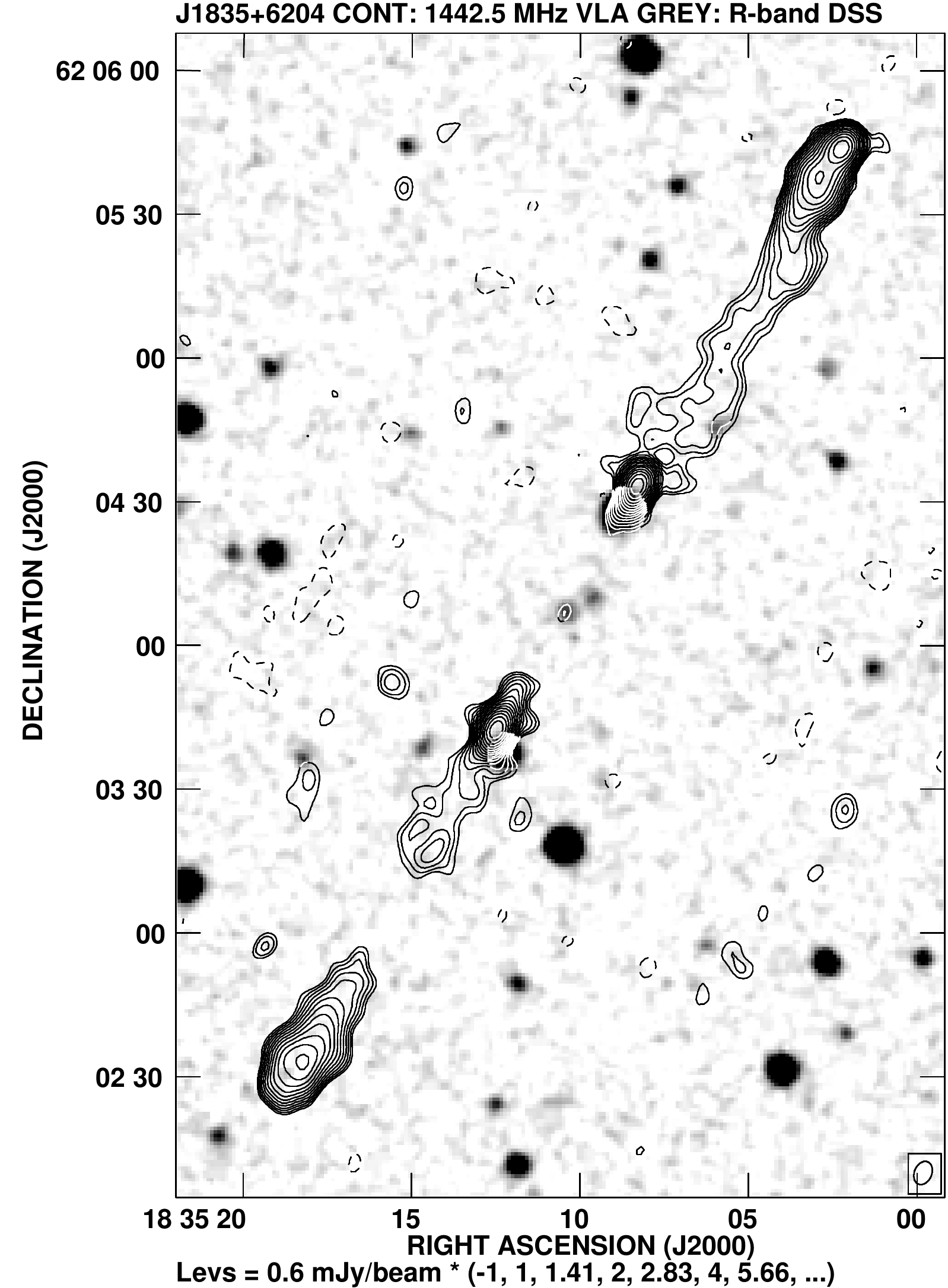} 
    \hspace{1cm}\includegraphics[height=5.1cm]{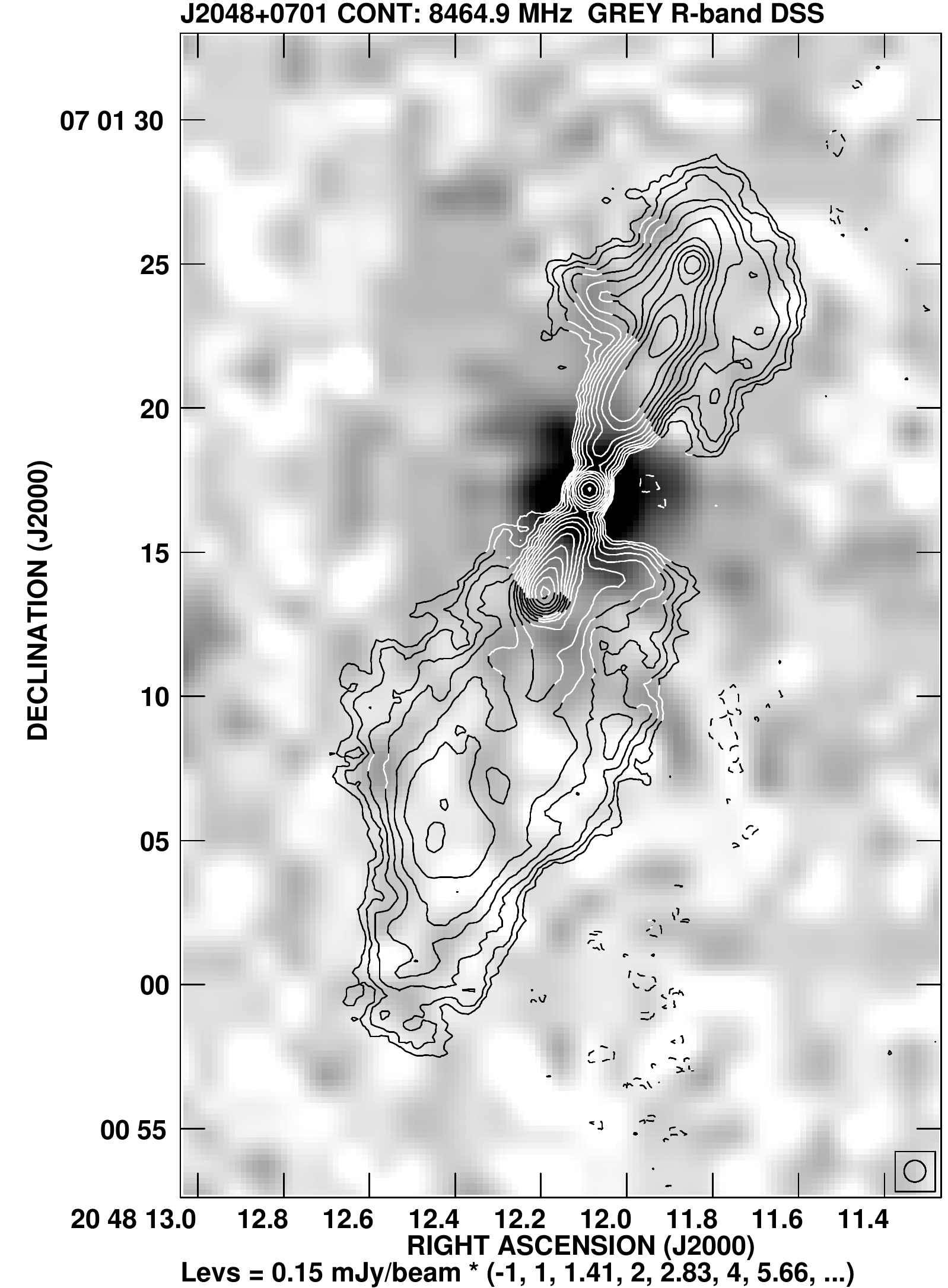}\\
\vspace{0.4cm} 
    \includegraphics[width=7cm]{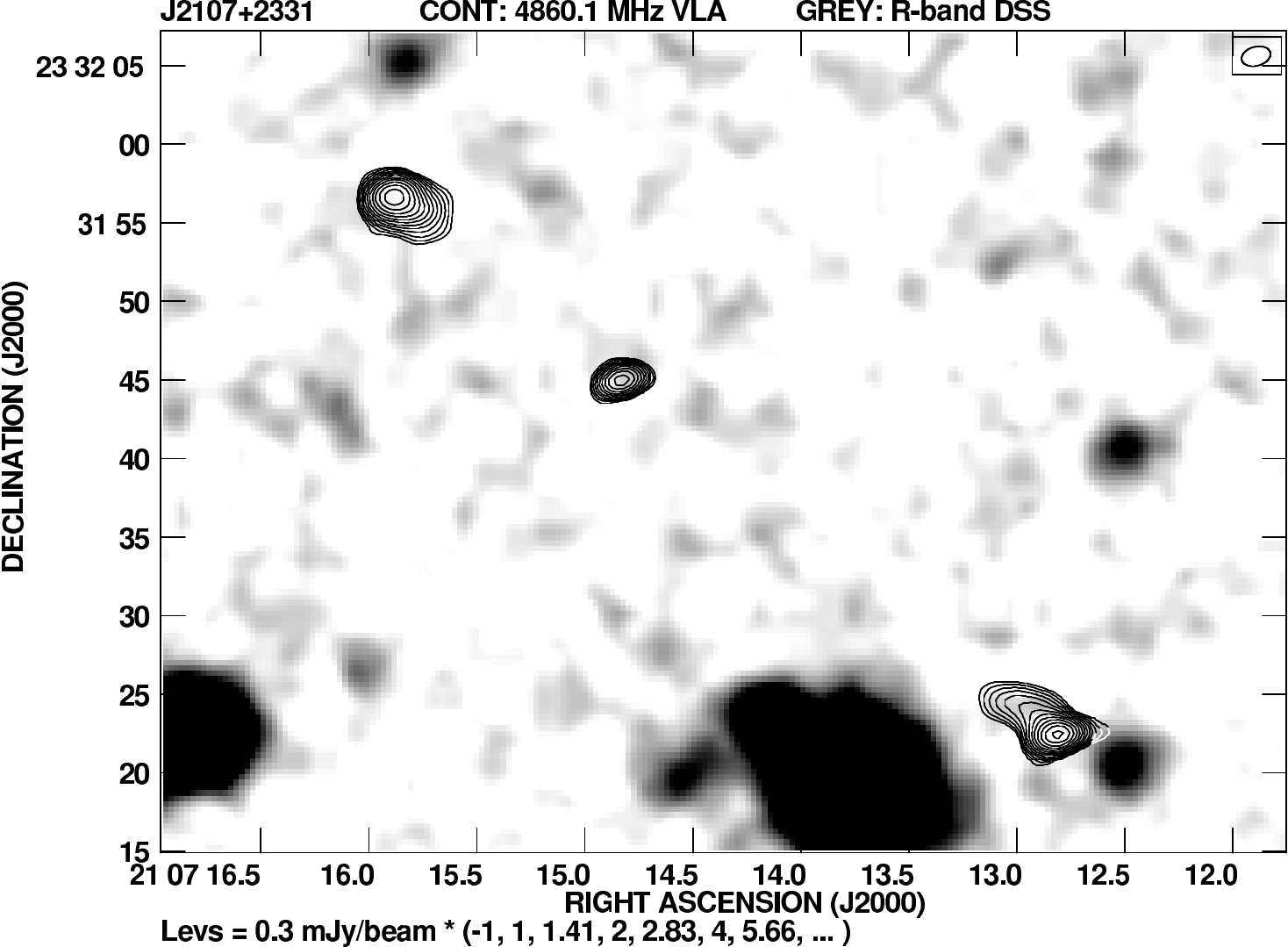}
    \hspace{1cm}\includegraphics[height=5.1cm]{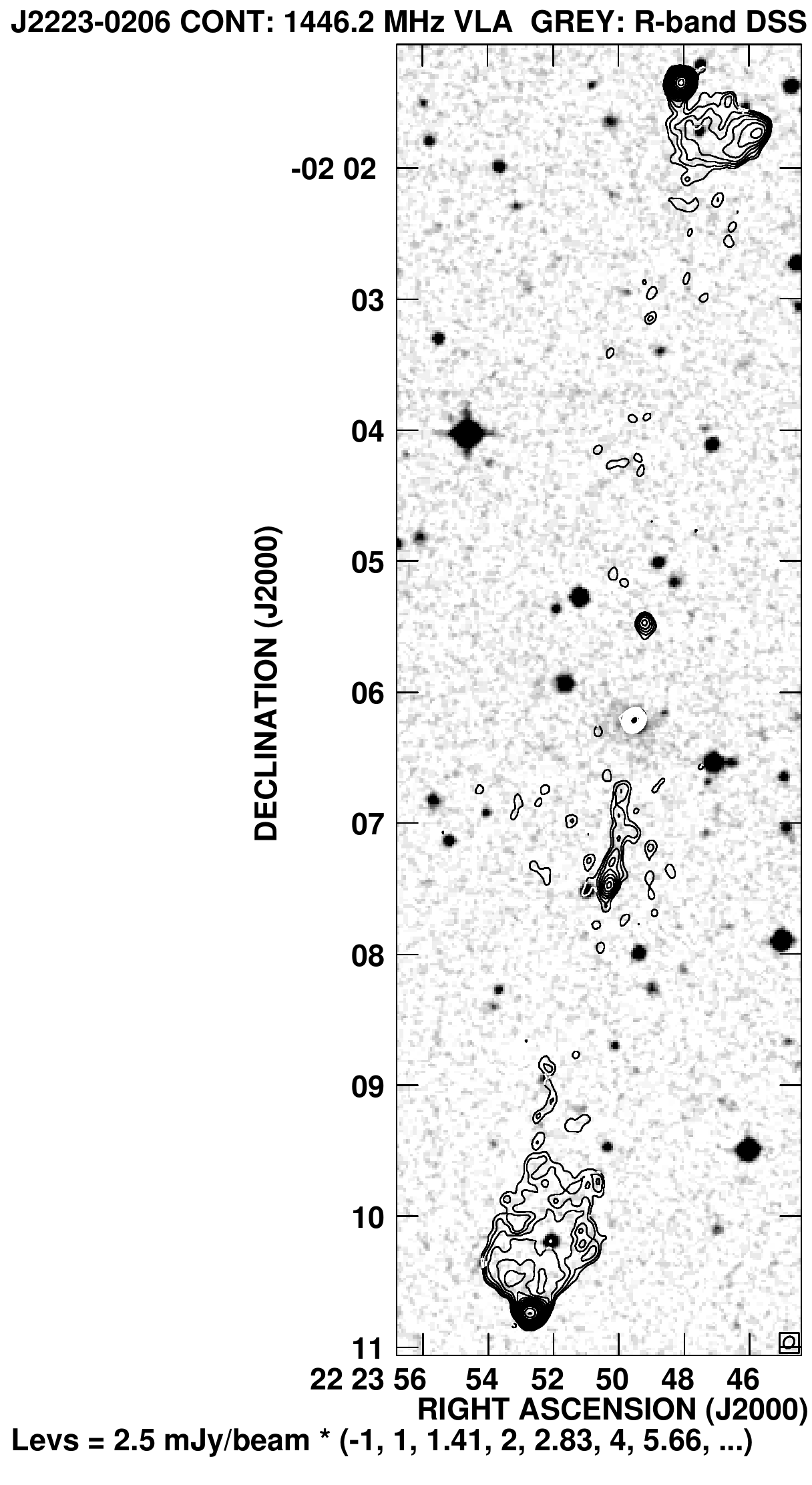} 
    \hspace{1cm}\includegraphics[height=5.7cm]{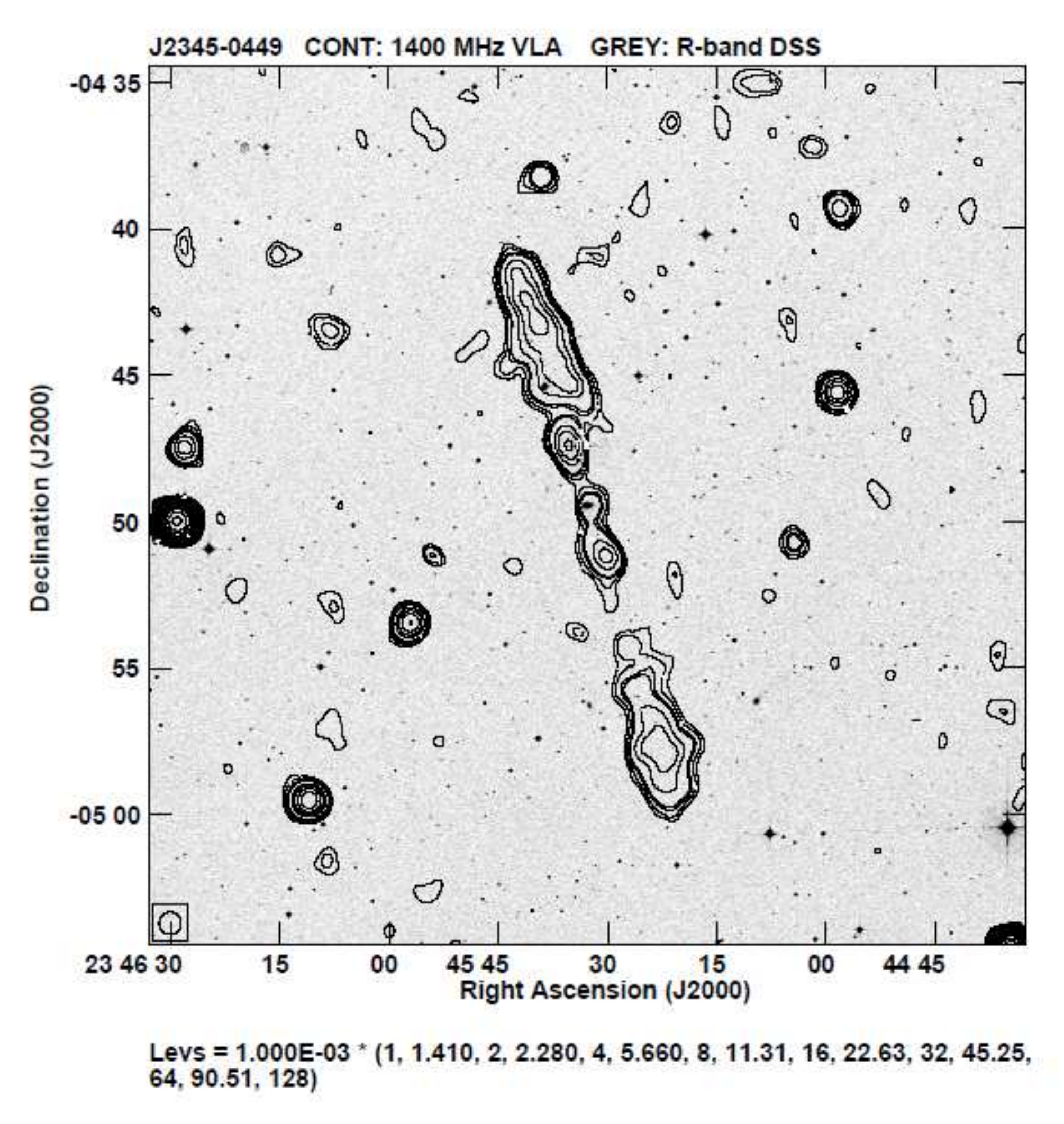} 
\end{figure}
\end{onecolumn}
\end{onecolumn}
\begin{onecolumn}
\newpage
\section{Radio maps of restarting radio sources -- class B}
This appendix presents the radio maps of restarting radio sources with prominent inner structures surrounded by diffuse outer structures -- class B. All of the radio maps presented here were obtained with the Very Large Array at 1.4 GHz. The map of J0301$+$3512 was taken from \citet{Shulevski2012}. In all the maps radio contours are overlaid onto the R-band optical image from Digital Sky Survey (DSS).  
\begin{figure}
    \includegraphics[height=4.7cm]{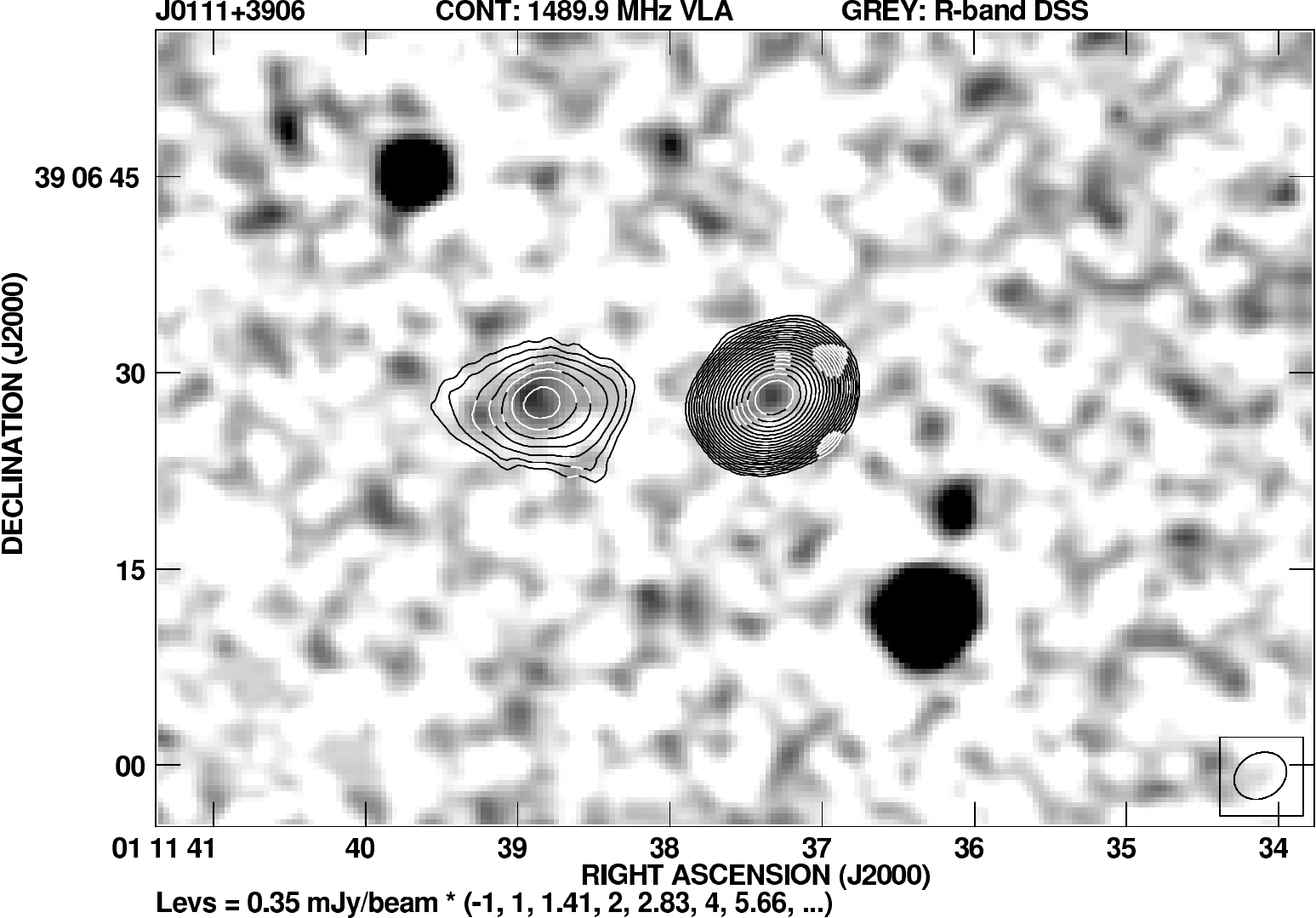} 
    \includegraphics[height=5.3cm]{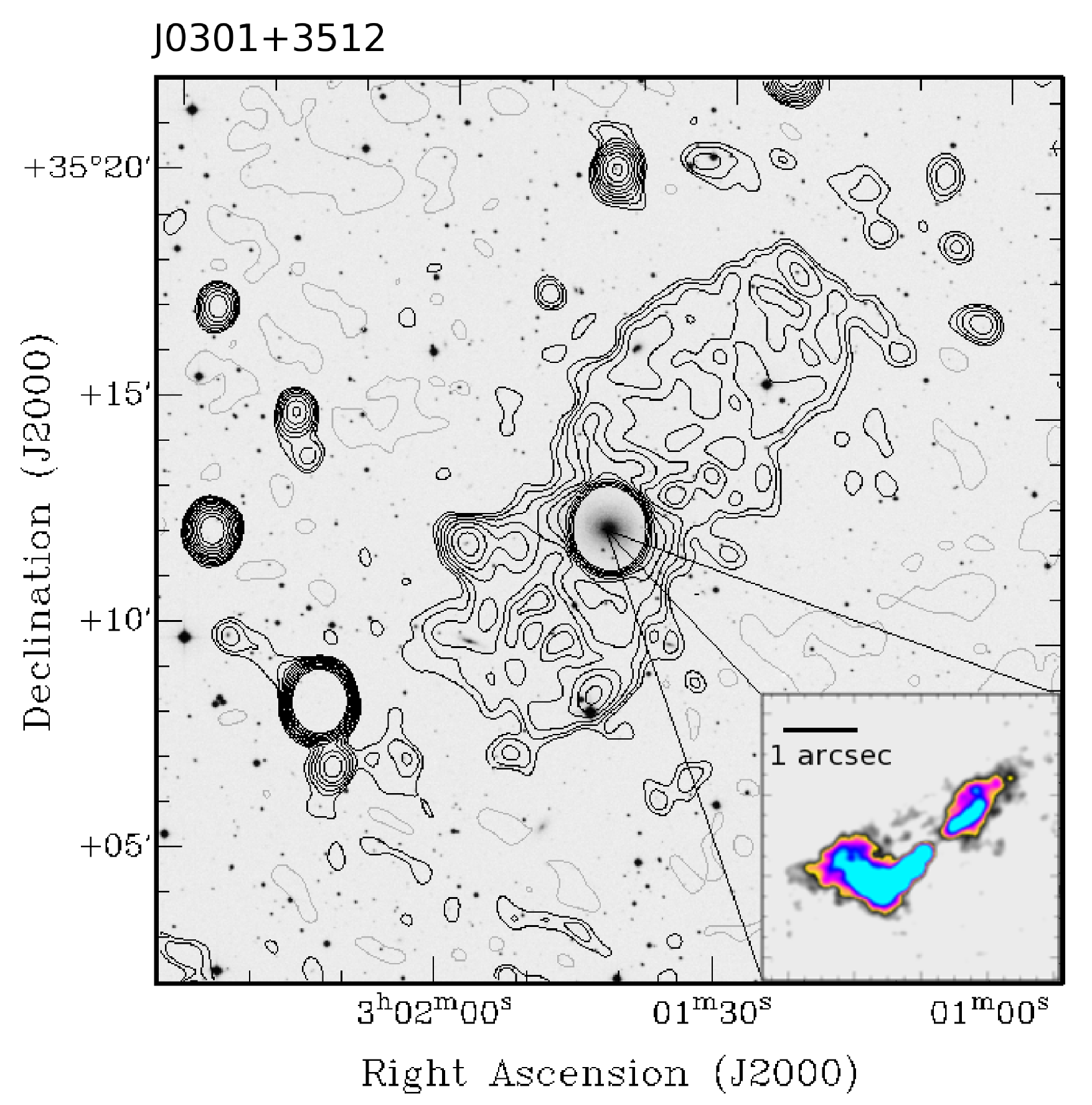} 
    \includegraphics[height=5cm]{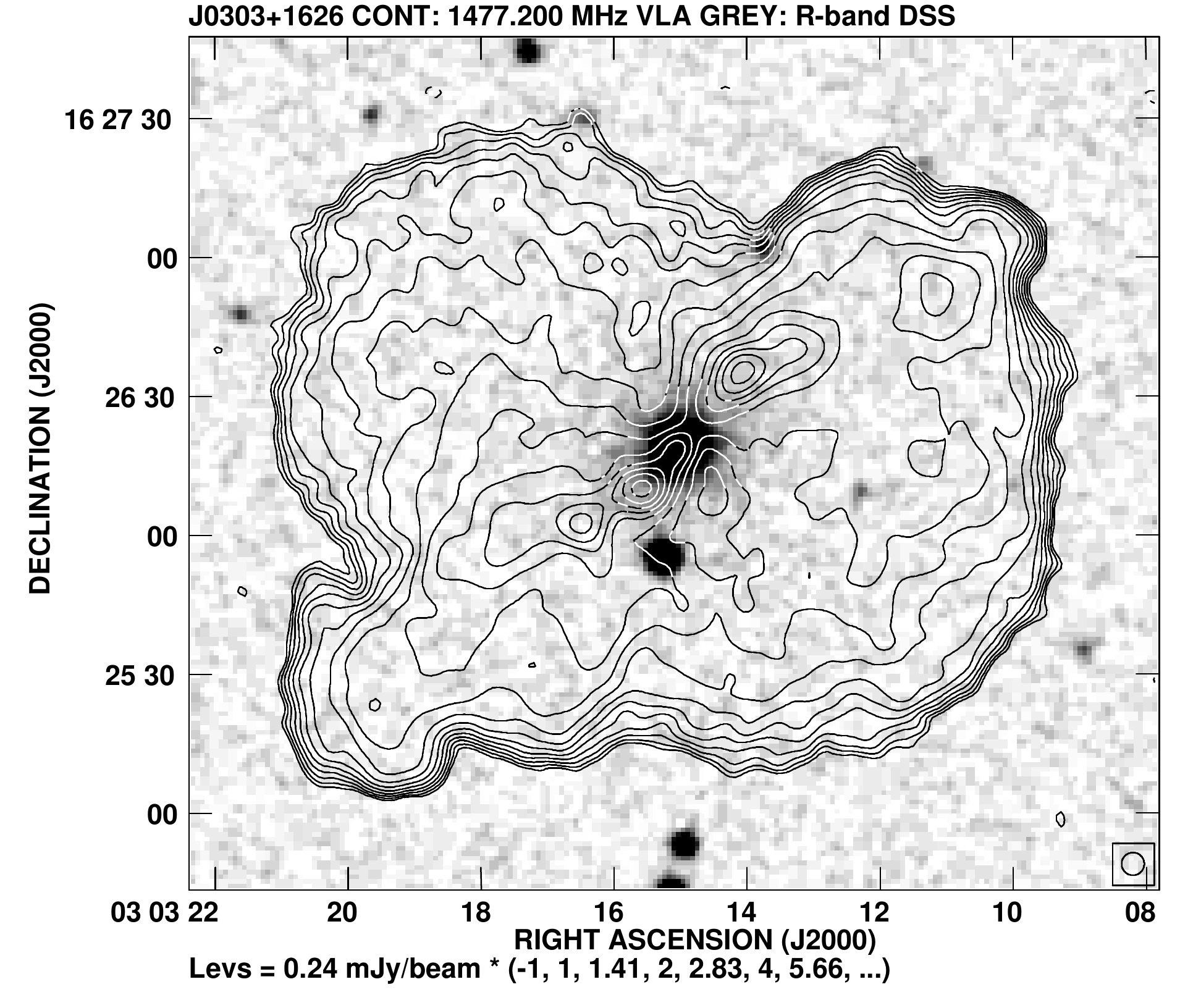}\\
\vspace{0.4cm}
    \includegraphics[height=5.5cm]{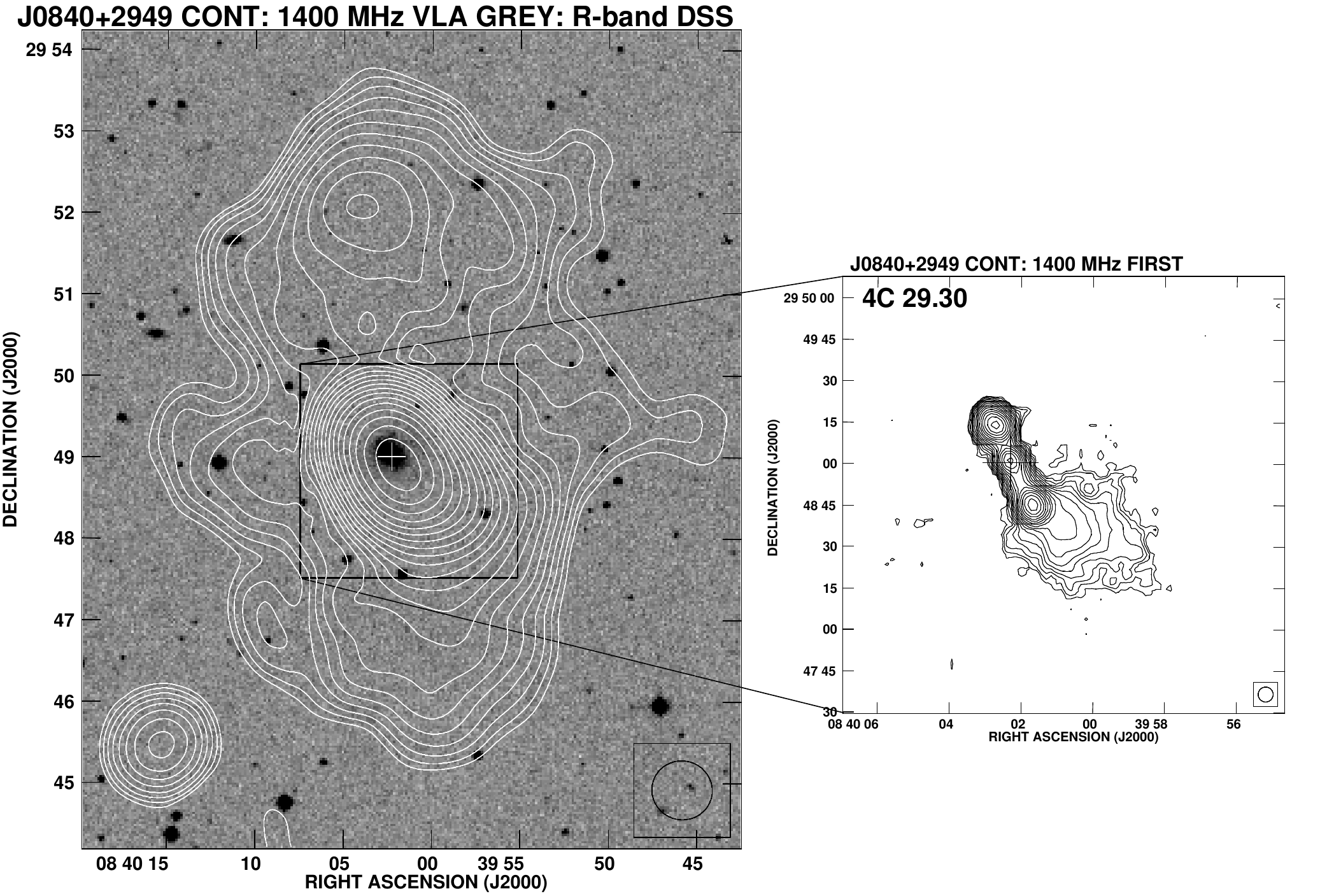}
    \hspace{0.5cm}\includegraphics[height=5cm]{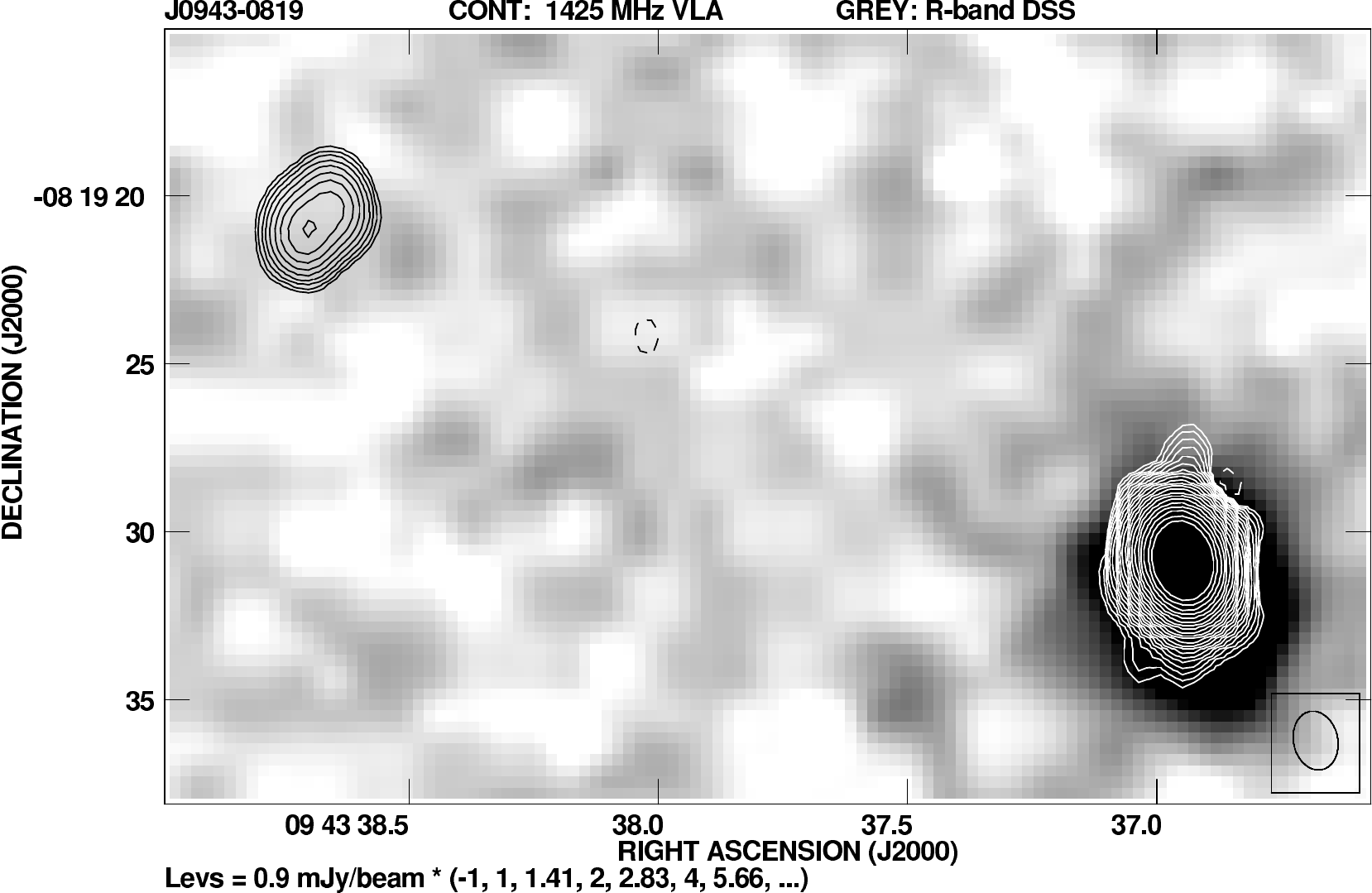}\\
\vspace{0.4cm}
    \includegraphics[height=5.1cm]{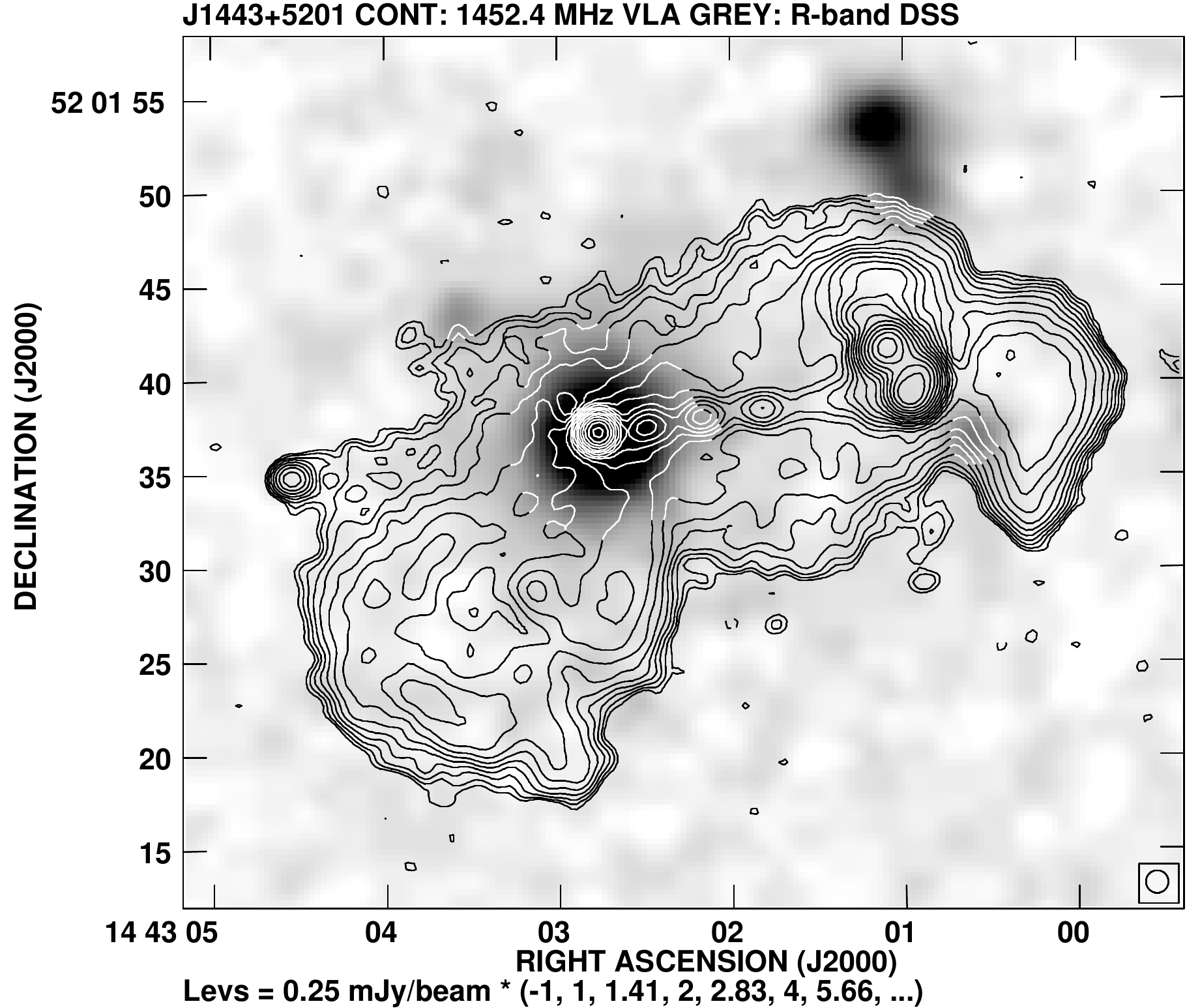}
    \hspace{0.5cm}\includegraphics[height=5.1cm]{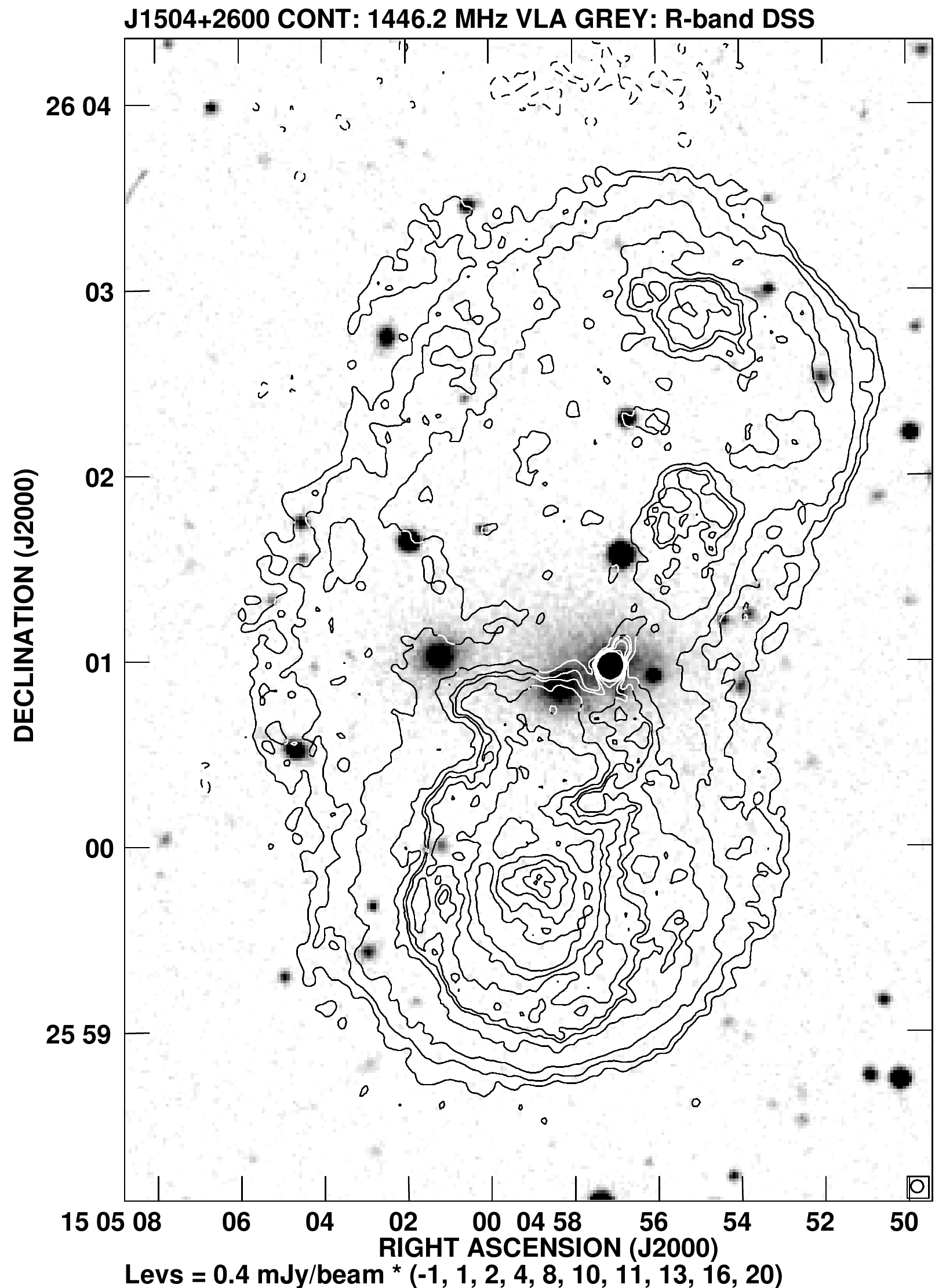}
    \hspace{0.5cm}\includegraphics[height=5.5cm]{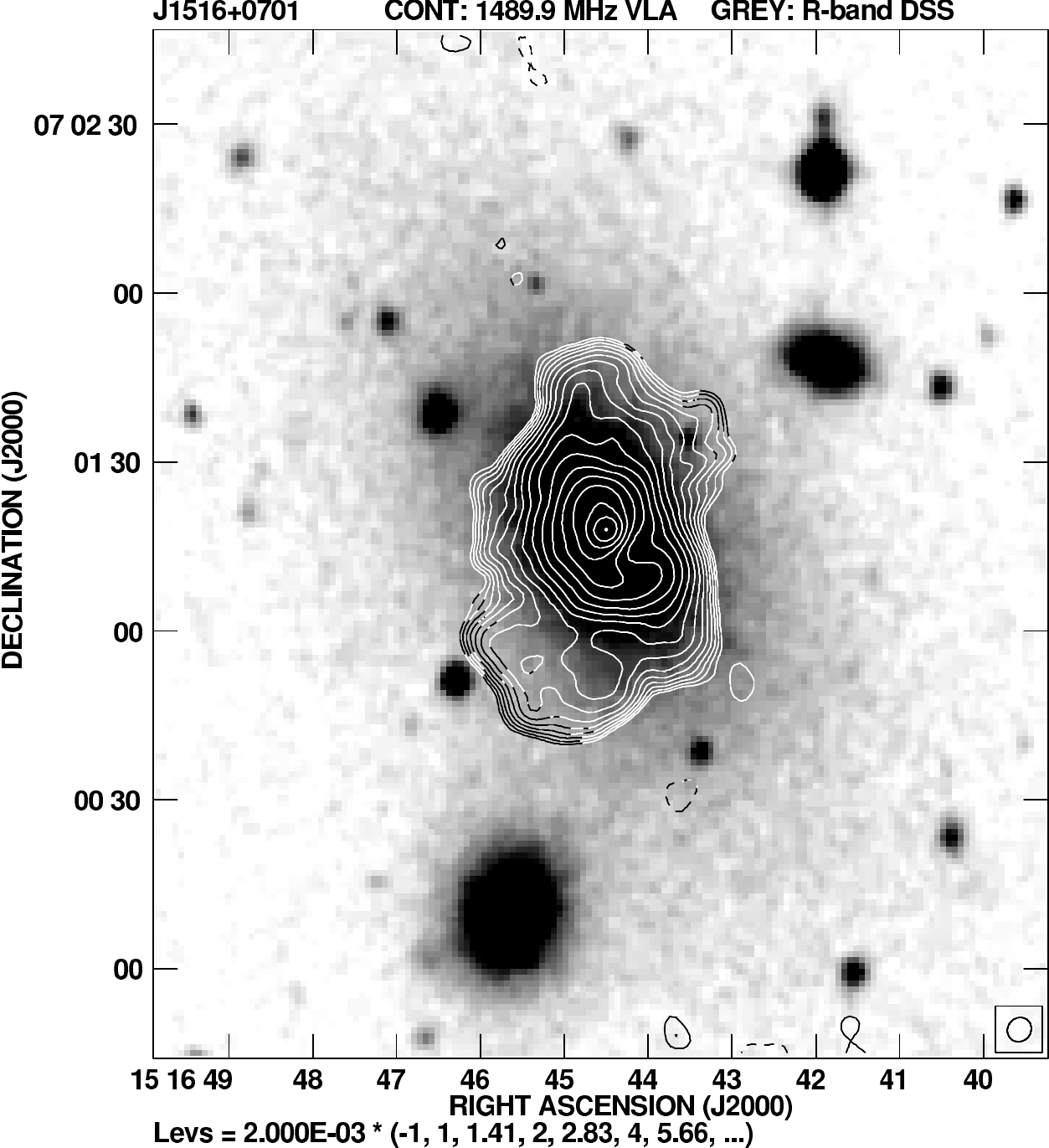}\\
\vspace{0.4cm}
   \includegraphics[height=5.5cm]{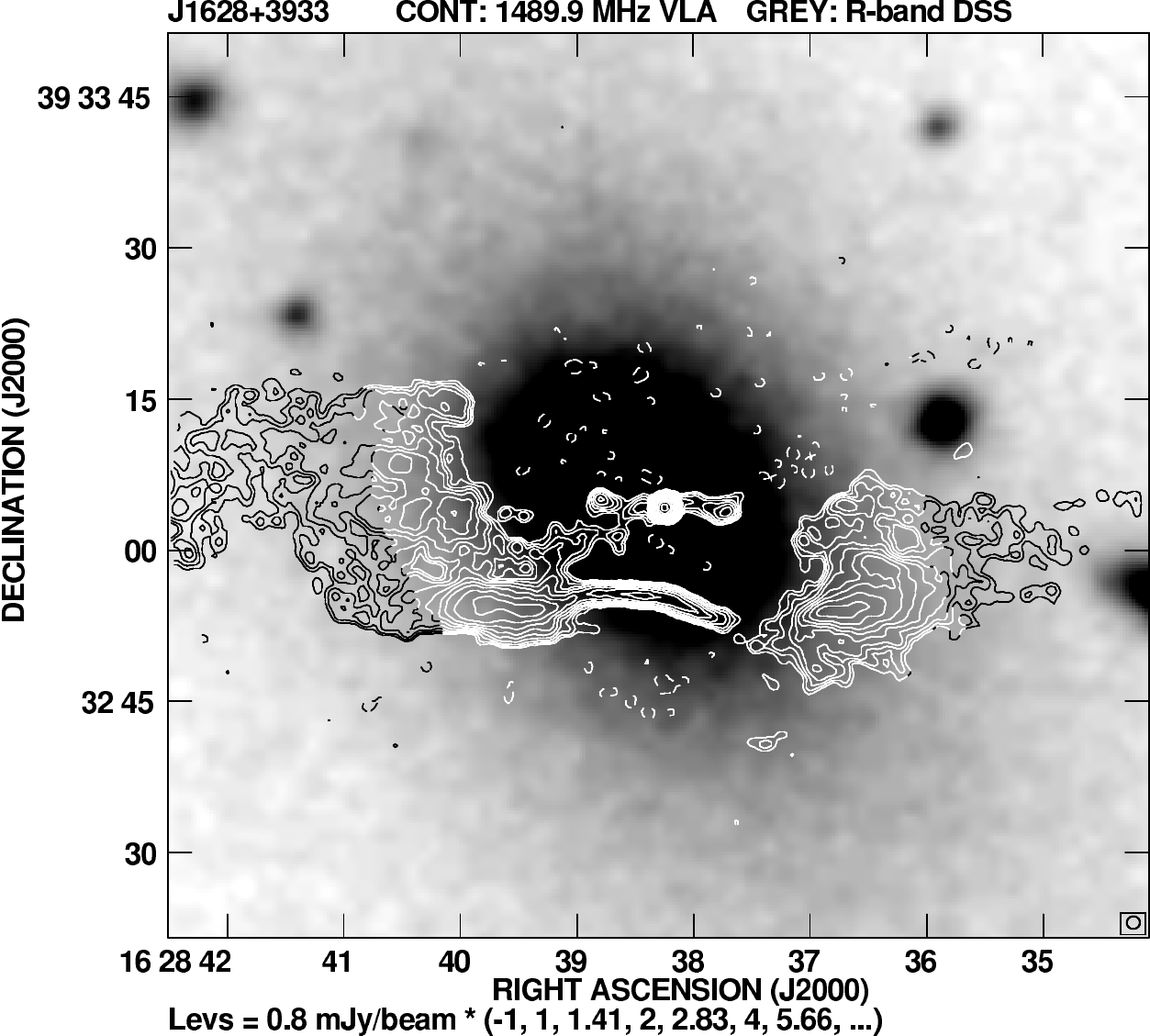}
   \includegraphics[height=5.1cm]{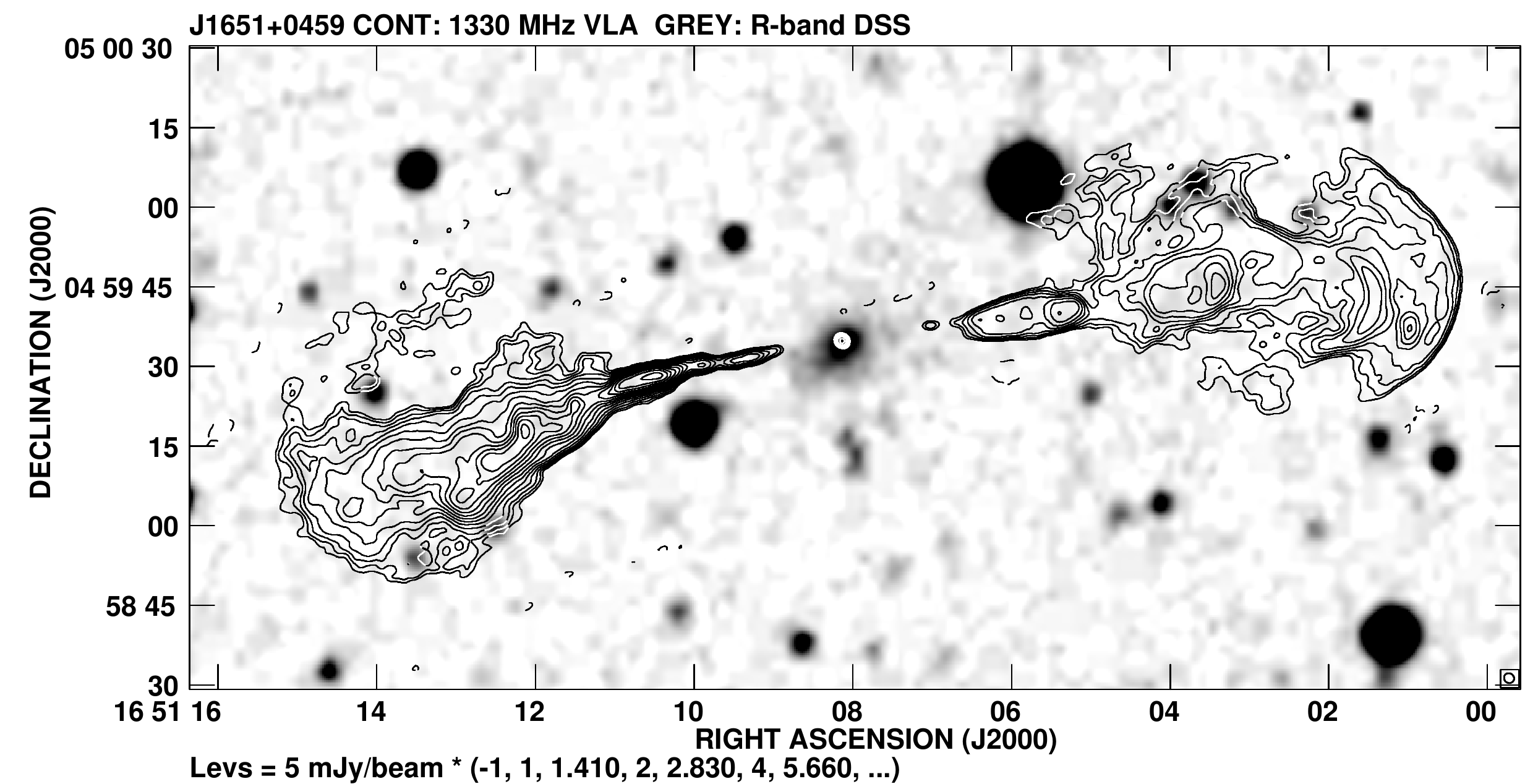} \\
\end{figure}
\begin{figure}
    \includegraphics[height=5.5cm]{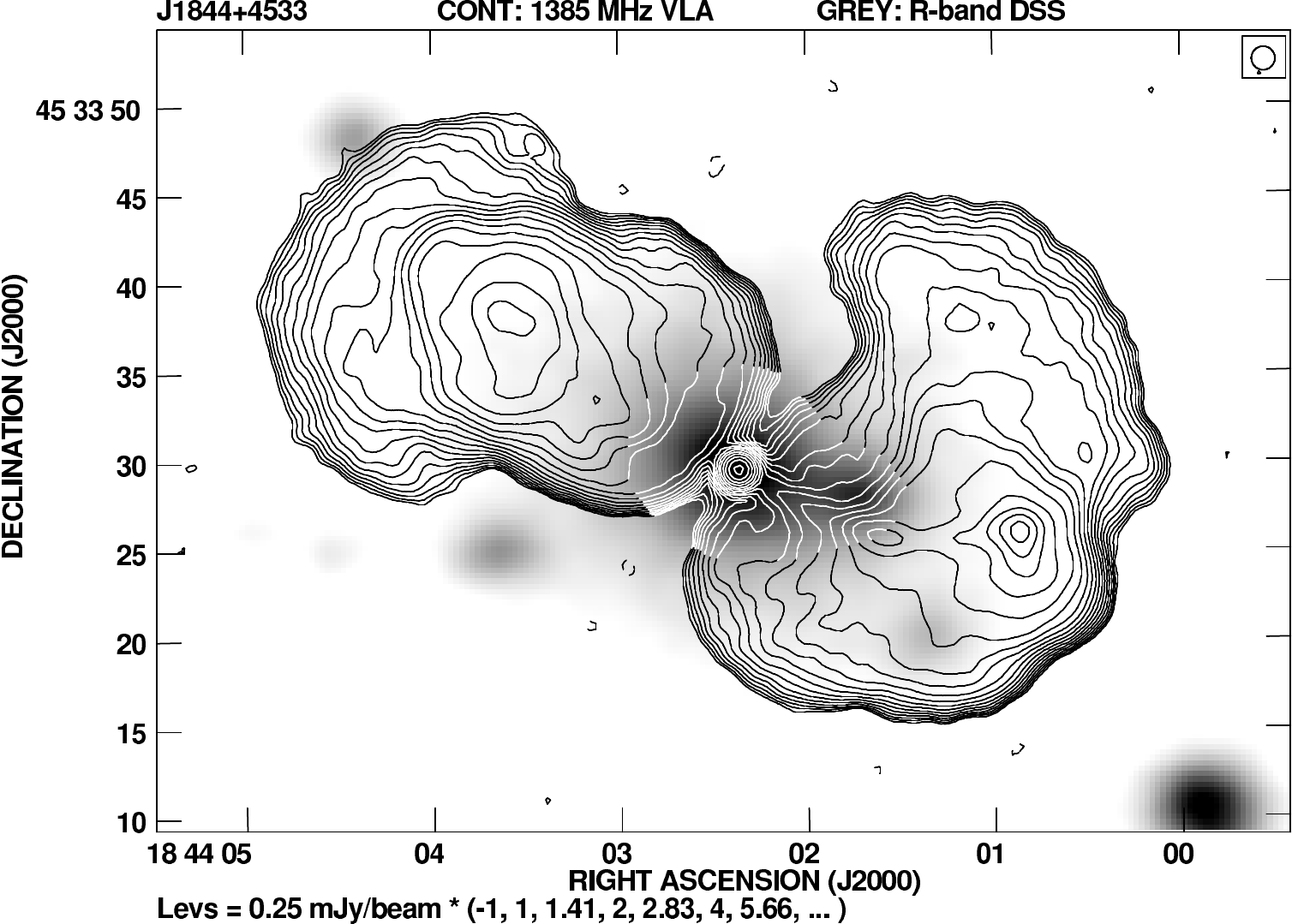}
\end{figure}

\end{onecolumn}

\bsp	
\label{lastpage}
\end{document}